\documentclass[bibyear]{aa} 
\pdfoutput=1
\usepackage{natbib}
\usepackage{txfonts}
\usepackage{verbatim} 
\usepackage{rotating} 
\usepackage{graphicx}
\usepackage[utf8]{inputenc}

\def\f#1   {Fig.~\ref{#1}}
\def\s#1   {Sect.~\ref{#1}}
\def\tab#1   {Table~\ref{#1}}
\def\t#1   {Table~\ref{#1}}

\def\comm#1   {{\tt (COMMENT: #1) }}

\def\smo               {Smol\v{c}i\'{c}}

\def\meanrms   {2.3~$\mu$Jy~beam$^{-1}$} 
\def\ncomp   {10,899}
\def\nsource   {10,830}
\def\nmult  {67}
\def\jybeam {Jy~beam$^{-1}$}
\def\ccor   {completeness and bias}
\def\Ccor   {Completeness and bias}

\begin{document}

\authorrunning{Smol\v{c}i\'{c} et al.}
\titlerunning{VLA-COSMOS 3~GHz Large Project}

\title{The VLA-COSMOS 3~GHz Large Project: \\ Continuum data and source catalog release}

\author{
        V.~Smol\v{c}i\'{c}\inst{1}, 
        M.~Novak\inst{1}, 
        M.~Bondi\inst{2},
        P.~Ciliegi\inst{3},
        K.~P.~Mooley\inst{4},
        E.~Schinnerer\inst{5},
G.~Zamorani\inst{3},
        F.~Navarrete\inst{6}, 
                S.~Bourke\inst{4},
        A.~Karim\inst{6},
        E.~Vardoulaki\inst{6},
        S.~Leslie\inst{5},
J.~Delhaize\inst{1},
C.~L.~Carilli\inst{7},
S.~T.~Myers\inst{7},
N.~Baran\inst{1},
I.~Delvecchio\inst{1},
O.~Miettinen\inst{1},
J.~Banfield\inst{8,9},
M.~Balokovi\'c\inst{4},
F.~Bertoldi\inst{6},
P.~Capak\inst{10},
D.~A.~Frail\inst{7}, 
G.~Hallinan\inst{4},
H.~Hao\inst{11},
N.~Herrera~Ruiz\inst{12},  
A.~Horesh\inst{13}, 
O.~Ilbert\inst{14},
H.~Intema\inst{7},
V.~Jeli\'c\inst{15,16,17},
H-R.~Kl\"ockner\inst{18,19},
J.~Krpan\inst{1},
S.~R.~Kulkarni\inst{4}, 
H.~McCracken\inst{20},
C.~Laigle\inst{19},
E.~Middleberg\inst{12},
E.~J.~Murphy\inst{21},
M.~Sargent\inst{22},
N.~Z.~Scoville\inst{4},
K.~Sheth\inst{21}
                }
\institute{
Department of Physics, Faculty of Science, University of Zagreb,  Bijeni\v{c}ka cesta 32, 10000  Zagreb, Croatia
\and
Istituto di Radioastronomia di Bologna - INAF, via P. Gobetti, 101, 40129, Bologna, Italy 
\and
NAF-Osservatorio  Astronomico di Bologna, Via Ranzani 1, I - 40127 Bologna, Italy
\and
California Institute of Technology, MC 249-17, 1200 East California Boulevard, Pasadena, CA 91125 
\and
Max-Planck-Institut f$\ddot{u}$r Astronomie, K$\ddot{o}$nigstuhl 17, D-69117 Heidelberg, Germany
\and
Argelander Institut for Astronomy, Auf dem H\"{u}gel 71, Bonn, 53121, Germany
\and
National Radio Astronomy Observatory, P.O. Box 0, Socorro, NM 87801, USA
\and
CSIRO Australia Telescope National Facility, PO Box 76, Epping, NSW 1710, Australia
\and
Research School of Astronomy and Astrophysics, Australian National University, Weston Creek, ACT 2611, Australia
\and
Spitzer Science Center, 314-6 Caltech, Pasadena, CA 91125, USA
\and
Smithsonian Astrophysical Observatory, 60 Garden St, Cambridge, MA 02138, USA
\and
Astronomisches Institut, Ruhr-Universit\"{a}t Bochum, Universit\"{a}tsstr. 150, 44801, Bochum, Germany
\and
Benoziyo Center for Astrophysics, Weizmann Institute of Science, 76100 Rehovot, Israel
\and
Aix Marseille Universit\'{e}, CNRS, LAM (Laboratoire d'Astrophysique de Marseille), UMR 7326, 13388, Marseille, France
\and
Kapteyn Astronomical Institute, University of Groningen, PO Box 800, 9700 AV, Groningen, The Netherlands
\and
ASTRON - The Netherlands Institute for Radio Astronomy, PO Box 2, 7990 AA, Dwingeloo, The Netherlands
\and
Ru{\dj}er Bo\v{s}kovi\'{c} Institute, Bijeni\v{c}ka cesta 54, 10000 Zagreb, Croatia
\and
Subdepartment of Astrophysics, University of Oxford, Denys-Wilkinson Building, Keble Road, Oxford OX1 3RH, UK
\and
Max-Planck-Institut f\:{u}r Radioastronomie, Auf dem H:{u}gel 69, D-53121 Bonn, Germany
\and
Institut d'Astrophysique de Paris, UMR7095 CNRS, Universit\;{e} Pierre et Marie Curie, 98 bis Boulevard Arago, 75014, Paris, France
\and
National Radio Astronomy Observatory, 520 Edgemont Road, Charlottesville, VA 22903, USA
\and
Astronomy Centre, Department of Physics and Astronomy, University of Sussex, Brighton, BN1 9QH, UK
}

   \date{Received ; accepted}

\abstract{
We present the VLA-COSMOS 3~GHz Large Project based on 384~hours of observations with the Karl G. Jansky Very Large Array (VLA) at 3~GHz (10~cm) toward the two square degree Cosmic Evolution Survey (COSMOS) field. 
The final mosaic reaches a median $rms$ of \meanrms \ over the two square degrees at an angular resolution of $0.75\arcsec$.
To fully account for the spectral shape and resolution variations across the broad (2~GHz) band, we image all data with a multiscale, multifrequency synthesis algorithm.
We present a catalog of \nsource \ radio sources down to $5\sigma$, out of which \nmult \ are combined from multiple components.
Comparing the positions of our 3~GHz sources with those from the Very Long Baseline Array (VLBA)-COSMOS survey, we estimate that the astrometry is accurate to $0.01\arcsec$ at the bright end (signal-to-noise ratio, S/N$_\mathrm{3GHz}>20$). 
Survival analysis on our data combined with the VLA-COSMOS 1.4~GHz Joint Project catalog yields an expected median radio spectral index of $\alpha=-0.7$.
We compute completeness corrections via Monte Carlo simulations to derive the corrected 3~GHz source counts. Our counts are in agreement with previously derived 3~GHz counts based on single-pointing ($0.087$~square degrees) VLA data.
In summary, the VLA-COSMOS 3~GHz Large Project  simultaneously provides  the largest and deepest radio continuum survey at high ($0.75\arcsec$) angular resolution to
date, bridging the gap between last-generation and next-generation surveys. 
}


\maketitle

\section{Introduction}
\label{sec:introduction}

One of the main quests in  modern cosmology is understanding  the formation of galaxies
and their evolution through cosmic time. In the past decade it has been demonstrated
that a panchromatic, X-ray to radio, observational approach is key to develop a consensus on galaxy formation
and evolution (e.g., \citealt{dickinson03, scoville07, driver09, driver11, koekemoer11, grogin11}). In this context, the radio regime offers an indispensable window toward star formation and
supermassive black hole properties of galaxies as radio continuum emission i) provides a dust-unbiased star formation
tracer at high angular resolution (e.g., \citealt{condon92,haarsma00,seymour08,smolcic09a, karim11}), and ii) directly probes those active galactic nuclei (AGN) that are hosted by the
most massive quiescent galaxies and deemed crucial for massive galaxy formation (e.g., \citealt{croton06,bower06,best06,evans06,hardcastle07,smolcic09b,smolcic09c,smolcic11,smolcic15}).

\begin{figure}[t]
\begin{center}
\includegraphics[bb= 0 0 432 432, width=\columnwidth]{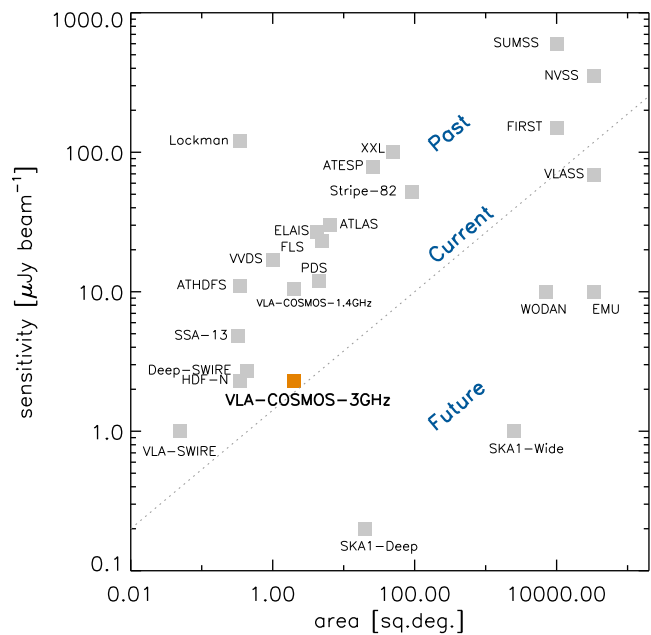}
      \caption{ 
      Sensitivity (at the observed frequency of the given survey) vs.\ area for past, current, and future radio continuum surveys. 
      }
   \label{fig:surveys}
\end{center}
\end{figure}

In recent decades, radio interferometers, such as the Karl G. Jansky Very Large Array (VLA), Australia Telescope Compact Array (ATCA), and Giant Meterwave Radio Telescope (GMRT), have surveyed fields of different sizes (ranging from tens of square arcminutes
to thousands of square degrees), depths (microjansky to Jansky), and multiwavelength coverage (e.g., \citealt{becker95, condon98, ciliegi99, georgakakis99,  bock99, prandoni01, condon03, hopkins03, schinnerer04, bondi03, bondi07, norris05, schinnerer07, schinnerer10, afonso05, tasse07, smolcic08, owen08, miller08, miller13, owen09, condon12, smolcic14, hales14}).  These past surveys have shown that deep observations at high angular resolution ($\lesssim1\arcsec$) with exquisite panchromatic coverage are critical to comprehensively study the radio properties of the main
galaxy populations, avoiding cosmic variance with large area coverage (e.g., \citealt{padovani09, padovani11a, smolcic08, smolcic09a, smolcic09b, smolcic09c, smolcic11, seymour08, bonzini12, bonzini13}). In this context, large area surveys down to unprecedented depths are planned with new and upgraded facilities (e.g., VLA, Westerbork, ASKAP, MeerKAT, and SKA; e.g., \citealt{jarvis12, norris11, norris13, norris15, prandoni15}).
Figure~\ref{fig:surveys} shows the $1\sigma$ sensitivity of each survey as a function of the area covered for past, current, and future radio continuum surveys.
The VLA-Cosmic Evolution Survey (COSMOS) 3~GHz Large Project bridges the gap between past and future radio continuum surveys by covering an area as large as two square degrees down to a sensitivity reached to date only for single pointing observations. This allows for individual detections of $>10,000$ radio sources, further building on the already extensive radio coverage of the COSMOS field at 1.4~GHz VLA (VLA-COSMOS Large, Deep and Joint projects; \citealt{schinnerer04,schinnerer07,schinnerer10}), 320~MHz VLA \citep{smolcic14}, 325~MHz and 616~MHz GMRT data (Karim et al., in prep.; Brady et al., in prep.), 6~GHz VLA (Myers et al., in prep.), and the deep multiwavelength X-ray to mm photometry   (\citealt{scoville07, koekoemor07, hasinger07, capak07, sanders07, bertoldi07, elvis09, ilbert13, mccracken12, scott08, aretxaga11, smo12, miettinen15, civano16, laigle16},  Capak et al., in prep.) and more than 97,000 optical spectroscopic redshifts (Salvato et al., in prep.; zCOSMOS, \citealt{lilly07, lilly09, trump07, prescott06, lefevre15, aihara11}; Nagao et al., priv. comm.). This further makes the survey part of one of the richest multiwavelength data sets currently available.

Radio continuum surveys at 3~GHz with the upgraded VLA are still sparse in the literature. \citet{condon12} performed single-pointing observations targeting the Lockman hole for 50-hours  on-source with the VLA in C-array configuration. The observations resulted in a confusion-limited map with an $rms$ of $1~\mu$Jy~beam$^{-1}$. 
Based on this they constrained the counts of discrete sources in the $1-10~\mu$Jy range via a $P(D)$ analysis.  A more complex $P(D)$ analysis using the same data was applied by \citet{vernstrom14} who probed the counts down to $0.1~\mu$Jy. 
Both results are qualitatively in agreement with the already well-known flattening of the radio source counts 
(normalized to the $N(S)\propto S^{-3/2}$ of a static Euclidian space)
 below flux densities of $S_\mathrm{1.4GHz}\approx1$~mJy, and a further decrease of the counts with decreasing flux density below $S_\mathrm{1.4GHz}\approx60~\mu$Jy. Such a shape of radio source counts is expected owing to the cosmic evolution of galaxy populations (e.g., \citealt{hopkins00,wilman08,bethermin12}), but this shape is contrary to that obtained based on i) the previous Lockman hole observations at 1.4~GHz \citep{owen08}, and ii) a comparison of the sky brightness temperature measured by the ARCADE~2
experiment \citep{fixsen09} with that derived from the integral of the observed radio source counts \citep{vernstrom11}. The latter results instead point 
 to a rise of the counts with decreasing flux density at these levels. To investigate this further, we here derive the radio source counts using our VLA-COSMOS 3GHz Large project data, yielding the  deepest radio counts derived to date based on direct source detections.

In Sect.~\ref{sec:obs} we describe the VLA 3~GHz observations, calibration, and imaging. We present the catalog extraction in Sect.~\ref{sec:cat}, an analysis of the radio spectral indices in Sect.~\ref{sec:alpha},  the radio source count corrections in Sect.~\ref{sec:completeness} and the radio source counts in Sect.~\ref{sec:counts}.
We summarize our products and results in Sect.~\ref{sec:summary}. We define the radio spectral index $\alpha$ as $S_\nu \propto \nu^\alpha$, where $S_\nu$ is flux density at frequency $\nu$.

\section{Observations and data reduction}
\label{sec:obs}

\subsection{Observations}
A total of 384 hours of observations toward the COSMOS field were taken in S band using the {\it S 3s full width} setup covering a bandwidth of 2048~MHz centered at 3~GHz, and separated into 16 $128$~MHz-wide spectral windows (SPWs hereafter), with full polarization, and a 3s signal-averaging time. 
The observations were taken from November 2012 to January 2013, June to August 2013, and February to May 2013 in A-array (324 hours) and C-array configurations (60 hours; Legacy ID AS1163). Sixty-four pointings, separated by $10\arcmin$ in right ascension (RA) and declination (DEC), corresponding to two-thirds of the half-power beam width (HPBW) at the central frequency of 3~GHz, were chosen to cover the full two square degree COSMOS field.  Three sets of 64 pointings in such a grid were used to achieve a uniform $rms$ over
the field; this resulted in a total of 192 pointings (shown in \f{fig:pointings} ). The
first set of pointings is nominal, the second  is shifted by $5\arcmin$ in RA
and DEC, while the third set is shifted by $-5\arcmin$ in RA.
Observing runs of 5 and 3 hours in length were conducted. In each observing run J1331+3030 was observed for flux and bandpass calibration for about 3-5 minutes on-source (J0521+166 was used only for the first day of observations) at the end of every run, J1024-0052 was observed every 30 minutes for 1m 40s on-source for gain and phase calibration, while the source J0713+4349 was observed for 5 minutes on-source at the beginning of each run for polarization leakage calibration. 
During the five hour observing runs each pointing was visited twice, 
while the order of the pointing coverage blocks during the fixed 5-hour observing blocks was changed between the different observing runs to optimize the $uv$ coverage. During the 3-hour observing blocks each pointing was visited once, and a good $uv$ coverage was assured via dynamic scheduling. 
Typically, 26 antennas were used during each observing run. The A-array configuration observations were mostly conducted under good to excellent weather conditions. 
The C-array configuration observations were partially affected by poor weather conditions (Summer thunderstorms), yielding on some days up to 30\% higher $rms$ than expected based on the VLA exposure calculator. 

\begin{figure}[t]
\begin{center}
\includegraphics[bb=  30 30 432 332, scale=0.75]{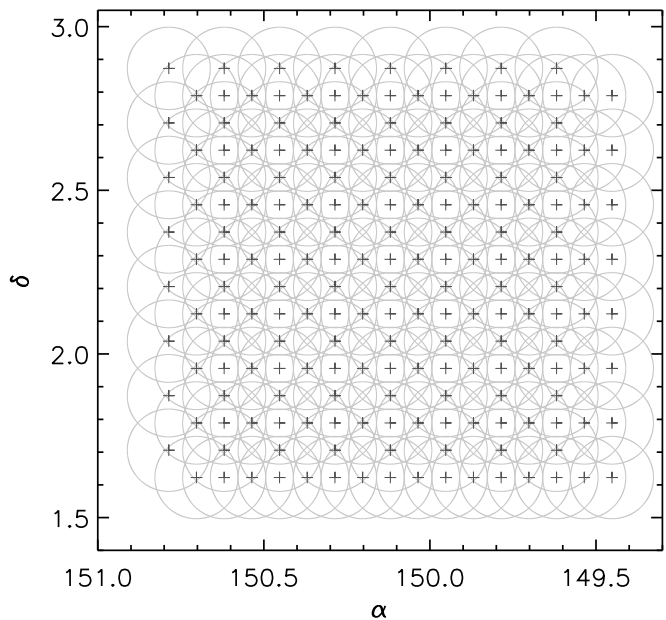}
      \caption{ \footnotesize{\baselineskip0.1cm{
      Pointing pattern used for the 3~GHz VLA-COSMOS Large Project. The centers of the 192 pointings are indicated by the plus signs. Circles indicate the primary beam of each pointing, represented here by the HPBW at 3~GHz ($15\arcmin$; the primary beam HPBW is a function of frequency and varies by a factor 2 between the lower and upper edge of the S band). 
      }}
   \label{fig:pointings}
}
\end{center}
\end{figure}

\subsection{Calibration}
\label{sec:calib}

Calibration of the data was performed via the Astronomical Image Processing System (AIPS) based data reduction pipeline AIPSLite \citep{bourke14} developed for the Caltech-National Radio Astronomy Observatory (NRAO) Stripe 82 Survey \citep{mooley16}. This pipeline was adapted for the VLA-COSMOS 3 GHz Large Project (as described below) and it follows, in general, the procedures outlined in Chapter E of the AIPS Cookbook\footnote{http://www.aips.nrao.edu/cook.html}.

\begin{figure*}[t]
\begin{center}
\includegraphics[scale=0.32]{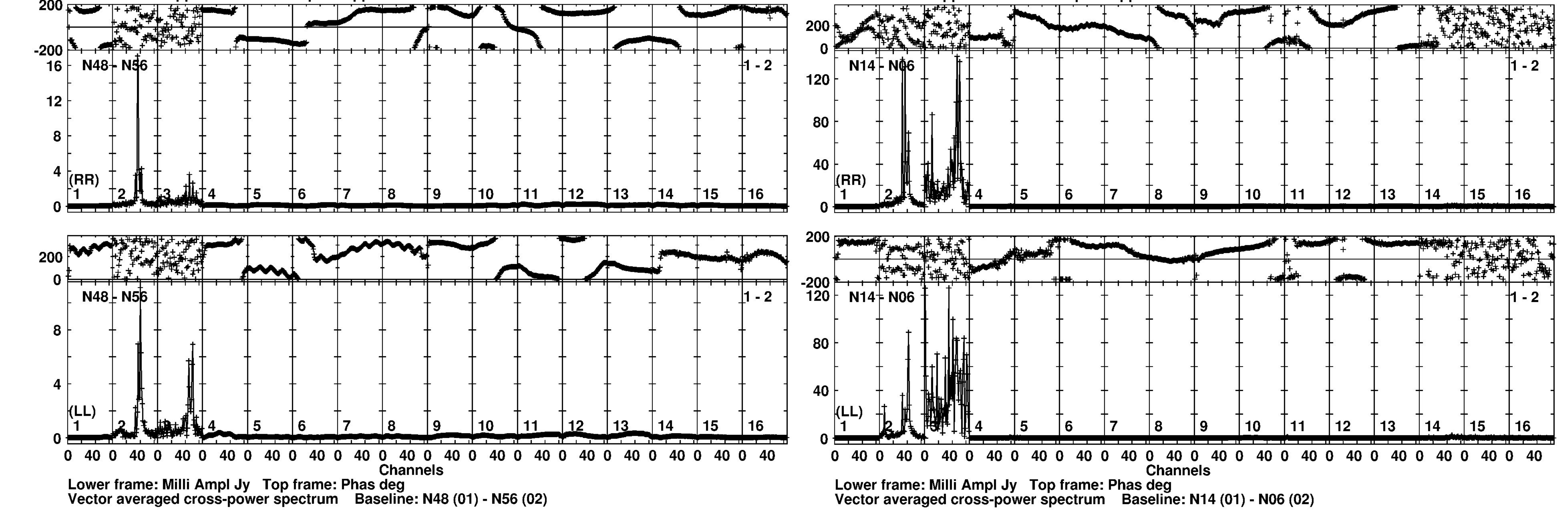}
     \caption{ 
Raw spectra of the gain calibrator source, i.e., phase vs. channel, (top frame in each of the four panels) and amplitude vs. channel (bottom frame in each of the four panels) for the right-right (RR) and left-left (LL) polarizations (top panels and bottom panels, respectively). The panels to the left are for one night of observation in the A-array configuration and the panels to the right are for a C-array observation. No calibration was applied. All baselines and all pointings of the gain calibrator source were combined to produce these plots. We note the RFI in sub-bands 2 and 3.
   \label{fig:spec-flag}
}
\end{center}
\end{figure*}

In brief, the data are first loaded with the Obit\footnote{{\tt http://www.cv.nrao.edu/\~{}bcotton/Obit.html }} task {\tt BDFIn}. Band edges, and to a larger extent IF edges, were then flagged with the task {\tt UVFLG}. SPWs 2 and 3, found to be irreparably corrupted by radio frequency interference (RFI) in all observations (see Fig.~\ref{fig:spec-flag}), were entirely flagged using the task {\tt UVFLG}. After flagging, {\tt FRING}, {\tt BPASS}, {\tt SETJY}, {\tt CALIB}, {\tt GETJY}, and {\tt CLCAL} tasks were used to derive the delay, bandpass, and complex gain solutions. Polarization calibration was performed using the tasks {\tt RLDLY}, {\tt PCAL}, and {\tt RLDIF} as detailed in Sect. 7 of Chapter E in the AIPS Cookbook. The task {\tt RFLAG} was used to flag all target pointings and the flags were applied using the {\tt UVCOP} task. The derived calibration was applied and the calibrated data set was produced with the {\tt SPLAT} task. Finally, the calibrated $uv$ data were saved to disk using the task {\tt FITTP}. During the pipeline process several diagnostic plots were generated to assess the quality of the calibration: bandpass solutions, antenna gains as a function of time, calibrated spectrum of the gain calibrator, and calibrated amplitude versus phase plots of the gain calibrator per pointing. In Table~\ref{tab:calib}\ we list the statistics for the amplitude of the phase calibrator in each SPW for all observing blocks. The average amplitude scatter around the mean is typically $~2-3$\%, with the exception of the highest frequency SPWs, for which it is higher than 10\%. 
The combined typical scatter around the mean is $\sim5$\%. This assures a good flux calibration.

The highest frequency SPWs marked 14, 15, and 16 have low amplitude RFI and the phases are significantly affected for some observations. The C-array configuration data at the upper end of the S band are mostly unusable due to this RFI. These data have been manually flagged, and we additionally ran the {\tt RFLAG} task on the rest of the C-array configuration data to further remove bad data and extend flags in frequency and time. The A-array configuration data for these SPWs are generally good. Our imaging tests show that the data from these SPWs generally improves the sensitivity, but limits the dynamic range for certain pointings. Looking at the overall imaging performance, we decided to retain these SPWs. Despite the data drop-outs, the median flux density values of the phase calibrator (J1024-0052; Table~\ref{tab:calib}) are consistent with the  spectral parameters inferred from the other SPWs. 
Through our tests we find that, in the majority of observations, RFI adversely affects the system temperature measurements, and hence we have left out the correction for the system temperature  from the calibration process. 

\begin{table}
\caption{Amplitude of the phase calibrator (J1024-0052) in each SPW for all observing blocks}
\centering
\small
\label{tab:calib}
\begin{tabular}{c c c c c }
\hline
SPW &  Frequency &  Mean flux &  Median flux &  Standard  \\
& (GHz)   & density (Jy) & density (Jy) & deviation (Jy)   \\
\hline
  1   & 2.060  & 0.739  & 0.735  & 0.029 \\
  4   & 2.444  & 0.707  & 0.704  & 0.025 \\
  5   & 2.572  & 0.700  & 0.696  & 0.023 \\
  6   & 2.700  & 0.684  & 0.680  & 0.023 \\
  7   & 2.828  & 0.668  & 0.665  & 0.024 \\
  8   & 2.956  & 0.652  & 0.648  & 0.026 \\
  9   & 3.084  & 0.645  & 0.642  & 0.020 \\
  10   & 3.212  & 0.635  & 0.632  & 0.020 \\
 11  &  3.340  & 0.625  & 0.622  & 0.021 \\
 12   & 3.468  & 0.615  & 0.611  & 0.020 \\
 13   & 3.596  & 0.603  & 0.600  & 0.021 \\
 14   & 3.724  & 0.539  & 0.579  & 0.129 \\
 15   & 3.852  & 0.525  & 0.569  & 0.153 \\
 16   & 3.980  & 0.535  & 0.566  & 0.124 \\
\hline 
\end{tabular} 
\end{table}

At this point the pipeline diverges in two directions to:
i) image the target fields and 
ii) produce and export a calibrated data set in preparation for mosaicking.
To image the target fields, they were split out with
calibration applied (using the task {\tt SPLIT}). The fields were then further auto-flagged (using the task {\tt RFLAG}), imaged
(using the task {\tt IMAGR}), and exported (using the task {\tt FITTP}) in parallel.
The calibrated data set was generated by applying {\tt RFLAG} and imaging the target
fields, including applying flags (using the task {\tt UVCOP}), calibration (using the task {\tt SPLAT}), and exporting the $uv$ data and maps (using the task {\tt FITTP}).

The pipeline performance and output were tested by i) manually reducing
separate blocks of VLA-COSMOS observations and comparing the results with
the pipeline output, and ii) comparing the output to the Common Astronomy Software Applications (CASA\footnote{CASA is developed by an international consortium of scientists
based at the NRAO, the European Southern Observatory (ESO), the National
Astronomical Observatory of Japan (NAOJ), the CSIRO Australia
Telescope National Facility (CSIRO/ATNF), and the Netherlands Institute
for Radio Astronomy (ASTRON) under the guidance of NRAO.
See http://casa.nrao.edu;  \citep{mcmullin07}})-based NRAO reduction pipeline for randomly selected data taken in the A- and C-array configurations. No obvious differences were found. As the pipeline used here was tailored specifically to the COSMOS field (e.g., it includes polarization calibration), after this verification it was further applied to the remaining VLA-COSMOS data sets. 

The calibrated $uv$ data sets output by the pipeline for each observing block were first run through the AIPS task {\tt UVFIX} to assure accurately computed positions. We note that applying {\tt UVFIX} at the end of calibration has the same effect as applying it at the beginning of calibration. They were then further processed in CASA by clipping each calibrated $uv$ data set in amplitude (above 0.4~Jy) using the task {\tt FLAGDATA}\footnote{In total, about 30-35\% of the data were flagged (using the tasks {\tt RFLAG} and {\tt FLAGDATA}).}, splitting the individual pointings using the task {\tt SPLIT}, and concatenating all existing observations of the same pointing using the task {\tt CONCAT}.  The concatenated $(u,v)$ data for each pointing were then imaged prior to being combined into the final mosaic, as described in detail in the next section.

\subsection{Self-calibration, imaging, and mosaicking}
\label{sec:imaging}

To image our data we used the multiscale multifrequency synthesis (MSMF) algorithm developed by \citet{rau11} and implemented in CASA. This method uses the entire 2~GHz bandwidth at once to calculate the monochromatic flux density at 3~GHz and a spectral index between 2 and 4~GHz. 
After extensive testing of various imaging methods (see \citealt{novak15}) we settled for the MSMF method as it allows for a combination of the best possible resolution, $rms$, and image quality. 
Because of the large data volume, joint deconvolution was not practical and we imaged each pointing individually and then combined them into a mosaic in the image plane. 

\begin{figure}[t]
\centering
\includegraphics[ width=0.4\columnwidth]{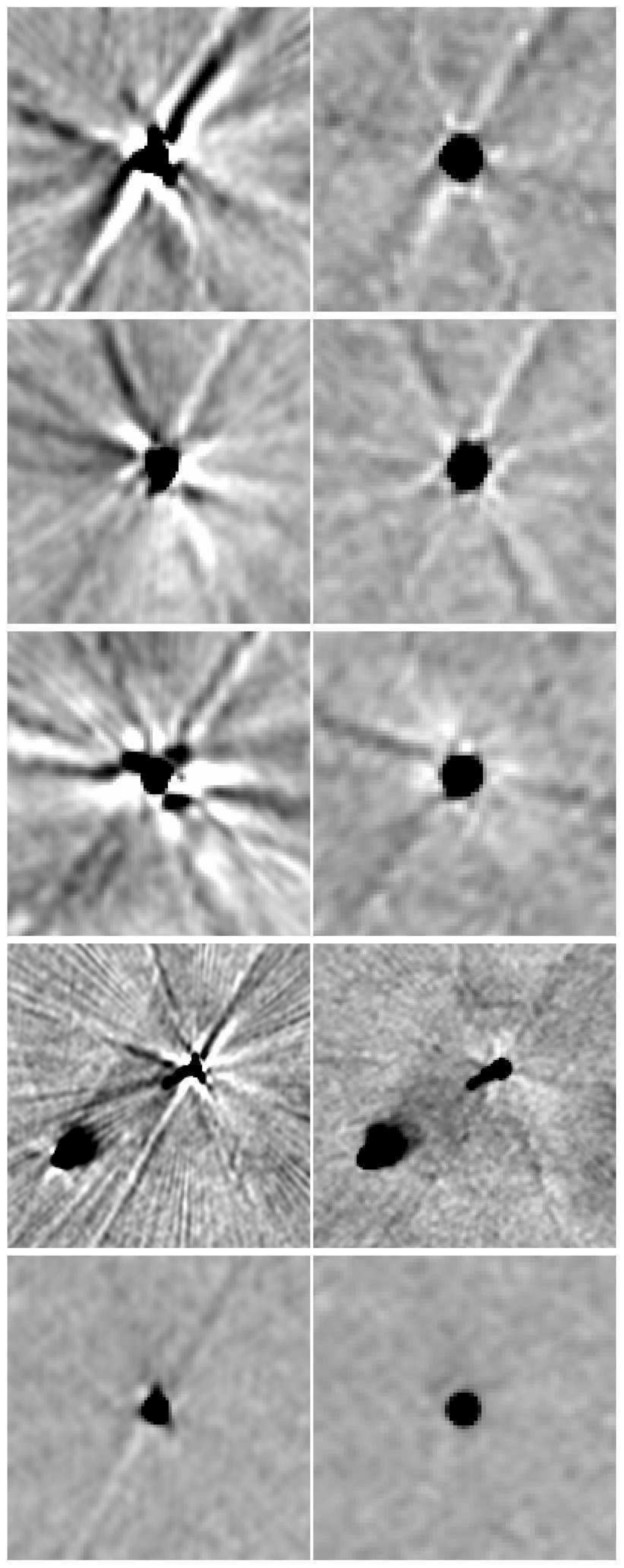}
\caption{ \footnotesize{\baselineskip0.1cm{
Artifacts around bright sources before (left panels) and after (right panels) applying self-calibration phase-gain solutions. The right column panels also have tapering applied that circularizes the beam shape. The top three rows show the same source with a peak surface brightness of around $S_p\approx 16~$mJy~beam$^{-1}$, but located inside three different pointings that were observed during different time epochs. The fourth row shows artifacts around the brightest source in our data ($S_p\approx 18~$mJy~beam$^{-1}$), which is also extended. The final row illustrates the improvement when applying self-calibration solutions only from the phase calibrator as this source with $S_p\approx 2~$mJy~beam$^{-1}$ has insufficient S/N for self-calibration (see text for details). 
}}}
\label{fig:selfcal}
\end{figure}

We found sources that were bright enough (peak surface brightness higher than 5~m\jybeam) to allow for self-calibration in 44 out of 192 pointings. To prevent artifacts affecting the model used for self-calibration small clean masks were centered around bright sources. An integration time of 3~min, which roughly corresponds to one scan length, was used to obtain phase-gain solutions for these pointings (i.e.,\ only the phase part of the complex gain was solved for and applied). It was typically not possible to find a solution using self-calibration  for 10\% of the data with the fraction increasing to 20\% for a few pointings, which was the maximum value we allowed. We applied gain solutions to the $uv$ data but did not apply the flags calculated in the self-calibration process as that usually increased the noise in the map. For the remaining pointings we applied phase gains obtained by self-calibrating the phase-calibrator J1024-0052, as it further reduced artifacts and sidelobes around brighter sources as illustrated in Fig.~\ref{fig:selfcal}.

We used the {\tt CLEAN} task with Briggs weighting scheme for gridding of visibilities with a robust parameter of 0.5 to obtain the best compromise between the resolution and the noise. Two Taylor terms (nterms=2: TT0 and TT1) were used for multifrequency synthesis, which allows the reconstruction of the total intensities and spectral slopes \citep{rau11}.
Each pointing was tapered with its own Gaussian to achieve a circular beam, where the difference between the major and the minor axis is 3\% at maximum (see \f{fig:beam} ). 
Prior to this step the beam was slightly elliptical, but the position angle changed considerably between different pointings. 
A cyclefactor of 3 was applied for a more robust deconvolution and to prevent artifacts in the map possibly caused by sidelobe intersections. Widefield imaging was necessary to produce correct astrometry far from the pointing center and we used 128 projection planes. We cleaned on three additional spatial scales corresponding to $2\times$, $5\times,$ and $10\times$ the synthesized beam size to better handle extended sources such as radio jets and lobes. A gain parameter of 0.3 was used to speed up this multiscale algorithm. Each pointing map was set to 8,000 pixels on-the-side with a pixel size of $0.2\times0.2$~arcsec$^2$.
Cleaning was performed down to $5\sigma$ in the entire map and further down to 1.5$\sigma$ using tight masks around  sources. These masks were defined manually across the entire observed field by visually inspecting the mosaic\footnote{A preliminary mosaic was generated with pointings cleaned down to $5\sigma$ and then used to define cleaning masks. Masks were usually circles with 0.7$\arcsec$ radius, but they were modified where necessary to accommodate larger (resolved) extended sources.  It was not necessary to set clean boxes around known strong sources outside of the imaged area. }. 
Synthesized beam size variations between different pointings were about $0.03\arcsec$, which was small enough to allow restoration of every cleaned pointing to an average circular beam of 0.75$\arcsec$. Finally, each pointing was corrected for the  frequency-dependent primary beam response down to a value of 20\% (corresponding to a radius of 10.5$\arcmin$) using the {\tt WIDEBANDPBCOR} task. The noise level in the phase center of an individual pointing was usually around 4-5~$\mu$\jybeam.

To construct the mosaic of all pointings, we used our custom IDL procedure combined with the  AIPS task {\tt FLATN} to carry out noise weighted addition of the individually imaged pointings. Every pixel in the sum was weighted by the inverse square of the local $rms$, which was determined in the pointing itself via the AIPS task {\tt RMSD} (see below). We mosaicked both Taylor terms individually using the noise weights calculated from the total intensity maps.
The 3~GHz continuum mosaic is shown in Fig.~\ref{fig:mosaic}, where we overplot Gaussian fits to the pixel surface brightness distributions across the mosaic. Cutouts of several extended sources and a mosaic zoom-in are presented in Fig \ref{fig:stamps}.
The visibility function showing the covered area at a given $rms$ is presented in Fig.~\ref{fig:vis}. In summary, the final mosaic has a resolution of $0.75\arcsec$, with a median $rms$ of 2.3 $\mu$Jy~beam$^{-1}$ over the COSMOS 2 square degrees.

\begin{figure}
\centering
\includegraphics[width=\columnwidth]{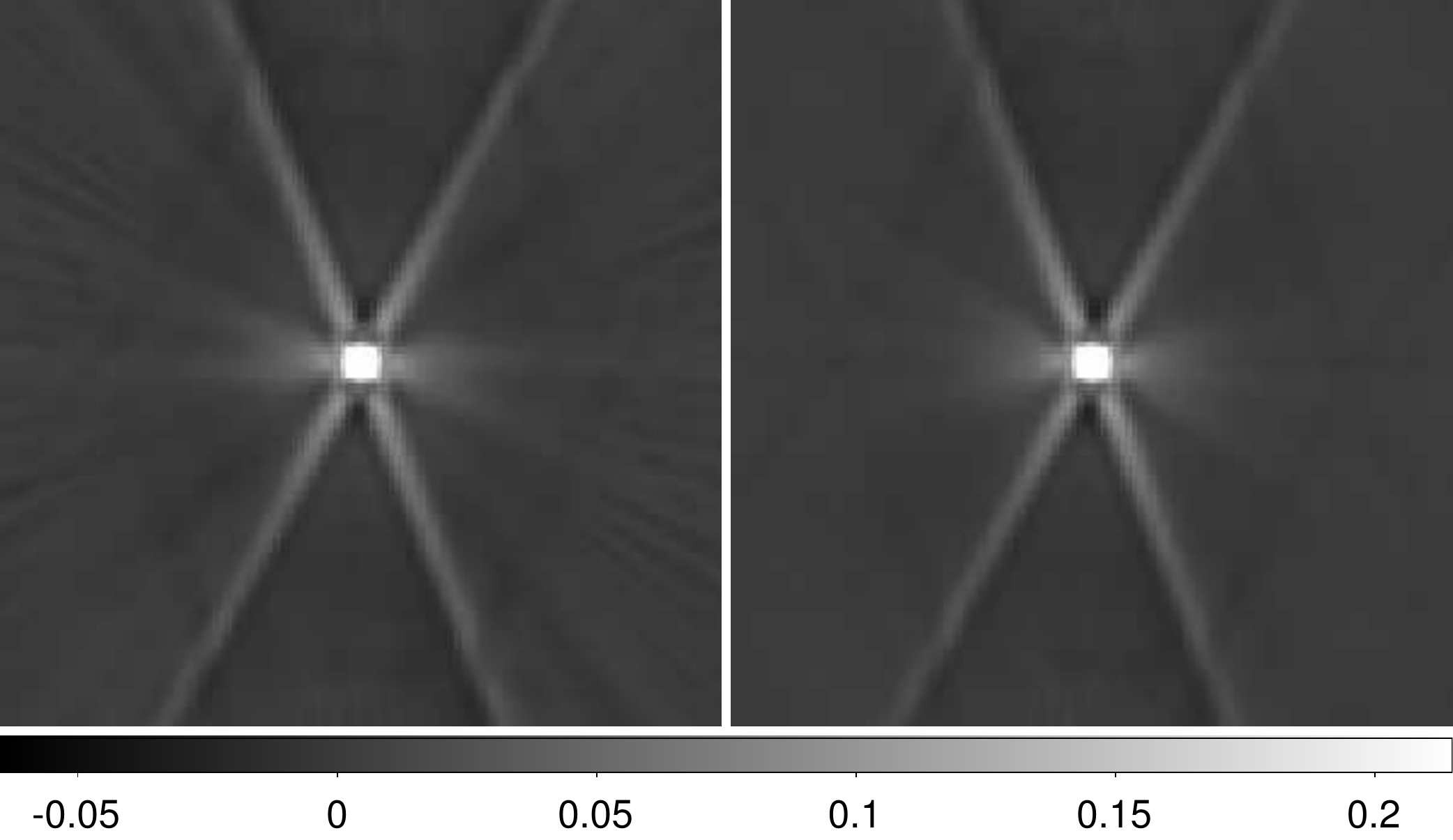}
\caption{ 
\footnotesize{\baselineskip0.1cm{
Left: final (A- plus C-array configuration combined) dirty beam of one pointing, after tapering. This beam was used in the cleaning of that pointing (see text for details). Right: mean stack of all 192 dirty beams. The contribution of radial sidelobes is 10\% at maximum. 
}}}
\label{fig:beam}
\end{figure}

\begin{figure*}[t]
\centering
\includegraphics[clip, trim=0.cm 0cm 0.cm 10cm, scale=0.9]{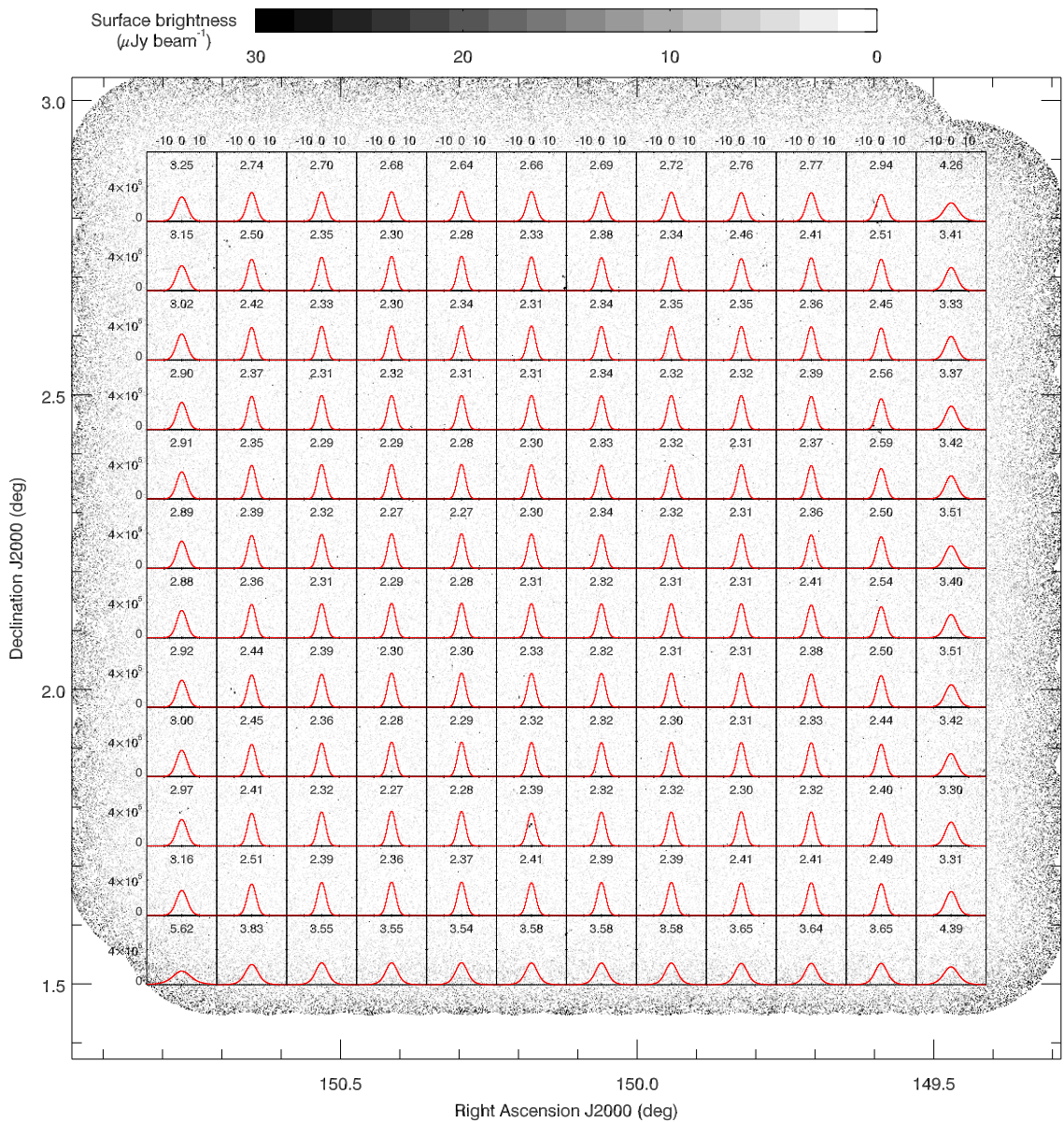}
\caption{ 
\footnotesize{\baselineskip0.1cm{Final VLA-COSMOS 3~GHz MSMF mosaic with overlaid Gaussian fits to the pixel surface brightness distributions in various mosaic sectors. The $rms$ obtained via the Gaussian fit (in units of $\mu$Jy~beam$^{-1}$) is indicated in each panel. The panels shown cover the full COSMOS two square degree field. 
The small-scale ($\sim1\arcmin$) $rms$ variations due to the pointing layout are less than 2\%. 
}}}
\label{fig:mosaic}
\end{figure*}

\begin{figure}[t]
\centering
\includegraphics[ width=\columnwidth]{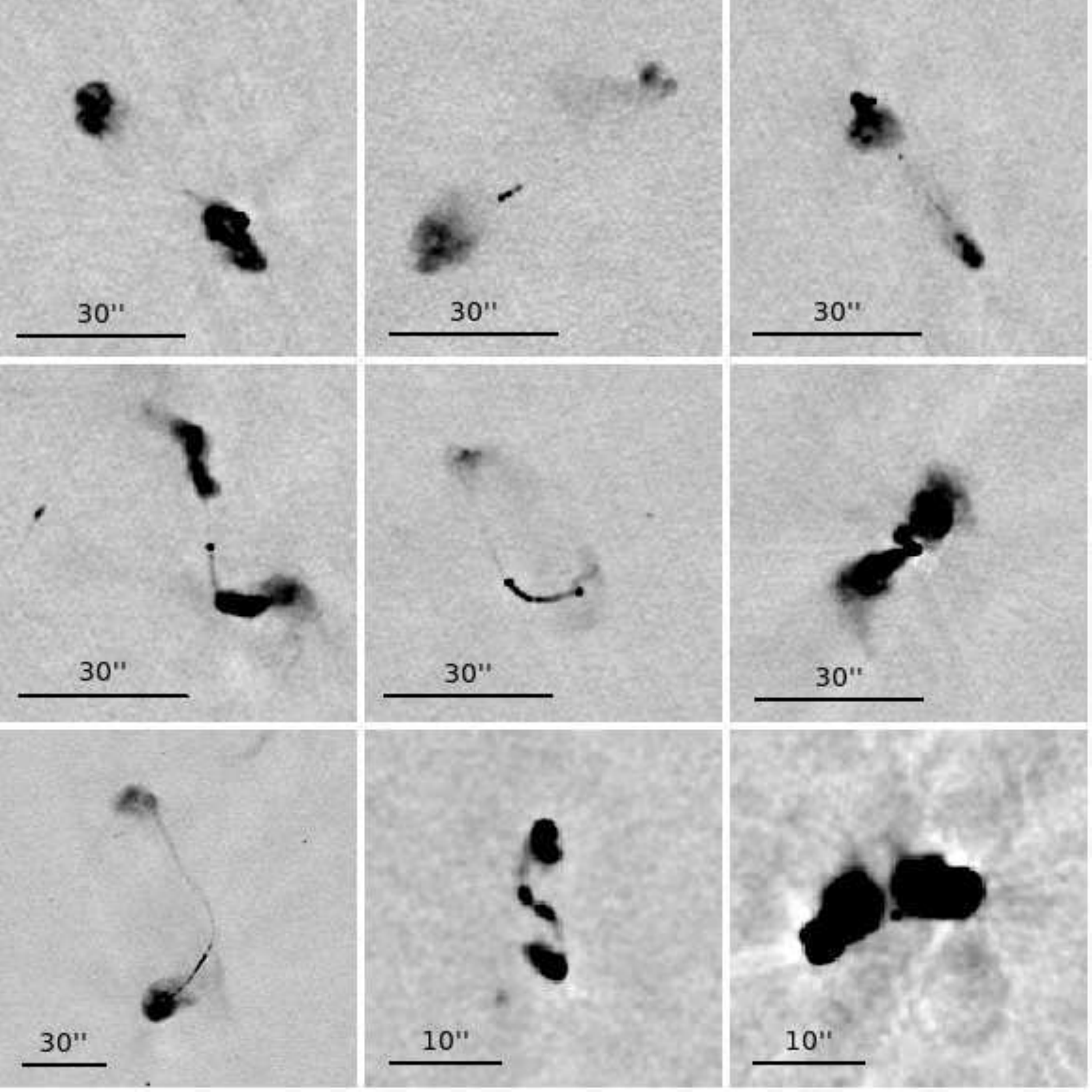}
\includegraphics[ width=\columnwidth]{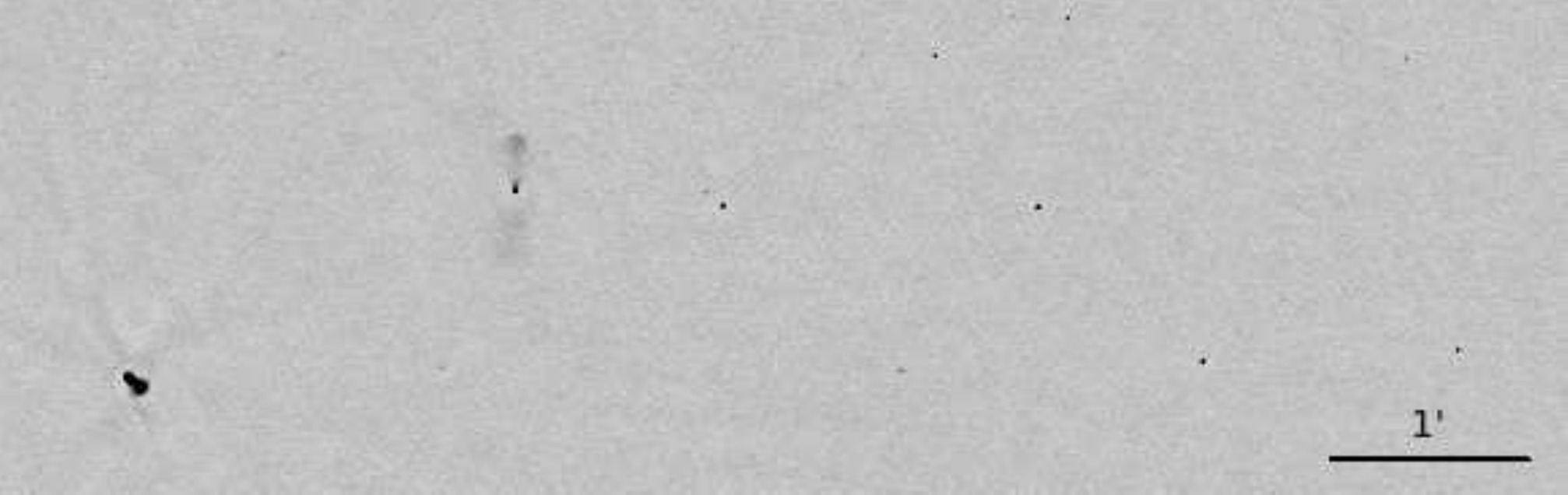}
\caption{\footnotesize{\baselineskip0.1cm{
 Stamps from the VLA-COSMOS 3~GHz continuum mosaic imaged with the MSMF algorithm showing examples of extended and compact radio sources.
 }}}
\label{fig:stamps}
\end{figure}

\begin{figure}[t]
\centering
\includegraphics[ width=\columnwidth]{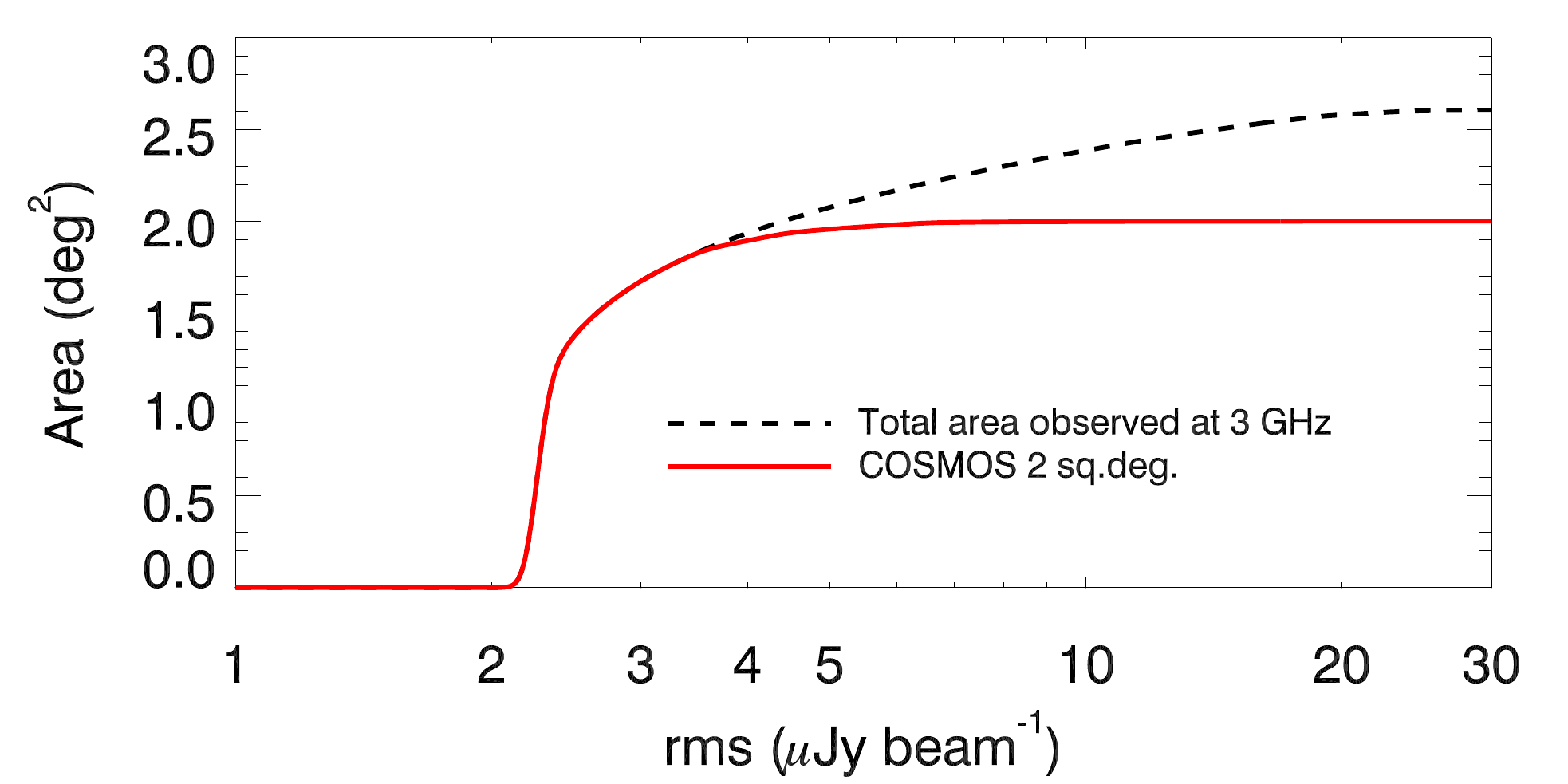}
\caption{\footnotesize{\baselineskip0.1cm{
 Visibility plot showing the total area covered down to a given noise level (black dashed line). Our data extend beyond the COSMOS two square degree field, which ensures more uniform noise inside it (red full line). The median noise level inside the COSMOS two square degrees is $\sigma=2.3~\mu$Jy~beam$^{-1}$. 
 }}}
\label{fig:vis}
\end{figure}

\section{Cataloging}
\label{sec:cat}

\subsection{Extracting source components}

To extract source components from the VLA-COSMOS MSMF mosaic and catalog their properties we employed {\tt blobcat} developed by \citet{hales12}. Extractor {\tt blobcat} 
uses the flood fill algorithm to find islands of pixels (blobs) above a certain signal-to-noise ratio (S/N) threshold. The local noise map used to evaluate the S/N at each pixel position was created from the total intensity mosaic with the AIPS task {\tt RMSD} with a circular mesh size of 100 pixels. Once  {\tt blobcat} locates islands, it measures the peak surface brightness ($S_p$)  by fitting a two-dimensional (2D) parabola around the brightest pixel, while the total flux density ($S_t$) is obtained by summing up the pixel values inside the island and dividing the sum by the beam size in pixels.
In the next step {\tt blobcat} takes into account a small positive peak surface brightness bias created by the presence of noise peaks in the map and also corrects for a negative integrated surface brightness bias caused by the finite island size used for integration. We used  default parameters when running {\tt blobcat} (as \citealt{hales12} ran extensive simulations to optimize them; see also \citealt{hales14}), where the required size of a blob is at least 3 pixels in RA and 3 pixels in DEC. This was necessary to detect low S/N sources, which would have otherwise been missed owing to our relatively coarse pixel grid.
With this setup we recovered \ncomp \ radio source components with local S/N greater or equal to 5 across the entire observed area. As detailed in \s{sec:mult} ,\ \nmult \ components have been merged into unique, multicomponent sources resulting in a total of \nsource \ radio sources.

\subsection{Resolved versus unresolved sources}

\begin{figure}[t] 
\centering
\includegraphics[ width=\columnwidth]{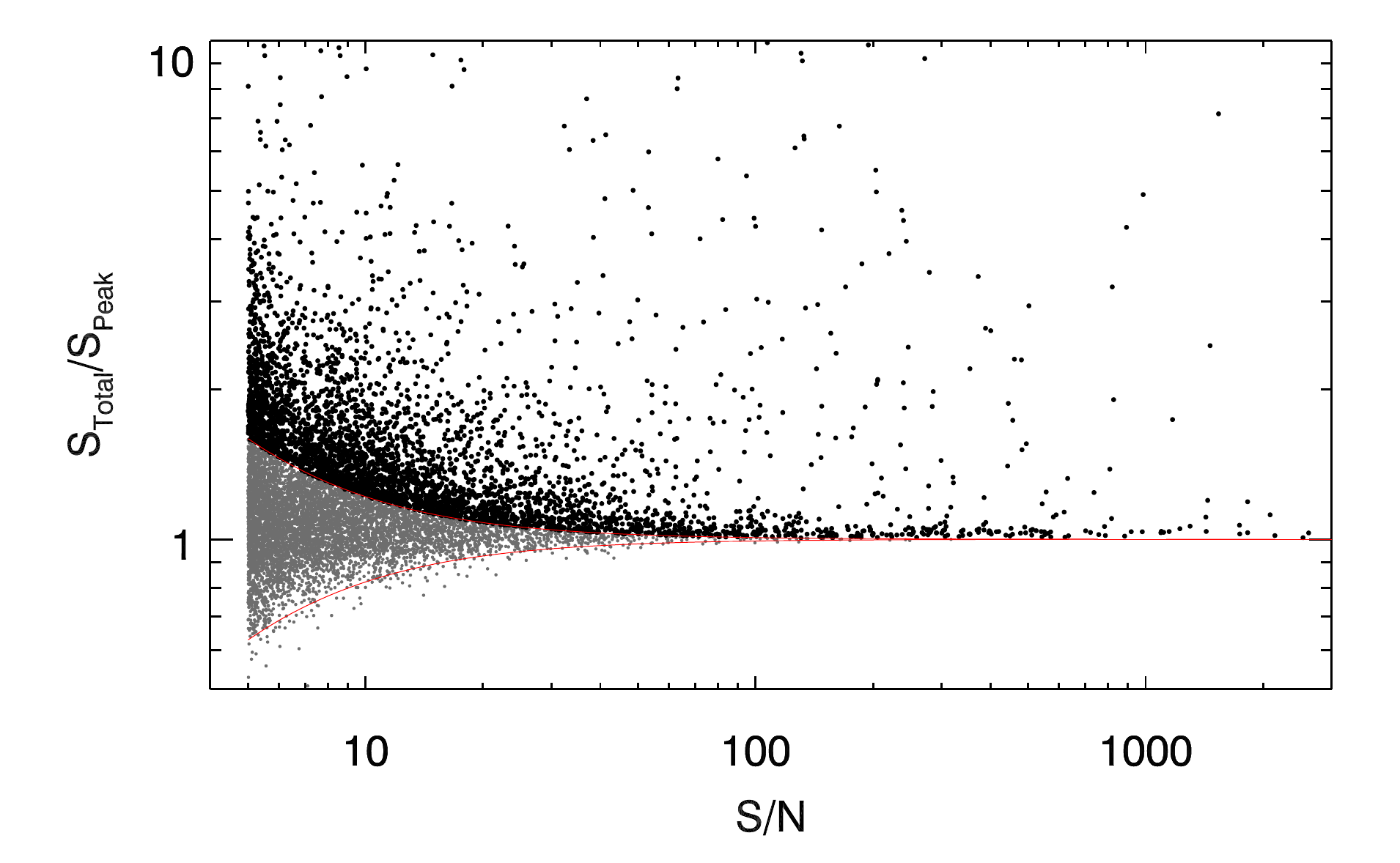}
\caption{\footnotesize{\baselineskip0.1cm{
 Ratio of total flux density to peak surface brightness as a function of S/N ratio.  Components within the envelopes (gray points) indicated by red solid lines are considered unresolved, while those above the upper envelope (black points) are considered resolved (see text for details).
   }}}
\label{fig:stsp}
\end{figure}

In order to determine whether our identified source components are resolved (i.e., extended, larger than the
synthesized beam) we make use of the ratio between total flux density ($S_t$) and peak surface brightness ($S_p$) as this is a direct measure of the extension of a radio source. The flux densitites were computed by {\tt blobcat} as described in the previous section. For a perfect Gaussian unresolved source, the peak surface brightness in Jy~beam$^{-1}$ equals the integrated flux density in Jy or $S_t/S_p=1$. 
The extension of a radio source increases its total flux density when compared to its peak surface brightness, however, background noise can lower the total flux density (see  \citealt{bondi03}).
Therefore, in Fig.~\ref{fig:stsp} we plot the ratio between the total flux density and the peak surface brightness as a function of the S/N (=$S_p/rms$) for all  \ncomp \ components in the catalog. To select the resolved components, we determined the lower envelope of the points in Fig.~\ref{fig:stsp}, which contains 95\% of the components with $S_t < S_p$ and mirrored it above the $S_t/S_p=1$ line (upper envelope in Fig.~\ref{fig:stsp}). The shape of the envelope was chosen following \citet{bondi08} and the fit to our data is given as $S_t/S_p=1+6\times(\mathrm{S/N})^{-1.44}$.
We consider the 3,975 components above the upper envelope as resolved. 
These resolved components were flagged in the catalog. For the unresolved components the total flux density was set equal to the peak surface brightness in the catalog. 

\subsection{Multicomponent sources}
\label{sec:mult}

Large sources with diffuse structures, such as\ radio galaxies (see \f{fig:stamps} ) or resolved star-forming disks, can be listed in a component catalog as multiple entries. This can happen for example if there is no significant radio emission between the two radio lobes, or if the local $rms$ noise is overestimated because of large-scale faint radio emission, which affects the ability of {\tt blobcat} to properly detect a contiguous blob.
We identified \ncomp \  components in our mosaic, as described above. In order to generate a source catalog rather than a source component catalog, we aimed to identify such sources and convert the multiple entries into one entry that described the entire source, i.e.,\ listing its proper total flux density and  position.
For this purpose we  visually inspected over 2,500 components. The inspected sample was a combination of the i) brightest 2,500 components, ii) all known multicomponent sources that were identified and listed in the 1.4~GHz joint catalog (126 components), and iii) sources with $R_\mathrm{EST} > 1+30\times(\mathrm{S/N})^{-1}$ (351 components).
The $R_\mathrm{EST}$ parameter is a size estimate reported by {\tt blobcat}, which can be used to find sources with non-Gaussian morphology; see \cite{hales12,hales14} for more details.
Following the procedure already applied to the VLA-COSMOS 1.4~GHz survey sources \citep{schinnerer07} these components were visually inspected with respect to the near-infrared (NIR) images, i.e.,\ the $z^{++}YJHK$ stacked maps \citep{laigle16}. In total, we identifed \nmult \ multicomponent sources. As for the previous VLA-COSMOS survey catalogs, we computed their total flux densities using the AIPS task {\tt TVSTAT} in the area encompassed by $2\sigma$ contours, where $\sigma$ is the local $rms$ measured as the average $rms$ from a 100-300 pixel wide area around the source, ensuring that the $rms$ is not biased by the influence of the strong sources. The source position is then taken to be the radio core or optical counterpart position (if identifiable) or the luminosity weighted mean. In our catalog we then excluded all the components combined into the multicomponent sources, and listed instead the multicomponent source with the above-defined position and total flux density, setting all other cataloged values to -99. A further column {\tt multi} was added designating the multicomponent sources ({\tt multi}=1 for a multicomponent source, and {\tt multi}=0 for a single-component source). We note that the number of multicomponent sources is smaller than that identified in the shallower VLA-COSMOS 1.4~GHz survey. This is due to the higher frequency of breaking-up large sources into multiple components within the latter as it used the AIPS {\tt Search and Destroy} source finding algorithm, when compared to the performance of the {\tt blobcat} algorithm used here. A full assessment of large sources in the survey will be presented by Vardoulaki et al.\ (in prep.). 

\subsection{Astrometric accuracy}

\begin{figure}[t]
\centering
\includegraphics[ width=\columnwidth]{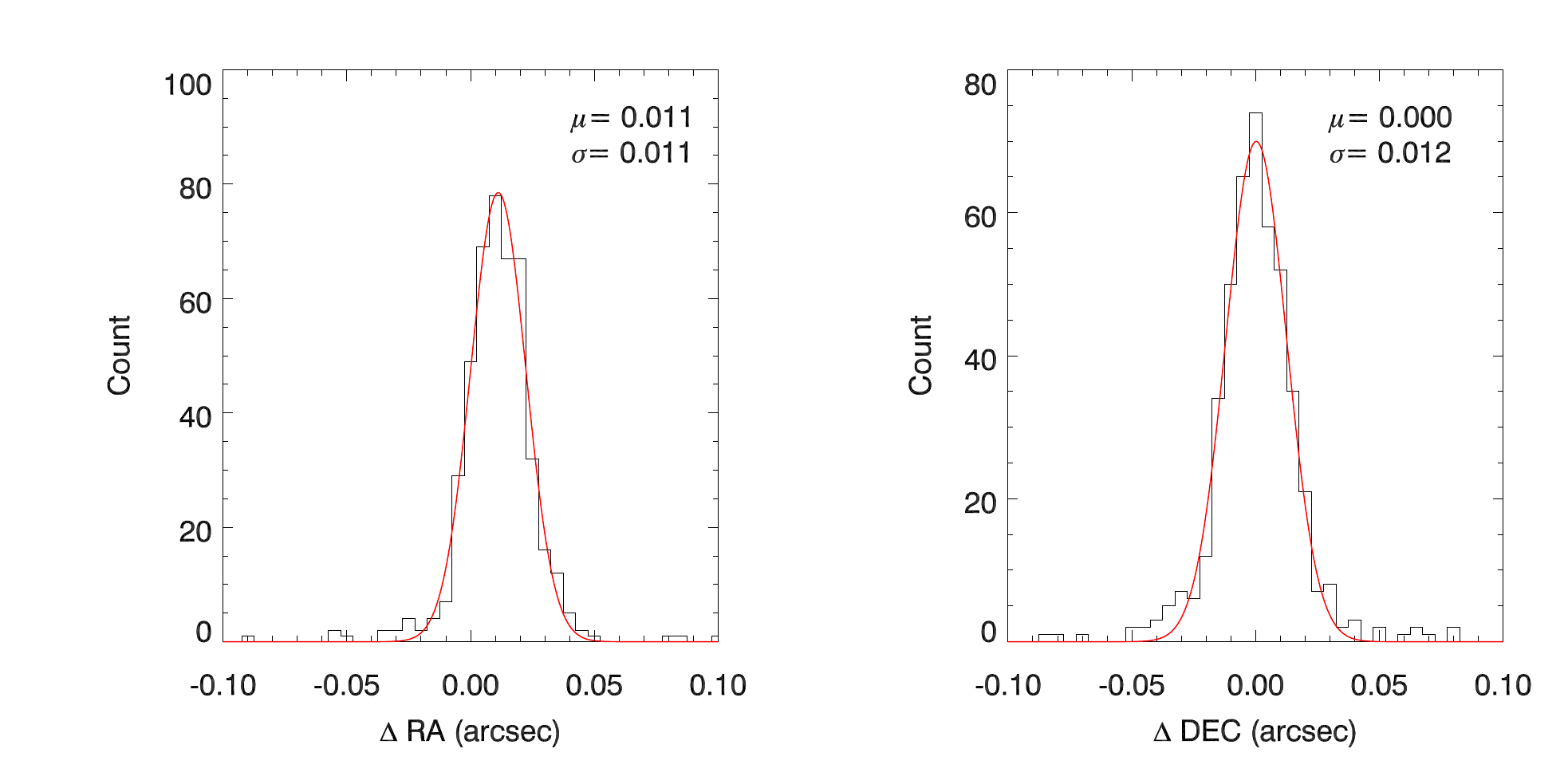}
\caption{ 
 \footnotesize{\baselineskip0.1cm{
 Astrometry comparison between 3~GHz and 1.4~GHz VLBA data for 443 VLBA sources (PI: Middelberg, N.~Herrera Ruiz et al. in prep.). 
 }}}
\label{fig:vlba}
\end{figure}

To assess our astrometric accuracy at the bright end we have compared the positions of 443 sources at 3~GHz with S/N $>$ 20, also detected in the Very Long Baseline Array (VLBA)-COSMOS 1.4~GHz survey (PI: Middelberg; N.~Herrera Ruiz et al., in prep.). 
The results, shown in Fig.~\ref{fig:vlba}, yield an excellent agreement with a  mean offset of $0.01\arcsec$ in $\Delta$~RA and $0.00\arcsec$ in $\Delta$~DEC and a standard deviation of 0.01$\arcsec$ for both. We note that we did not correct the catalog entries for the $0.01\arcsec$ offset in $\Delta$~RA. We took the standard deviation value (0.01$\arcsec$) as the calibration error in RA and DEC to compute the positional uncertainties for our sources using the equations reported in \citet{hales12}.  We note that these are estimated to be accurate for point-sources, but likely underestimated for resolved sources (see \citealt{hales12} and references therein for details).

\subsection{Bandwidth smearing}
\label{sec:bws}

Owing to the finite bandwidth of the antenna receiver, bandwidth smearing (BWS) occurs and radially smears peak surface brightness while conserving the integrated flux density. The effect is a function of distance from the phase center in a given pointing while it reaches a constant smearing value in the combined mosaic (see, e.g., \citealt{bondi08}). 
Although the bandwidth of the antenna receiver is large ($\sim4$~GHz), the relevant bandwidth for the smearing effect is only the $2$~MHz channel width used to image the data.

To empirically test BWS in our data, we selected 106 point-like ($0.9\leq S_\mathrm{t}/S_\mathrm{p}\leq1.1$) radio sources with S/N$>$200. Since each source can be observed in up to 11 neighboring pointings, we can compare the peak surface brightnesses obtained in various pointings ($S_\mathrm{P}$) relative to the peak surface brightness retrieved from the mosaic ($S_\mathrm{M}$) as a function of distance from the pointing center. If our data were affected by BWS, $S_\mathrm{P}/S_\mathrm{M}$ would exhibit a declining trend with increasing distance from the pointing center. This surface brightness ratio,  
obtained by fitting an inverted parabola at the 106 bright source positions in the individual pointings, and the mosaic
 is shown in the top panel of Fig.~\ref{fig:mos_p_dist}. 
The median ratio stays constant ($S_P/S_M\approx1$) across all distance ranges, with increasing scatter toward higher distances where the noise is amplified by the primary beam correction. This demonstrates that there are no empirical bandwidth-smearing issues. 
This is also in accordance  with theoretical expectations. 
A theoretical prediction for BWS can be made using the \citet{condon98} equation [12] for the reduction of peak response $I/I_0\approx1/\sqrt{1+0.46\beta^2}$, where $\beta=(\Delta\nu/\nu_0)\times(\theta_0 / \theta_{HPBW})$ equals fractional bandwidth times offset in synthesized beam-widths. Using the VLA channel width $\Delta\nu=2$~MHz, central frequency $\nu_0=3$~GHz, distance from the phase center $\theta_0=300\arcsec$, and beam size of $\theta_{HPBW}=0.75\arcsec$ the estimated peak reduction amounts to about 2\%. The distance was chosen as a minimal distance between two different pointing centers. 
This is illustrated in \f{fig:bws_new} \ where we show the peak over total flux density for $\mathrm{S/N}>200$ sources in different pointings. An offset of $\sim2.5\%$ is present in this diagram, however, it is not distance dependent, and thus unlikely to be related to  bandwidth smearing. 
Thus, for the reasons outlined above, we do not apply any corrections for the BWS effect.

\subsection{The 3~GHz VLA-COSMOS Large Project catalog}

A sample page of the catalog is shown in Table~\ref{tab:cat}. 
For each source, we report its ID, 3~GHz name,  RA and DEC position (weighted centroid) and error on the position,  total flux density with relative error\footnote{
The flux errors reported do not depend on the number of pixels used for integration, but scale with the source brightness (see  \citealt{hales12,hales14}).
}, 3~GHz $rms$ calculated at the position of the source,  S/N, number of pixels used in flux density integration, flag for resolved sources, and flag for multicomponent sources.  The peak surface brightness of resolved sources can be obtained by multiplying the S/N with the $rms$ value.
The catalog is available in electronic format in the COSMOS IRSA archive\footnote{http://irsa.ipac.caltech.edu/frontpage/}.

\section{Radio spectral indices}
\label{sec:alpha}

Given the wide bandwidth of our VLA-COSMOS 3~GHz survey and the existence of previous COSMOS radio surveys, we approached radio spectral index calculations in two ways. The first method uses the MSMF algorithm to construct spectral indices  directly from our observed data by fitting a two-term Taylor polynomial to amplitudes between 2 and 4~GHz (MSMF-based spectral index or $\alpha_{\scriptsize\mbox{MSMF}}$ hereafter). The second method 
uses the cataloged monochromatic flux densities at 3~GHz in combination with the values taken from the 1.4~GHz joint catalog \citep{schinnerer10} to calculate spectral indices between these two frequencies (1.4--3~GHz spectral index or $\alpha_{\scriptsize1.4-3~\mbox{GHz}}$ hereafter). In \s{sec:alphamsmf} \ we investigate systematics in the MSMF spectral index maps, and compare the differently derived spectral indices. In \s{sec:alpha1.4-3} \ we derive the 1.4--3~GHz spectral index distribution for the full sample of the 3~GHz sources. 

\subsection{  MSMF-based versus\ 1.4--3~GHz spectral indices}
\label{sec:alphamsmf}
We can calculate the MSMF-based spectral indices defined for each source using the wide bandwidth of our observations if the source  has a sufficient S/N between 2 and 4~GHz.
These spectral indices should be viable for point sources that have S/N$>$10, and for diffuse emission that has S/N$>$100.
To do so, a mosaic of spectral indices ($\alpha$-map) was generated by dividing the Taylor term~1 (TT1) mosaic by the Taylor term~0 (TT0) mosaic (see \citealp{rau11}). For each source, its  spectral index was extracted from the pixel in the $\alpha$ map that corresponds to the pixel containing the peak surface brightness in the total intensity mosaic.

\begin{figure}[t]
\centering
\includegraphics[ width=\columnwidth]{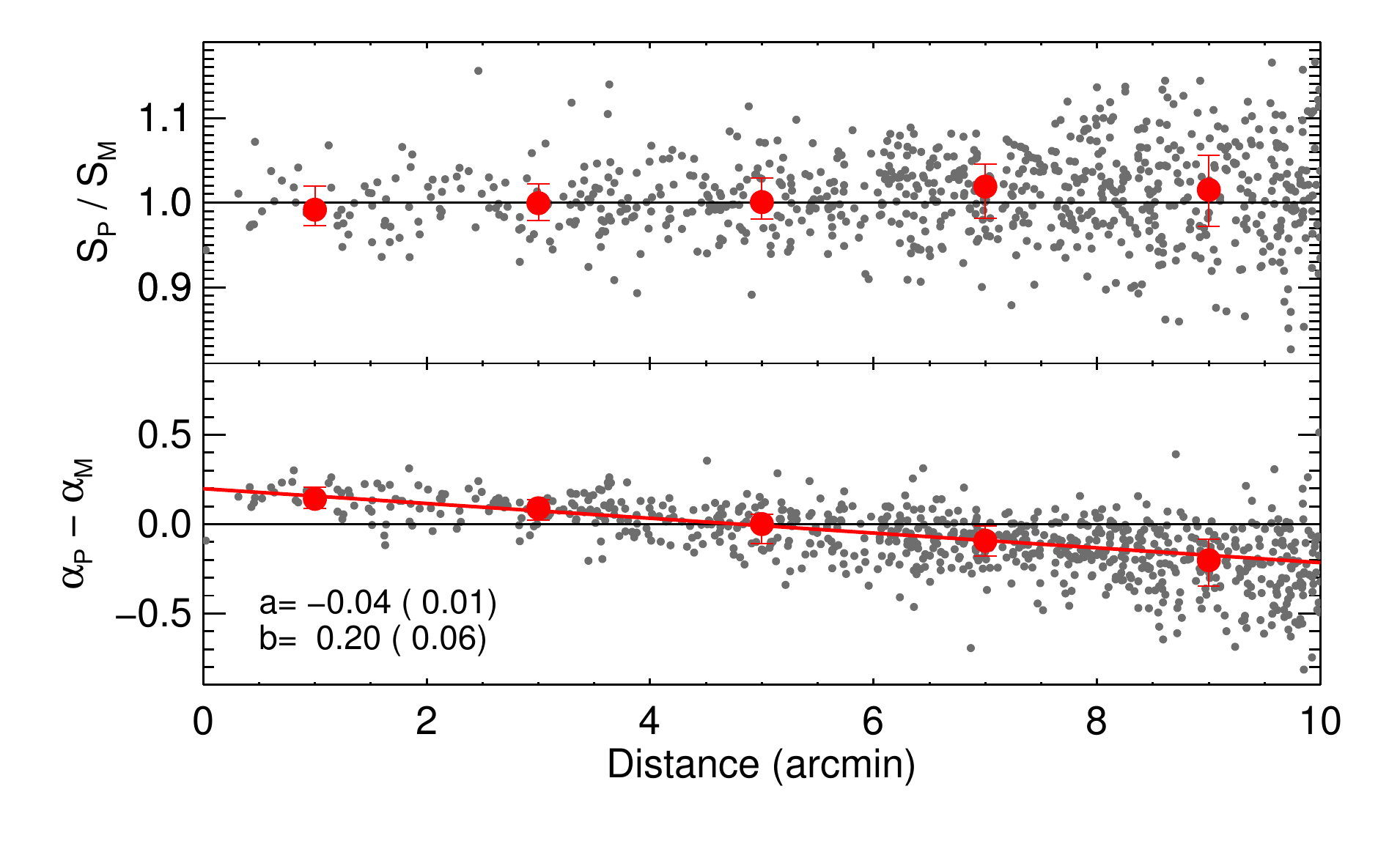}
\caption{ \footnotesize{\baselineskip0.1cm{
Comparison of peak surface brightnesses (top) and MSMF-based spectral indices (bottom) determined  inside the mosaic ($S_\mathrm{M}$, $\alpha_\mathrm{M}$) and individual pointings ($S_\mathrm{P}$, $\alpha_\mathrm{P}$)  as a function of distance from the pointing center for 106 bright, point-like sources ($0.9\leq S_\mathrm{t}/S_\mathrm{p}\leq1.1$, S/N$>$200) observed in up to  11 neighboring pointings at varying distances from the pointing center (gray points). In both panels the large red  points and their corresponding errors indicate median values and interquartile ranges inside 5 equally spaced distance bins. In the bottom panel a linear fit is performed on the median values to obtain the needed correction of the systematic trend across all distances (see text for details). 
 }}}
\label{fig:mos_p_dist}
\end{figure}

To investigate possible systematics in the $\alpha$ map due to wideband primary beam corrections we utilized the 106 bright, point-like sources introduced in \s{sec:bws} . We derived MSMF-based spectral indices both in the mosaic and individual pointings for these sources. In the bottom panel of Fig.~\ref{fig:mos_p_dist} we show the difference between such derived spectral indices as a function of distance from the pointing center.
The MSMF spectral indices show a systematic steepening  with increasing distance, which likely arises due to an  imperfect primary beam correction of TT1.\footnote{The MSMF algorithm is still in active development and the upcoming software versions should correct for this.} To correct for this effect a posteriori (as necessary here), we performed a linear fit to the trend. We then applied this distance-dependent correction to each  $\alpha$-map pointing 
prior to mosaicking to generate an  $\alpha$ mosaic corrected for this effect.

\begin{figure}
\centering
\includegraphics[ width=\columnwidth]{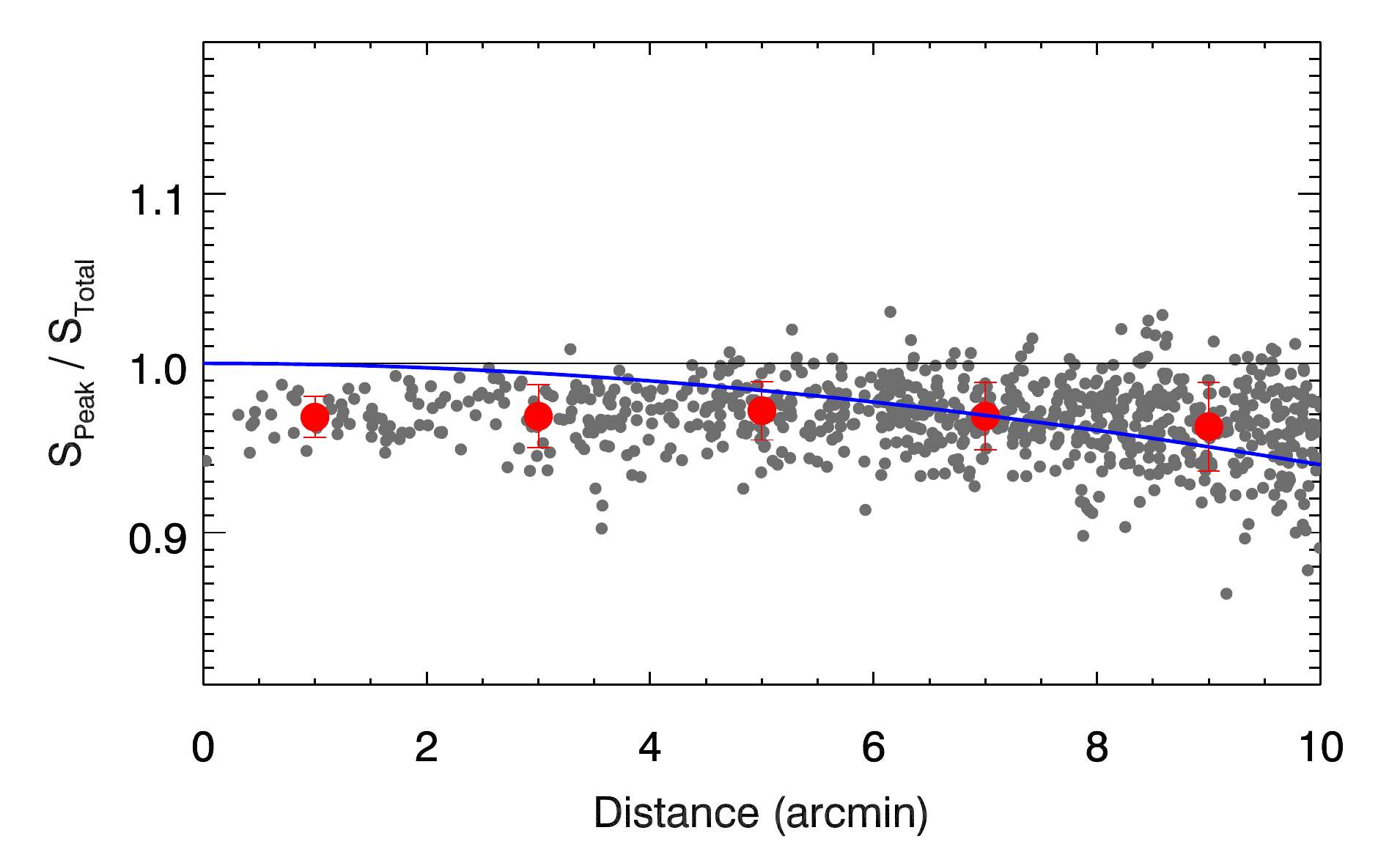}
\caption{ \footnotesize{\baselineskip0.1cm{
Comparison of peak surface brightnesses over total flux densities for 106 bright, point-like sources ($0.9\leq S_\mathrm{t}/S_\mathrm{p}\leq1.1$, S/N$>$200) observed in up to  11 neighboring pointings at varying distances from the pointing center (gray points). The large red points and their corresponding errors indicate median values and interquartile ranges inside 5 equally spaced distance bins. The theoretical prediction of the bandwidth smearing effect is shown by the blue curve (see text for details). 
 }}}
\label{fig:bws_new}
\end{figure}

\begin{figure}[t]
\centering
\includegraphics[ width=\columnwidth]{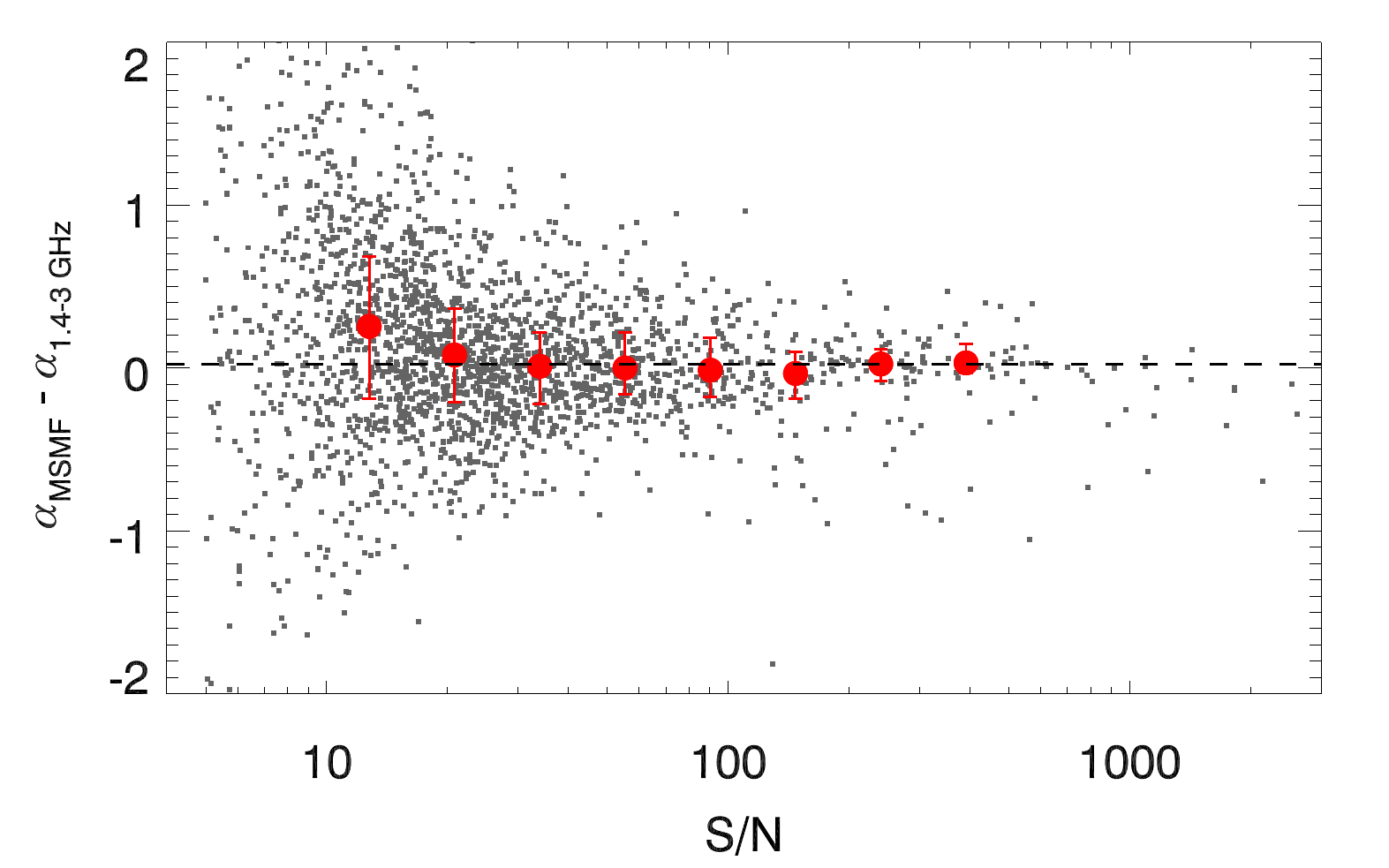}
\caption{ \footnotesize{\baselineskip0.1cm{
Comparison between MSMF-based and 1.4--3~GHz derived spectral indices, where the first were corrected for the observed systematic trend illustrated in \f{fig:mos_p_dist} . 
Red symbols and the corresponding errors denote median spectral indices and interquartile ranges, respectively, for sources in different S/N bins (10$<$S/N$<$500). The black dashed line indicates the median value of the red circles set at $0.02$. 
 }}}
\label{fig:alpha_snr}
\end{figure}

In Fig.~\ref{fig:alpha_snr} we compare the corrected spectral indices from MSMF with those derived from the cataloged flux densities at 3~GHz and 1.4~GHz (joint catalog, \citealt{schinnerer10}). The catalogs were cross-matched using a search radius of 1.3$\arcsec$, which is half of the beam size of the lower resolution (1.4~GHz) survey. The sample was further limited to single-component sources with S/N$>5$ in the 1.4~GHz catalog yielding a total of 2,191 sources. 
Although there are no systematic offsets within the error margins, there is a rather large scatter between the spectral indices obtained with these two methods.  A non-negligible portion of this spread is due to the large uncertainty on the in-band (i.e., MSMF derived) spectral indices; a point-like source with S/N $\sim50$ and $\alpha = 0.7$ has an uncertainty of $\sim0.1$ in its in-band spectral index \citep[see][]{condon15}.
If the MSMF spectral indices had not been corrected, there would have been a systematic offset of -0.2 across the entire S/N range. 

In summary, the MSMF-based spectral indices require further corrections after PB corrections are applied to the data. As a result of this, and the large scatter observed between the MSMF-based and 1.4--3~GHz spectral indices, we do not include the MSMF-based spectral indices in the final catalog. New CASA software versions should intrinsically  correct for this. For the further analysis of spectral indices presented here we have, therefore, used only the values based on flux density measurements at 3 and 1.4~GHz. 

\subsection{ 1.4--3~GHz spectral indices}
\label{sec:alpha1.4-3}

A high percentage of 3~GHz sources do not have a counterpart in the 1.4~GHz survey because of the better sensitivity of our 3~GHz survey. 
We employed the survival analysis to properly constrain the distribution of  spectral indices for our 3 GHz selected sample  without introducing any bias due to neglecting sources not detected in one of the surveys. This is a statistical method that takes into account both direct detections as well as upper (or lower) limits (see \citealt{feigelson85} and \citealt{schmitt85} for details). 

\begin{figure}[t]
\centering
\includegraphics[ width=\columnwidth]{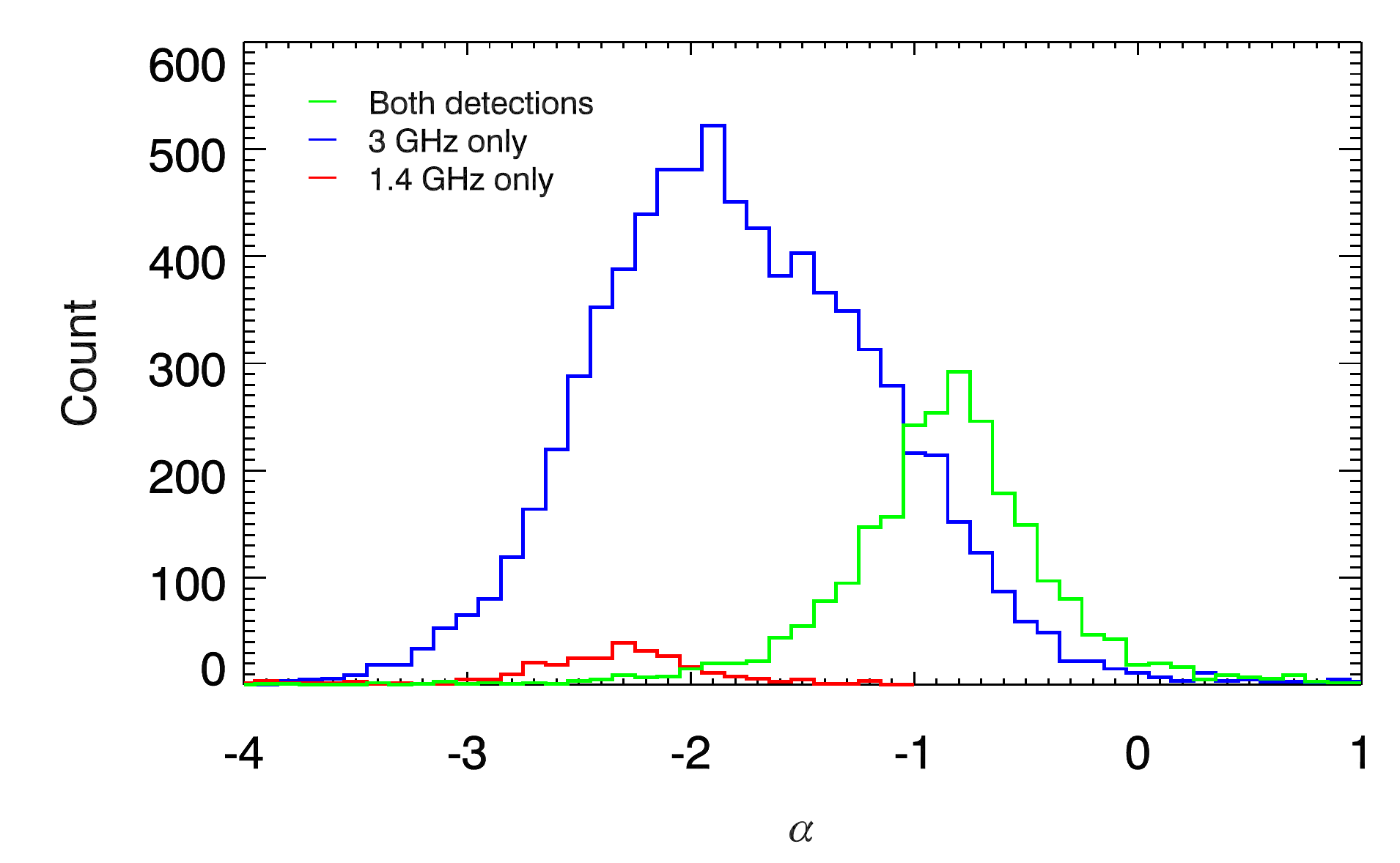}
\includegraphics[ width=\columnwidth]{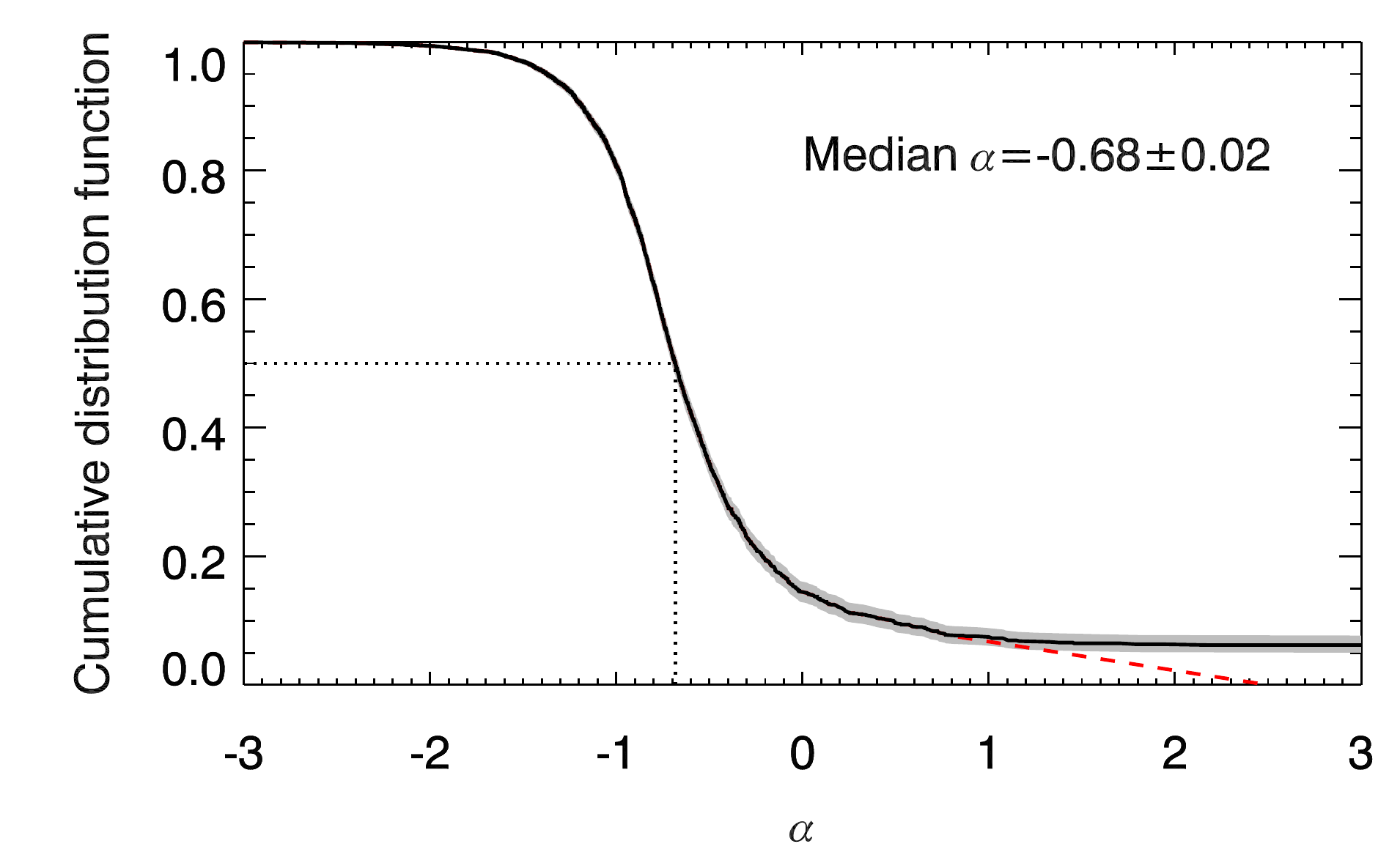}
\includegraphics[ width=\columnwidth]{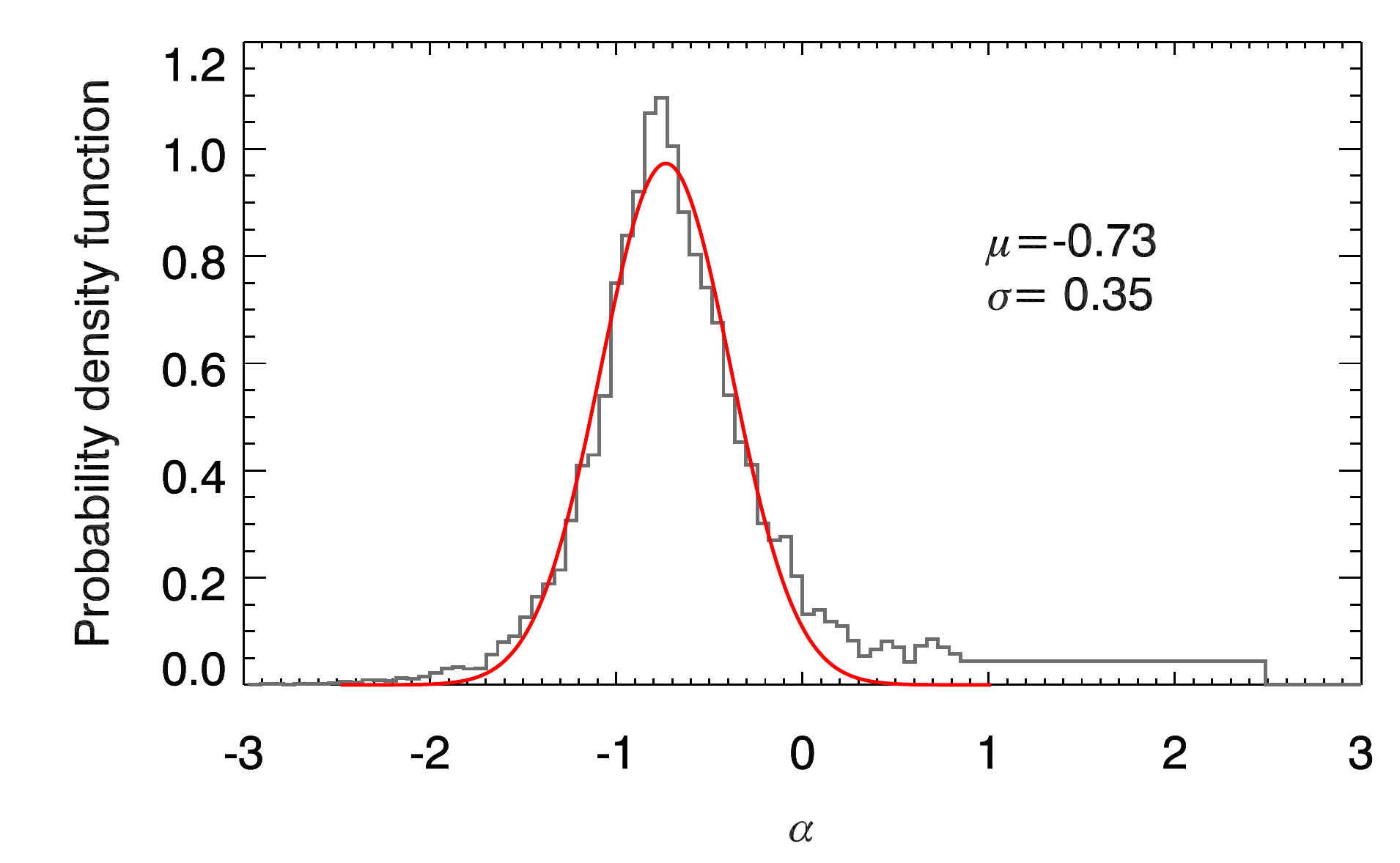}
\caption{ \footnotesize{\baselineskip0.1cm{
Top panel: distribution of 1.4--3~GHz spectral indices for sources detected at both frequencies (green line),  and only at 3~GHz (lower limits, blue line) or  1.4~GHz (upper limits, red line).
Middle panel: cumulative distribution function (CDF; black line) and error estimate (gray shaded area) of spectral indices calculated using the survival analysis also taking lower limits into account.  The red dashed line shows a linear extrapolation of the distribution to zero assuming the maximum theoretically attainable spectral index of $\alpha=2.5$ (see text for details).
Bottom panel:  probability density function (PDF) for spectral indices calculated as a first derivative of the CDF  extrapolated to 0 at high end (middle panel).  A Gaussian fit to the distribution is also shown (red curve) and its mean and standard deviation are indicated in the panel.
 }}}
\label{fig:survival}
\end{figure}

We first cross-correlated and combined our 3~GHz catalog with the 1.4~GHz joint catalog \citep{schinnerer10} with a maximum separation of 1.3$\arcsec$, but also including sources without counterparts in one survey or the other. We then removed all sources that fell outside of the  area observed at 1.4~GHz as the area observed at 3~GHz is larger (2.6 square degrees.). This was performed to ensure the same area for both surveys. We also removed 1.4~GHz multicomponent sources (80) and their 3~GHz counterparts. The final sample contains 10,523 entries out of which 23\% were detected in both surveys, 74\% were detected only at 3~GHz, and 3\% were detected only at 1.4~GHz, as illustrated in Fig.~\ref{fig:survival} (top panel). If a source was not cataloged in one of the surveys we used five times the local $rms$ value at the coordinates of the source as an upper limit on the flux density. Each nondetection at 1.4~GHz yielded one lower limit on spectral index, and similarly, each nondetection at 3~GHz yielded one upper limit. 

A Gaussian fit to the distribution of spectral indices detected in both surveys (green line in Fig.~\ref{fig:survival}, top panel) results in the peak at $\alpha=-0.84$ and a standard deviation of $\sigma=0.35$. As this result is valid only for the subsample of 3~GHz sources also detected at 1.4~GHz, we employed the survival analysis to account for the full 3~GHz detected sample. 
We therefore ran the survival analysis on a single-censored data set that only included detections in both surveys and lower limits. The method assumes that limits follow the same distribution as direct detections and generates a cumulative distribution for all sources in the sample. This is shown in the middle panel of Fig.~\ref{fig:survival}. There was enough overlap between direct detections and lower limits enabling the survival analysis to properly constrain  the median of the total distribution to $\alpha=-0.68\pm0.02$, even though there were three times more limits than detections. This method however cannot constrain all lower limits and the cumulative function does not converge to 0, yielding a total of 6\% unconstrained sources. To constrain these (as needed to derive the probability density function, PDF, for spectral indices; see below)
we  employed a physical argument that a radio source exhibiting standard synchrotron self-absorption cannot have a spectral index higher than $\alpha_{\mbox{\scriptsize max}}=2.5$ (\citealt{rees67}; unless it is extremely rare and exotic; for example see \citealt{krishna14}). Our data can also constrain the distribution of spectral indices only up to value of $\alpha=0.8$, since this interval contains 99.5\% of sources detected at both 1.4 and 3~GHz. With these limits we can at best assume a flat probability that unconstrained sources have $0.8<\alpha<2.5$, and we can formally extrapolate the cumulative distribution function to 0 (red dashed line in Fig.~\ref{fig:survival}, middle panel).
Having constrained this, we then derived the PDF for spectral indices of our 3~GHz sources 
by calculating the first derivative of the cumulative distribution function extrapolated to 0. The PDF is shown in the bottom panel of Fig.~\ref{fig:survival}. The best-fit Gaussian  to the PDF yields a mean of $\alpha=-0.73$ and a standard deviation of $\sigma=0.35$. Both the median of the distribution and the mean of the Gaussian fit agree very well with previous work carried out on spectral indices (e.g., \citealt{condon84,lisenfeld00,kimball08}) and we conclude that our catalog flux densities do not show any significant systematics.

\section{Radio source counts corrections}
\label{sec:completeness}

A well-established approach to estimate the combined effects of noise
bias, source extraction and flux determination systematics,
inhomogeneuos noise distribution over the imaged field, and resolution
bias on the measured source counts (\ccor \ corrections hereafter) is to rely on mock samples of radio
sources, as described in \s{sec:mock} . 
As these corrections do not take into account the fraction of spurious sources (as the mock sources are always inserted into the same mosaic) in \s{sec:falsedet} ,\  we separately derive the false detection rate. The combination of the two corrections then yields the net radio source count corrections.

\subsection{\Ccor \ corrections}
\label{sec:mock}

We here describe the Monte Carlo simulations used to derive the \ccor \ corrections.
Mock sources were injected over the imaged field and then
recovered using the same technique adopted for the real radio sources, as detailed in \s{sec:cc} . The flux density and size distributions assumed are described in \s{sec:flux} \ and \s{sec:size} , respectively. The results of the final simulations yielding the adopted \ccor \ corrections are detailed in \s{sec:finalsimul} , and a summary of the 
 effects taken into account by the \ccor \ corrections is given in \s{sec:bias}  .

\subsubsection{Retrieval of mock radio sources}
\label{sec:cc}

The procedure adopted to insert and retrieve mock sources in and from the mosaic is  as follows. 
Since {\tt blobcat} does not produce a residual map, we inserted mock sources (Gaussian in shape) directly into the continuum map avoiding already cataloged components. This procedure was limited to the central two square degrees of the mosaic. For each mock source, a square shape with a width of $6\sigma+21$~pixel  on the side was required to be free of any cataloged emission (real or mock), where $\sigma$ is the standard deviation along the Gaussian major axis. The positions were randomly chosen until this was satisfied. At a resolution of 0.75$\arcsec$ the continuum map is mostly empty of sources and confusion is negligible. 
We did not observe any systematic clustering of mock sources toward the less populated parts of the mosaic (more noisy parts closer to the edge for example) by requiring no overlap between the components. After all mock sources were inserted, we ran {\tt blobcat} with the exact parameters as performed for the real sources. Since the extraction was carried out on a map containing both real and mock emission, all the \ncomp \  real components were always recovered and then removed from the extracted catalog, prior to further processing.
To generate realistic mock catalogs of radio sources, however, we needed to assume i) a flux density distribution
in (and below) the range tested by the observations and ii) an angular
size distribution of the radio sources. This is described in detail in the following sections.

\subsubsection{Flux density distribution}
\label{sec:flux}

We  simulated the flux density distribution using both a 
simple power-law model (PL model)   and a multinode power law (MPL model)
that better reproduces the observed source counts below $500~\mu$Jy.
In the former case we used the 1.4 GHz source counts from previous surveys
scaled to 3 GHz (see \citealt{bondi08}).
The multinode power-law model is that derived by \citet{vernstrom14} (see their Table 4, Zone 1). For both models the mock catalogs were generated down to a total flux density of 5~$\mu$Jy and contained more than 40,000 (65,000) objects in the PL (MPL) model. This also allowed us to count
 sources with flux densities below the S/N threshold as
positive noise fluctuations might lead these to have a measured peak flux density above our source detection threshold. As shown below, the results of our simulations do not yield differences between the two models, and we adopted the MPL model for our final simulations.

\subsubsection{Angular size distribution}
\label{sec:size}

We needed to assign an intrinsic angular size to each mock source. Unfortunately, a satisfying description of the intrinsic source angular size
distribution at sub-mJy flux density is still missing and we needed to rely
on extrapolations from higher flux densities.
\citet{bondi08} used a simple power-law parametrization distribution
of the angular sizes of the sources  as a function of their total flux density. We followed the same method with some adaptations, as described below.

The angular size ($\theta$) distribution was simulated, assuming a power-law relation between angular size and flux density ($\theta\propto S^{n}$).
This relation was normalized using the cumulative angular size distribution
derived at $\sim 1$ mJy from the VVDS 1.4 GHz
observations with a resolution of $6\arcsec$ \citep{bondi03}. The relatively low
resolution of the VVDS survey allowed us to avoid bias against sources with angular sizes of 
up to $15\arcsec$  in our simulations
\citep{bondi03}.
We explored the range of $n$ values between 0.3 and 1.0 in steps of 0.1.
To infer the best $n$ value, the angular size distribution of the sources from the catalog in a specific
total flux density range was compared with the corresponding distribution derived
from the mock samples with different $n$ values. The value of $n$ that
gave the best match between the angular size distribution of observed and mock 
sources was then chosen as the
best approximation for the intrinsic source size versus total flux density
relation.

Since the observed source angular sizes are not provided by {\tt blobcat}, these were estimated using the
relation between the ratio of the total flux density and peak surface brightness and angular sizes,

\begin{equation}
\label{eq:eqgeneral}
{S_t \over S_p} = {\sqrt{\theta^2_M + \theta^2_b} \sqrt{\theta^2_m +
\theta^2_b}\over \theta_b^2}
,\end{equation}

where $S_t$ is the total flux density, $S_p$ is the peak surface brightness, $\theta_b$
is the FWHM of the circular beam ($0.75 \arcsec$ in our observations), 
$\theta_M$ and $\theta_m$
are the major and minor FWHM intrinsic (deconvolved) angular sizes;
see \citealt{bondi08}, where  the same
approach was used to derive a size estimate of sources affected by bandwidth smearing.
In doing so we needed to make some assumptions on the morphology of the sources
and in particular on how the sources are, eventually,  resolved.
We considered two limiting cases as follows:

\begin{enumerate}
\item
{\em Elongated geometry}: sources are resolved in only one direction. This
implies  that $\theta_m=0$ and 

\begin{equation}
\label{eq:elong}
{S_t \over S_p} = \sqrt{{\theta^2_M + \theta^2_b}\over \theta^2_b}
.\end{equation}

The simulated mock sources were accordingly generated as sources extended in
one direction and eq.~\ref{eq:elong} is the appropriate relation between 
$S_t/S_p$ and the angular size.  

\item
{\em Circular geometry}: sources are uniformly resolved in all directions. This implies that
$\theta_M=\theta_m$ and 

\begin{equation}
\label{eq:circ}
{S_t \over S_p} = {{\theta^2_M + \theta^2_b}\over \theta^2_b}
.\end{equation}

The simulated mock sources were accordingly generated as sources uniformly
extended in
all directions and eq.~\ref{eq:circ} is the appropriate relation between 
$S_t/S_p$ and the angular size.  

\end{enumerate}

Mock catalogs were generated for each combination of the 2 source count models
(PL and MPL), the 8 different $n$ values (0.3--1.0 in steps of 0.1), and
the  2 different source geometries ($\theta_m=0$ or $\theta_m=\theta_M$).
For each of these 32 combinations we generated and merged 10 different mock samples.
 Then, we derived for each of the 32 different mock catalogs the 
$S_t/S_p$ distributions for sources with $S_t < 100$ $\mu$Jy,
splitting them into two subranges: $S_t\le 40$ $\mu$Jy,
and $40 < S_t \le 100$  $\mu$Jy.
This range in total flux density is more affected by the choice on
the intrinsic source size distribution and therefore is the best suited
for a comparison between the  $S_t/S_p$ distribution of the real cataloged sources
and that derived from the mock samples reprocessed as the observed catalog.
Using $S_t/S_p$  as a proxy for the angular size of the radio sources
has the advantage that we do not need to deal with upper limits in 
the measured source sizes because of sources classified as unresolved.

The results of this comparison can be summarized as follows.
No significant differences were found using the PL or MPL
distributions for the source counts. For this reason, we were able to adopt either
of the two models in the following analysis and we decided to
use the MPL model, which provides a more realistic and detailed description of the observed
source counts.
However, none of the 16 combinations of $n$ value and source geometry provided a satisfying match between
the $S_t/S_p$ distribution of the reprocessed mock catalog and that of
the observed catalog, in the
flux density range $S_t< 100$ $\mu$Jy.
While some combinations of parameters provided a reasonable match for sources
with $S_t \gtrsim 40 $ $\mu$Jy, they all failed to reproduce the observed distribution 
of $S_t/S_p$ below this
threshold. In particular, the mock samples showed lower values of 
 $S_t/S_p$
than the catalog for $S_t \lesssim 40 $ $\mu$Jy.
This is shown in \f{fig:thetamin} \ where we plot in the two panels the $S_t/S_p$ distribution
for sources with $S_t < 40$ $\mu$Jy and sources with $40 < S_t \le 100$ $\mu$Jy,  respectively.
Together with the observed distribution derived from the sources in the catalog
we also plot the distribution obtained from our original simulation using
$n=0.6$ and elongated geometry. The two distributions are clearly shifted and this effect is found in all the simulations.

This result is not completely unexpected,
The extrapolation to very low flux density of our power-law relation 
between angular size and flux density, which has been previously tested only for sources more than one order of magnitude
brighter, produces mock samples of radio sources dominated by extremely compact objects
at the faint end of the total flux density distribution.
For instance, for the simulations shown in \f{fig:thetamin} ,\ 45\% of 
 all the sources
with $S_t\le 40$ $\mu$Jy have $S_t/S_p < 1$. 
This result is at odds with the distribution of the observed catalog, where only 26\% 
of the observed sources fainter than 40 $\mu$Jy have $S_t/S_p < 1$.

The simplest way to decrease the number of extremely compact objects at the faint end of the flux distribution in our mock sources, without modifying the adopted power-law relation between angular size and flux density, is to apply a minimum angular size to the faint mock sources. We tested the following expression for $\theta_\mathrm{min}$:

\begin{equation}
\theta_{min}= k_1 e^{-(S_t/k_2)^2}
\label{eq:cut-off}
,\end{equation}

in which the exponential part is motivated by the fact that, on the basis of the analysis shown in the lower panel of \f{fig:thetamin} , no $\theta_\mathrm{min}$ is required at flux densities $\gtrsim40$ microJy.
We included the minimum angular size requirement in our 
procedure to generate the mock samples of radio sources and repeated
the simulations, extraction process, and comparison of the
 $S_t/S_p$ distributions. By varying the parameters $k_1$ and $k_2$ we found that the best value for the $k_2$ parameter
is $k_2 =40$ $\mu$Jy, while for the normalization factor 
$k_1$ is equal to 0.3 (for the elongated geometry) and 0.2
(for the circular geometry). The different normalization 
is necessary because for a given intrinsic source size
the area covered by a circular source is larger
 and derived from eq.~\ref{eq:elong} and \ref{eq:circ}.
As shown in \f{fig:thetamin} , 
this time we found a very good agreement between the observed and simulated distributions
of $S_t/S_p$  at low flux densities as well. In particular, for the simulation shown in \f{fig:thetamin} , 
introducing a minimum angular size as a function of the
total flux density reduces the percentage of faint sources ($S_t\le 40$ $\mu$Jy) with 
$S_t/S_p< 1$ from 45\% to 30\% close to the observed value of 26\%. Thus,  we adopted the above-described size parametrization for our final simulations used to derive the \ccor \ corrections. 

We further found that the geometry of the radio sources has some effects on the results  we obtained.
We obtained the best match between 
the  $S_t/S_p$ distributions
for $n=0.5-0.6$ for elongated geometry
 ($\theta_m=0$),  and we obtained the best match for $n=0.6-0.7$ for circular geometry
($\theta_m=\theta_M$). 
We note that both the assumptions made on the source geometry are clearly simplistic
and real sources will consist of a mix of elongated and circular sources. Thus, to compute our final \ccor \ corrections
for the adopted MPL flux density distribution models
we computed the \ccor \ correction factors using the average of
those from the four best-matched simulations: i) elongated sources with $n=0.5$, 
and $n=0.6$, and ii) circular sources with $n=0.6$, and $n=0.7$, 
as described in more detail in the next section.

\begin{figure}[h!]
\includegraphics[clip, trim=0.cm 5cm 0.cm 2cm, width=80mm,angle=0]{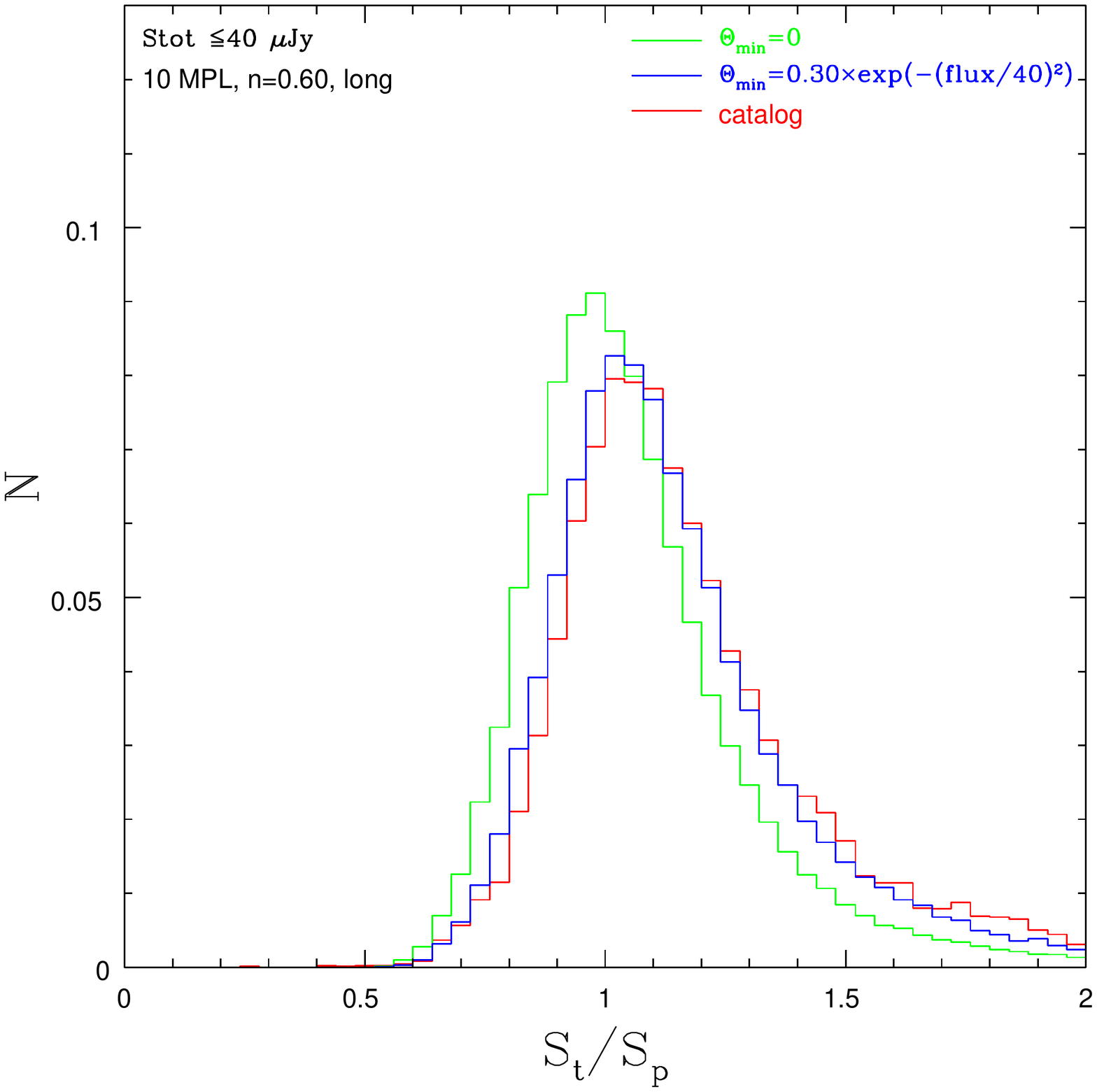}
\includegraphics[clip, trim=0.cm 5cm 0.cm 3cm, width=80mm,angle=0]{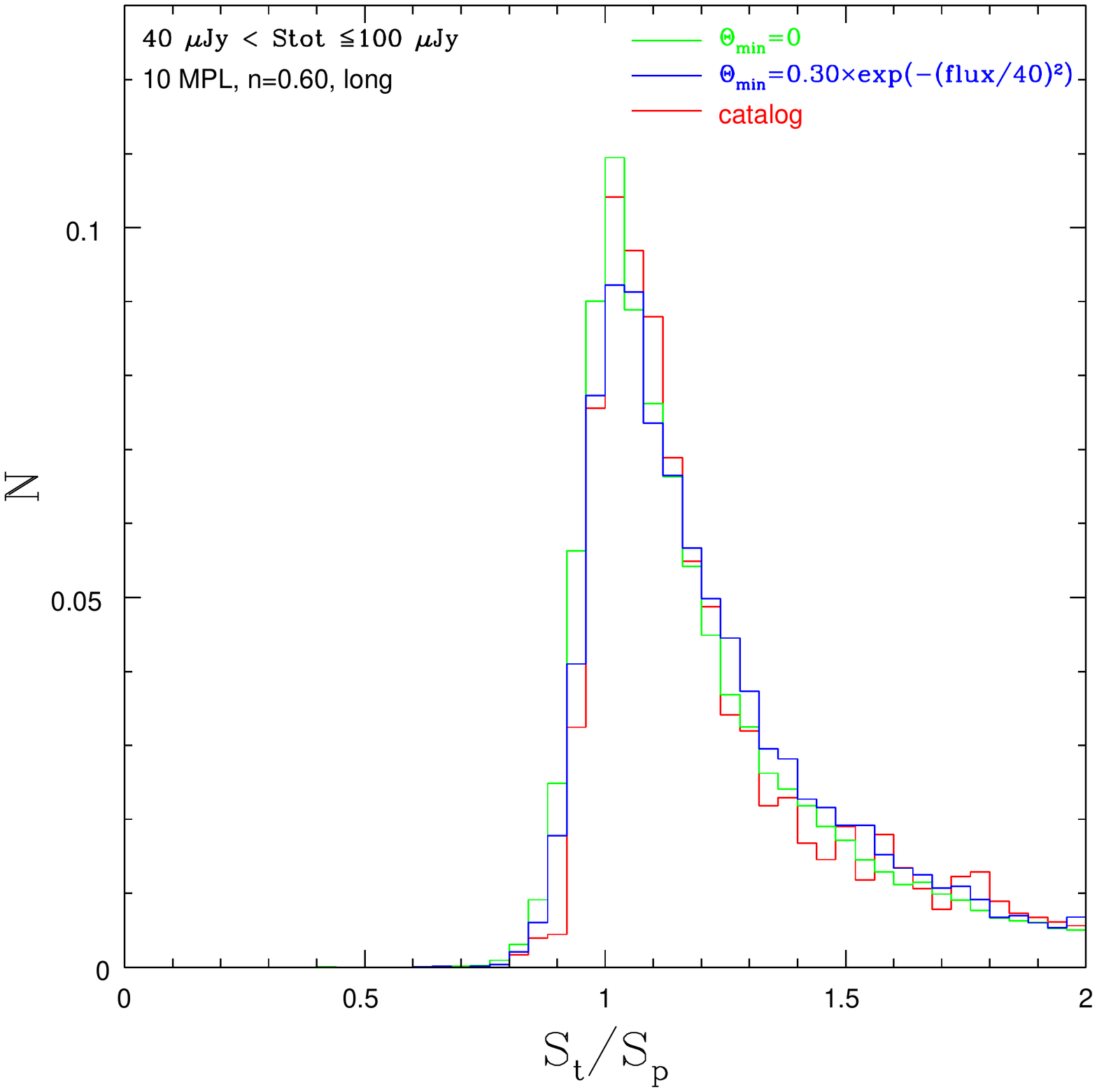}
\caption{Total-to-peak flux density ratio distributions in two total flux density
ranges: $S_t \le 40$ $\mu$Jy (top panel) and $40 < S_t \le 100$ $\mu$Jy
(bottom panel). Each panel shows the distribution of the observed sources
(red histogram) that derived from the 10 sets of simulations using
an elongated geometry and $n=0.6$, no minimum angular size (green histogram)
and with a minimum angular size  $\theta_{min}= 0.3 e^{-(S_t/40)^2}$ (blue
histogram).}
\label{fig:thetamin}
\end{figure}

\subsubsection{Derivation of \ccor \ corrections}
\label{sec:finalsimul}

We generated 60 mock catalogs using the parameterization as described above (\s{sec:flux} \ and \s{sec:size} ; see also below). 
The mock sources were inserted into the mosaic and retrieved as described in \s{sec:cc} .
The retrieved mock sources were then cross-correlated with the input mock catalog and their measured total flux density chosen to be either their integrated flux density if resolved, or peak surface brightness if unresolved. For this, the same $S_t/S_p$ envelope was used as described in  Sect.~\ref{sec:cat}. Lastly, successfully extracted mock sources and original mock sources were binned separately in flux densities. The ratio of their numbers in each flux density bin represents the \ccor \ correction factor.

In \f{fig:complet} \ we show the net result of the above-described Monte Carlo simulations for the MPL model and best-matched simulations, i.e.,\ i) elongated sources with $n=0.5$ and $0.6$, and ii) circular sources with $n=0.6$ and $0.7$. We take the average of these simulations as the \ccor \ correction with a confidence interval that
takes into account the differences within the six sets of simulations. This is tabulated in \t{tab:comp} . For reference, in \f{fig:complet} \ we also show the average \ccor \ corrections obtained using only the elongated and circular geometry approximations. The net curve yields values of about 55\% at $12~\mu$Jy (S/N=5.2),  and rather constant up to $20~\mu$Jy (S/N~$<$9), beyond which they rise to a 94\% completeness above $40~\mu$Jy (S/N$\geq$16).
The mean error of the \ccor \ corrections is 5\%. The two (elongated and circular geometry) approximations are consistent up to $\sim30~\mu$Jy, beyond which they start diverging with the circular approximation being systematically lower at higher flux densities. However, beyond this limit both curves saturate at fairly constant values ($\sim0.92$ for circular and $0.98$ for elongated morphology), implying average values of over  95\% for fluxes higher than $\sim40$~$\mu$Jy.

\begin{figure}
\includegraphics[ width=\columnwidth]{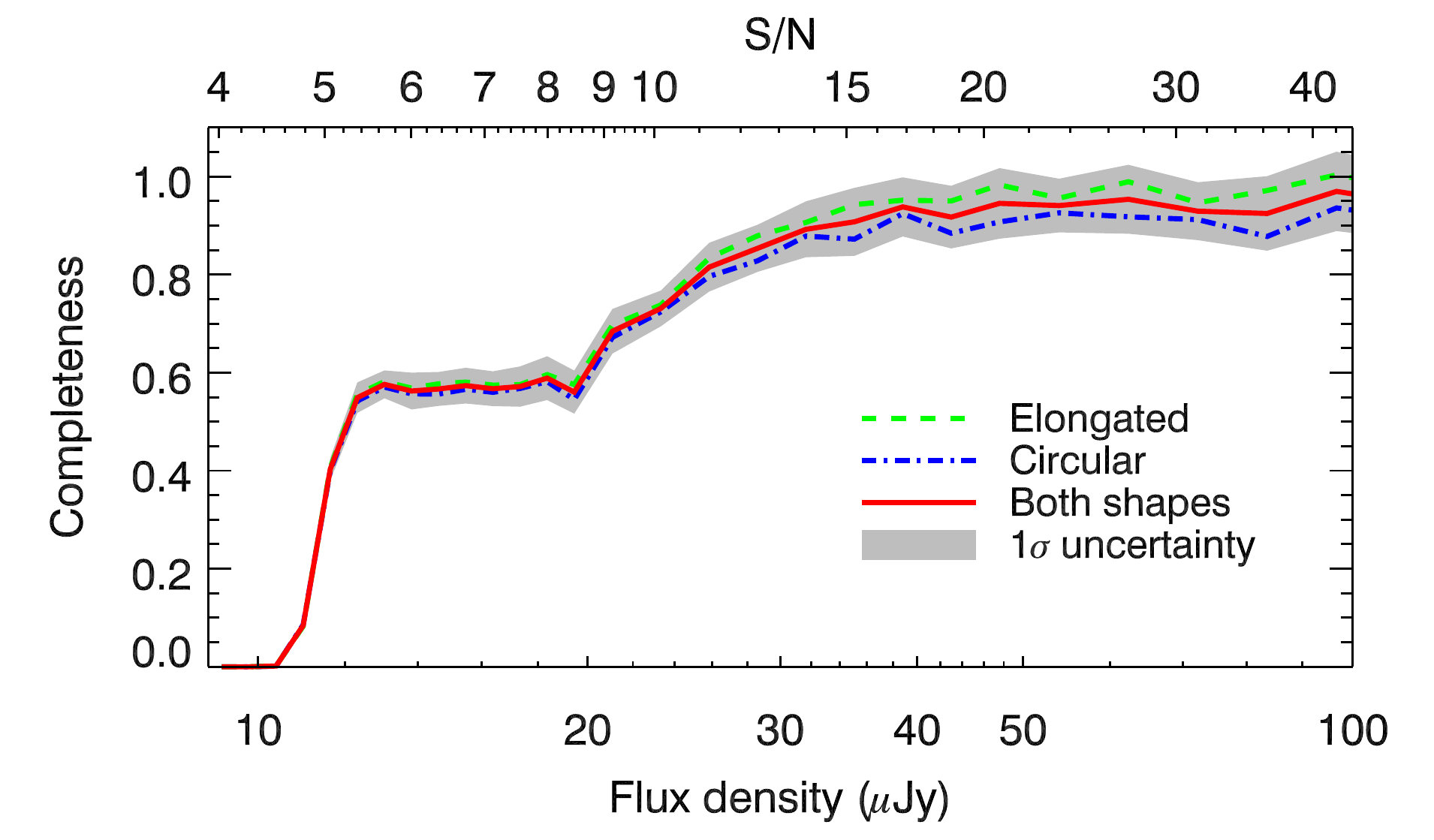}
\caption{ \footnotesize{\baselineskip0.1cm{
Completeness of our 3~GHz source catalog as a function of flux density and S/N. The mean completeness of all Monte Carlo runs (red line) and its standard deviation (gray shaded area) are shown. Also shown are the corrections when elongated (dash-dotted line) and circular (dashed line) geometries are assumed.
}}}
\label{fig:complet}
\end{figure}

\begin{table}
\caption{\Ccor \ correction factors for the VLA-COSMOS 3~GHz catalog as a function of flux density }
\label{tab:comp}
\begin{tabular}{c c c}
\hline
Flux density & \Ccor \ & Error \\
 ($\mu$Jy) & correction factor (C$_\mathrm{compl}$) &  \\
\hline  
$<$ 10.4 & 0 & - \\
11.0 & 0.08 & 0.01 \\
11.6 & 0.40 & 0.02 \\
12.3 & 0.55 & 0.03 \\
13.0 & 0.58 & 0.03 \\
13.8 & 0.56 & 0.04 \\
14.6 & 0.57 & 0.03 \\
15.5 & 0.57 & 0.04 \\
16.4 & 0.57 & 0.04 \\
17.3 & 0.57 & 0.04 \\
18.4 & 0.59 & 0.04 \\
19.4 & 0.56 & 0.04 \\
21.1 & 0.68 & 0.05 \\
23.3 & 0.73 & 0.04 \\
25.8 & 0.82 & 0.05 \\
28.6 & 0.85 & 0.05 \\
31.7 & 0.89 & 0.06 \\
35.1 & 0.91 & 0.07 \\
38.8 & 0.94 & 0.06 \\
43.0 & 0.92 & 0.06 \\
47.6 & 0.95 & 0.07 \\
53.9 & 0.94 & 0.05 \\
62.4 & 0.95 & 0.07 \\
72.2 & 0.93 & 0.06 \\
83.5 & 0.92 & 0.08 \\
96.7 & 0.97 & 0.08 \\
$> 100^\mathrm{a}$ & 1.00$^\mathrm{a}$ & 0.05$^\mathrm{a}$ \\
\hline
\end{tabular}
\vspace{\baselineskip}\\
$^\mathrm{a}$Assumed corrections for fluxes $>100~\mu$Jy.\\
\end{table}

\subsubsection{Biases addressed}
\label{sec:bias}

There are several effects and biases that occur in the cataloging process that we addressed through our simulations.
Firstly, an incompleteness in the extracted catalog exists as real sources on the sky are not be detected if i) their peak surface brightness falls below the chosen threshold of $5\sigma$  because of fluctuations in the local $rms$, or ii) they are extended enough for their peak surface brightness to fall below the detection threshold, even though their integrated flux density is well above it. 
Secondly, a contamination effect is also present. If a source is detected, its flux density might be wrongly computed  because of the presence of a noise peak. Statistically, this effect is mostly symmetric around the mean flux density. However, when we set the total flux density of an unresolved source to its peak surface brightness we may introduce an asymmetric bias toward smaller flux densities. Some sources with $S_t>S_p$ within the envelope in Fig.~\ref{fig:stsp} might truly be resolved, however noise variations do not allow us to determine this with sufficiently high accuracy leading to a potential bias. The final result is that a source can jump into a flux density bin where it does not belong, thus increasing its contamination. The combination of completeness, which always decreases with decreasing flux density, and the significant number of sources that move from their original flux density bin to another  owing to errors in flux measurement at faint flux densities, produces the flat distribution of the completeness and bias correction factor seen at flux densitites of $\sim12 -20~\mu$Jy in \f{fig:complet} .

In summary, the simulations we performed account for both the fraction of nondetected sources (incompleteness), and also the redistribution of sources between various flux density bins. Thus, in principle,  its value can be larger than one if the contamination is high. 
These corrections, however, do not take into account the fraction of spurious sources as a function of flux density, which are   derived separately in the next section.

\subsection{False detection rate}
\label{sec:falsedet}

To assess the false detection rate of our source extraction we ran {\tt blobcat} on the inverted (i.e., multiplied by $-1$) continuum map with the same settings used for the main catalog. Since there is no negative emission on the sky, every source detected in the inverted map is per definition a noise peak (i.e.,\ a false detection). The source extraction returned 414 negative detections with S/N~$\geq5$ across the entire observed field, 95 of which were outside 
 the central two square degrees, demonstrating that 23\% of false detections lies at the edge of the mosaic. 
 
\begin{figure}[t]
\centering
\includegraphics[ width=\columnwidth]{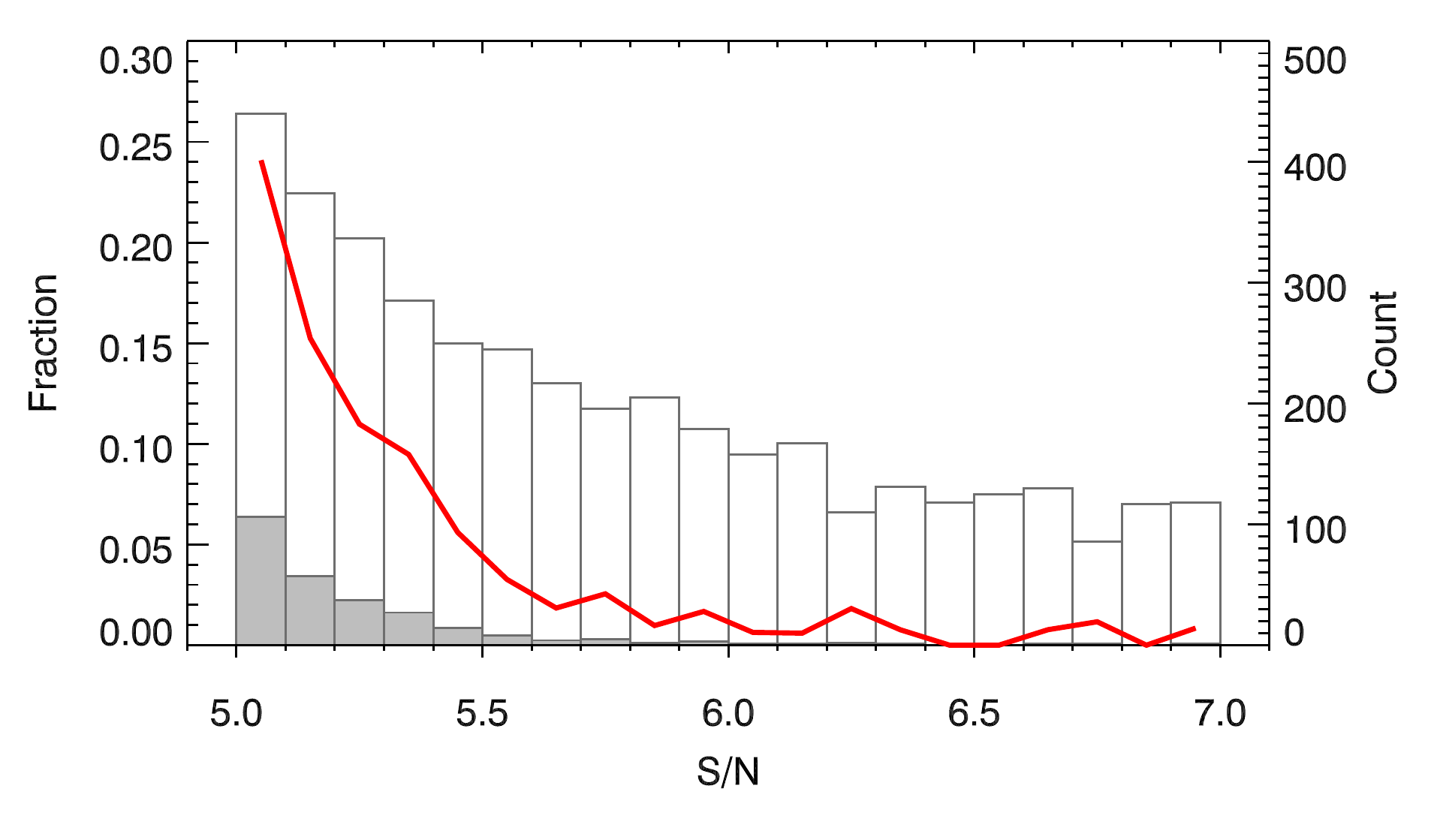}
\includegraphics[ width=\columnwidth]{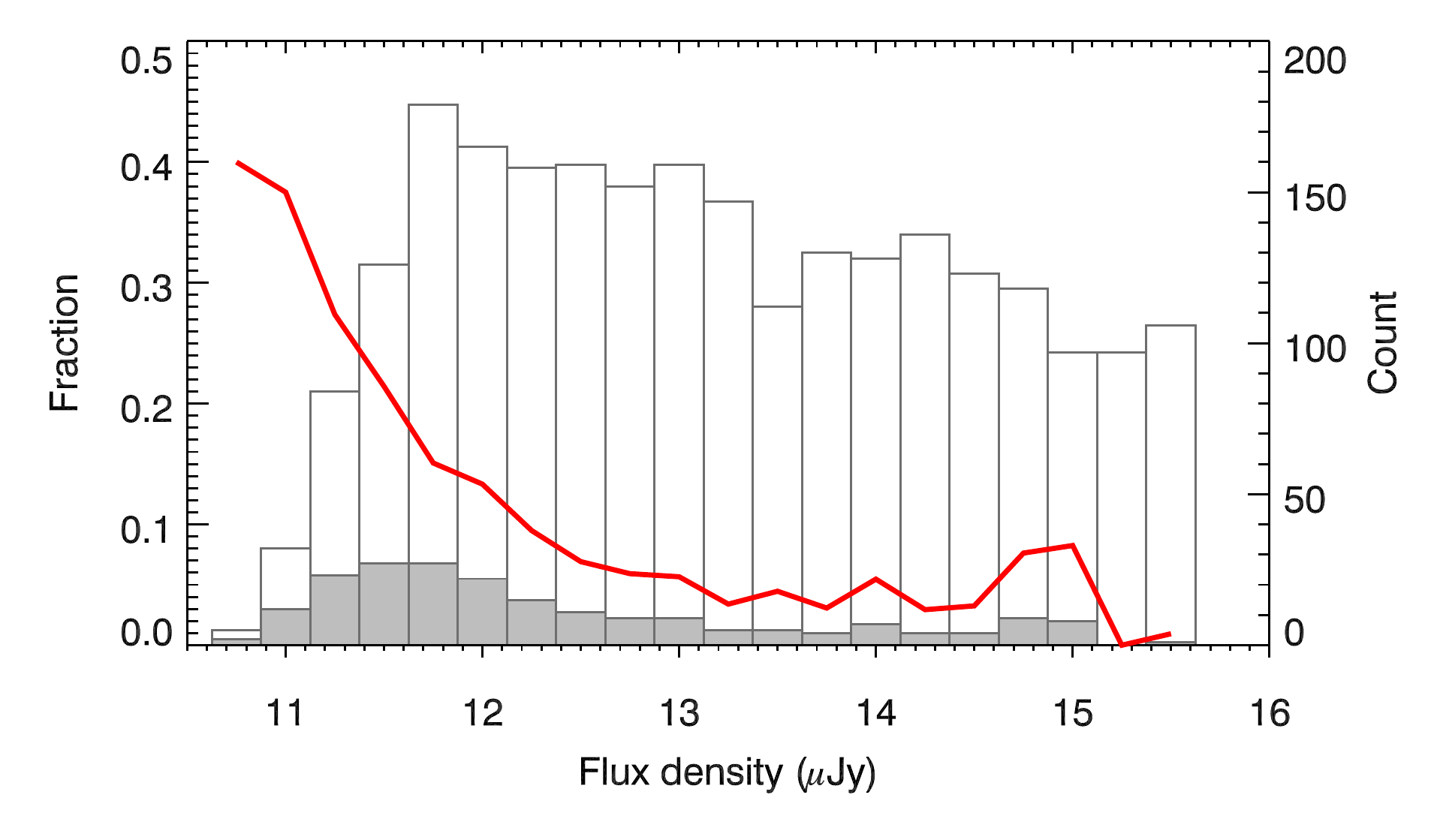}
\caption{ \footnotesize{\baselineskip0.1cm{
Fraction of false detections (red line) as a function of S/N (top panel) and flux density (bottom panel). The open (filled) histogram shows the number of components cataloged in the observed 3~GHz mosaic (detected in the inverted map), and limited to the central two square degrees. The data are also listed in Table~\ref{tab:false_det}. 
}}}
\label{fig:false_det}
\end{figure}

The highest S/N negative detections were predominantly located around true bright sources as they suffer from artifacts; up to six negative components could be found around a single bright object due to the VLA synthesized beam shape (see also Sect. 7.1.1.\ in \citealt{vernstrom14} for an explanation of this effect). Since extraction of real emission does not exhibit this behavior, 
we removed all negative components that were less than 3$\arcsec$ away from a real source with  S/N$>100$. This step  removed further 40 components. We additionally removed four sources with catastrophic peak estimates, which increased their S/N by more than a factor of four owing to poor parabola fits. We note that there were no such sources in the catalog of real emission. The remaining 275 negative detections within the inner two square degrees were then classified into resolved and unresolved using the same envelope as was carried out
for the real data. Finally, they were binned in S/N and flux densities alongside true detections to enable direct comparison. The results are shown in Fig~\ref{fig:false_det} and also listed in Table \ref{tab:false_det}. As expected, only the lowest S/N bins have a noticeable portion of false detections ($24\%$ for S/N$=5.0$--$5.1$), which quickly decreases to less than 3\% for any S/N bin at S/N$>5.5$.  The estimated fraction of spurious sources over the entire catalog above S/N$> 5\, (5.5)$ drawn from the inner two square degrees is only 2.7 (0.4)\% .

\begin{table}[t]
\caption{False detection probability as a function of S/N and flux density in the COSMOS two square degree field}
\centering
\label{tab:false_det}
\begin{tabular}{c c}
\hline
S/N & Fraction\\
 & \\
\hline  
5.05 & 0.24 \\
5.15 & 0.15 \\
5.25 & 0.11 \\
5.35 & 0.09 \\
5.45 & 0.06 \\
5.55 & 0.03 \\
5.65 & 0.02 \\
5.75 & 0.03 \\
5.85 & 0.01 \\
5.95 & 0.02 \\
6.05 & 0.01 \\
6.15 & 0.01 \\
6.25 & 0.02 \\
6.35 & 0.01 \\
6.45 & 0.00 \\
6.55 & 0.00 \\
6.65 & 0.01 \\
6.75 & 0.01 \\
6.85 & 0.00 \\
6.95 & 0.01 \\
\hline\end{tabular}
\quad
\begin{tabular}{c c}
\hline
Flux density & Fraction\\
($\mu$Jy) & (F$_\mathrm{false-det}$) \\
\hline  
10.75 & 0.40 \\
11.00 & 0.38 \\
11.25 & 0.27 \\
11.50 & 0.21 \\
11.75 & 0.15 \\
12.00 & 0.13 \\
12.25 & 0.09 \\
12.50 & 0.07 \\
12.75 & 0.06 \\
13.00 & 0.06 \\
13.25 & 0.03 \\
13.50 & 0.04 \\
13.75 & 0.03 \\
14.00 & 0.05 \\
14.25 & 0.03 \\
14.50 & 0.03 \\
14.75 & 0.08 \\
15.00 & 0.08 \\
15.25 & 0.00 \\
15.50 & 0.01 \\

\hline\end{tabular}

\vspace{\baselineskip}
\end{table}

\section{Radio source counts}
\label{sec:counts}

In this section we present our 3~GHz radio source counts (\s{sec:ourcounts} ), and compare them to 3~GHz and 1.4~GHz counts available in the literature (Sects.~\ref{sec:count3} and \ref{sec:count1.4}, respectively).

\subsection{ VLA-COSMOS 3~GHz radio source counts}
\label{sec:ourcounts}

\begin{figure}[h!]
\centering
\includegraphics[ width=\columnwidth]{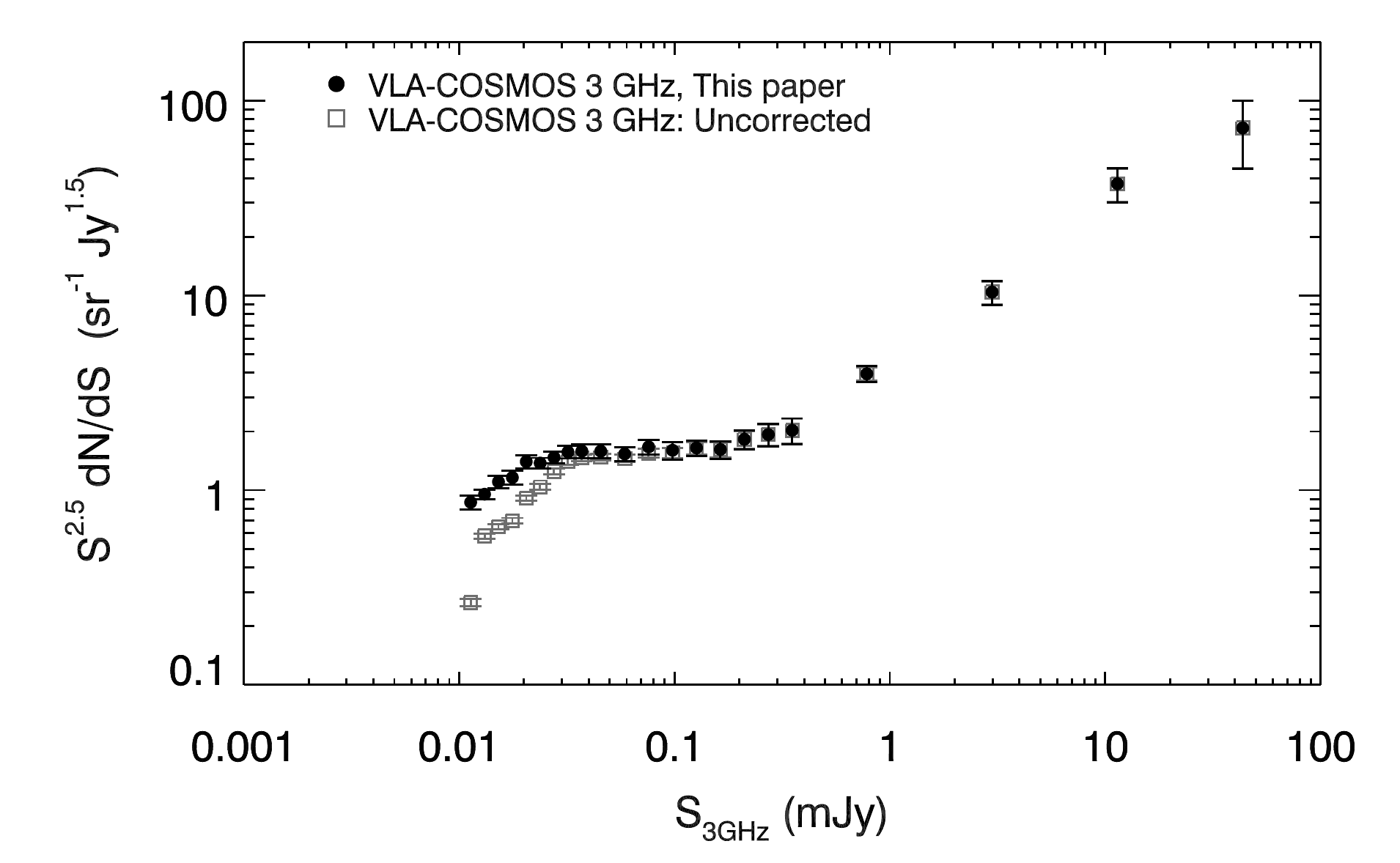}
\includegraphics[ width=\columnwidth]{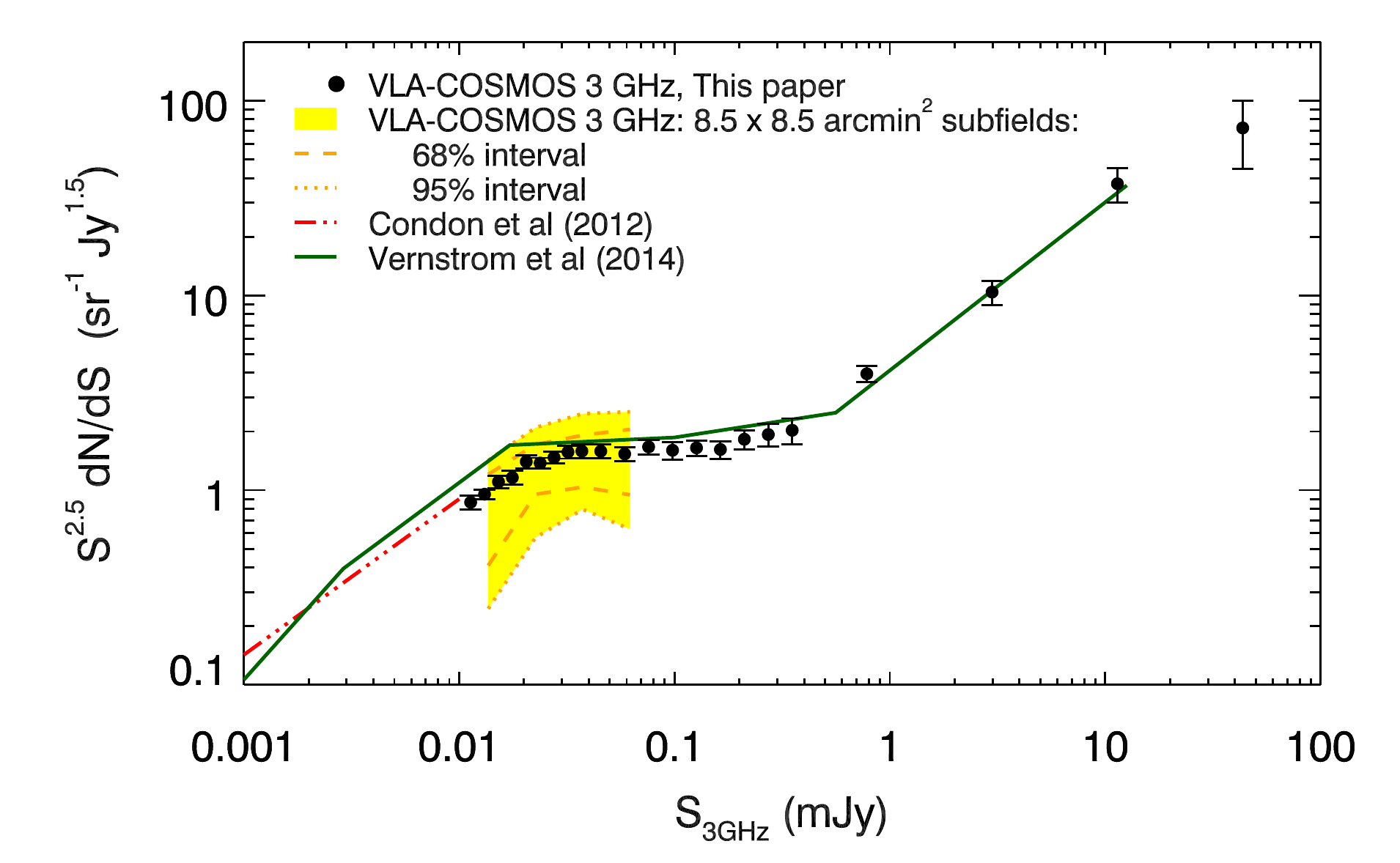}
\includegraphics[ width=\columnwidth]{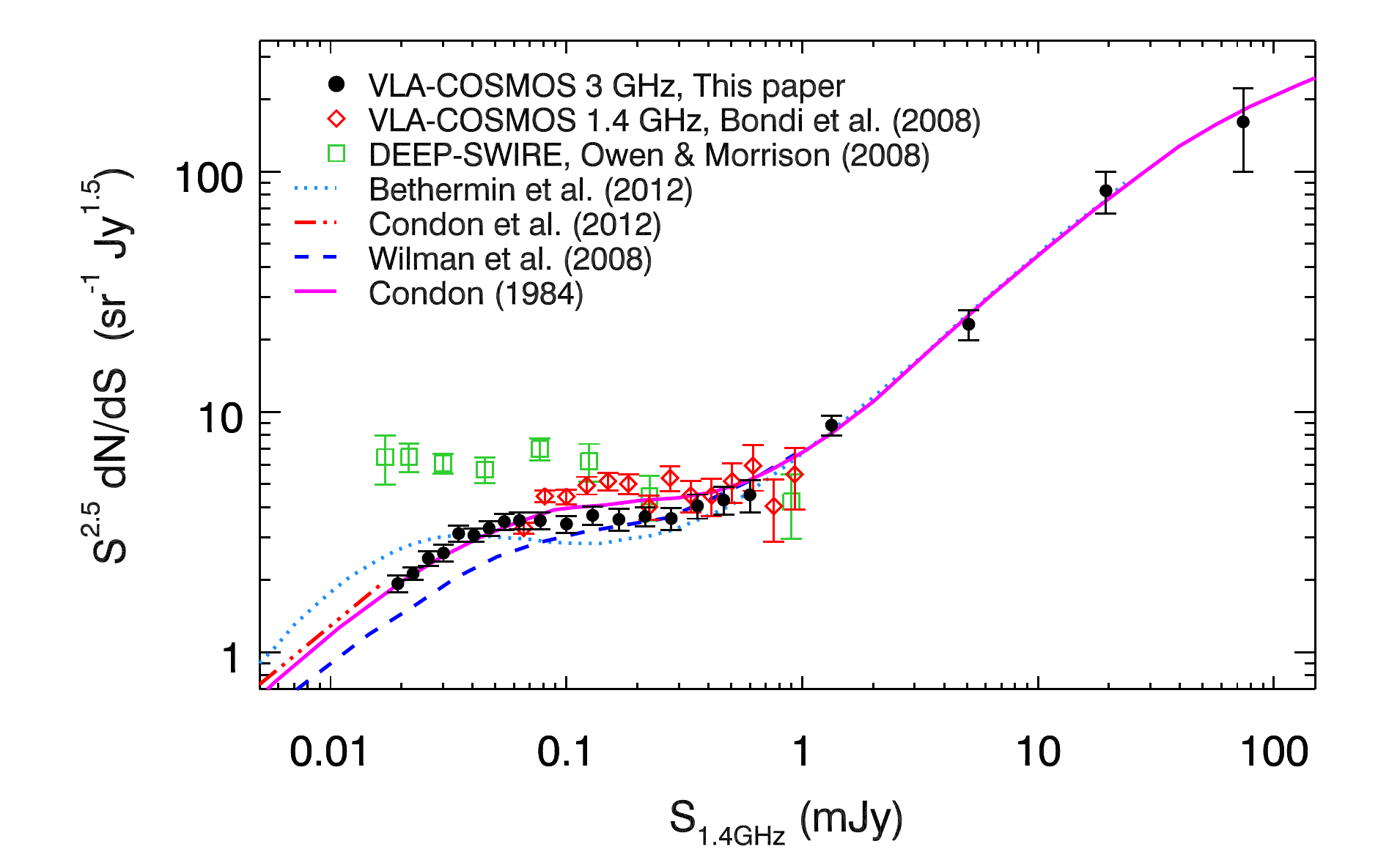}

\caption{ \footnotesize{\baselineskip0.1cm{
Top panel: VLA-COSMOS 3~GHz Euclidean-normalized radio source counts,  corrected using the \ccor \ and false-detection correction factors (black filled points) and without corrections (gray squares).
Middle panel: VLA-COSMOS 3~GHz radio source counts compared to \cite{condon12} $P(D)$ analysis with a single power law (dot-dashed red line) and \cite{vernstrom14} $P(D)$ analysis with multiple power laws  (green line) at 3~GHz. 
The yellow shaded area contains 95\% of different source counts obtained from 100 square and nonoverlapping ($8.5 \times  8.5$~arcmin$^2$) subfields of the COSMOS field, thus demonstrating the effect of cosmic variance  on fields with sizes similar to those analyzed by \cite{condon12} and \cite{vernstrom14}. The dashed orange line shows the 68\% interval of different source counts (obtained from 16th and 84th percentile in each flux density bin).
Bottom panel: counts of the same sources, but shifted to the 1.4~GHz observed frame using a spectral index of  $\alpha=-0.7$ prior to binning (black filled points). A selection of existing 1.4~GHz source counts in the literature is also shown, as indicated in the legend.
}}}
\label{fig:counts}
\end{figure}

We present our 3~GHz radio source counts normalized to Euclidean geometry,  both corrected and uncorrected, in the top panel of Fig.~\ref{fig:counts}.  In Table~\ref{tab:counts}  we  list the counts, errors, number of sources, and radio source count corrections (i.e., \ccor \ and false detection fraction corrections) in each flux density bin. The source count errors take into account both the Poissonian errors as well as \ccor \ correction uncertainties. Most of our sources are located at low flux densities (below 0.5~mJy) with more than 500 sources in each flux density bin below 60~$\mu$Jy resulting in small Poissonian errors.
As evident from the plot, our source counts at 3~GHz exhibit a flattening at about 0.3~mJy that continues one order of magnitude in flux densities down to 30~$\mu$Jy, steepening further at fainter flux densities.

\begin{table}[t]
\caption{Radio source counts at 3~GHz within the  COSMOS two square degree field, normalized to Euclidean geometry}
\label{tab:counts}
\begin{tabular}{c c c c c}
\hline
Flux density & Counts$^\mathrm{a}$ & Error$^\mathrm{b}$ & N & Correction \\
(mJy) & (Jy$^{1.5}$sr$^{-1}$) &  (Jy$^{1.5}$sr$^{-1}$)  & &  factor$^\mathrm{a}$ \\
\hline  
0.011 & 0.866 & 0.068 & 631 & 3.27 \\
0.013 & 0.952 & 0.056 & 1109 & 1.64 \\
0.015 & 1.10 & 0.078 & 991 & 1.70 \\
0.018 & 1.16 & 0.094 & 849 & 1.67 \\
0.020 & 1.40 & 0.11 & 888 & 1.54 \\
0.024 & 1.38 & 0.086 & 811 & 1.33 \\
0.028 & 1.47 & 0.10 & 780 & 1.18 \\
0.032 & 1.57 & 0.12 & 702 & 1.12 \\
0.037 & 1.59 & 0.13 & 587 & 1.08 \\
0.045 & 1.58 & 0.13 & 753 & 1.07 \\
0.059 & 1.53 & 0.12 & 505 & 1.05 \\
0.076 & 1.67 & 0.14 & 366 & 1.08 \\
0.098 & 1.60 & 0.17 & 250 & 1.03 \\
0.13 & 1.65 & 0.15 & 181 & 1.00 \\
0.16 & 1.62 & 0.17 & 121 & 1.00 \\
0.21 & 1.82 & 0.21 & 93 & 1.00 \\
0.27 & 1.93 & 0.25 & 67 & 1.00 \\
0.35 & 2.03 & 0.31 & 48 & 1.00 \\
0.78 & 3.95 & 0.37 & 159 & 1.00 \\
3.0 & 10.4 & 1.5 & 56 & 1.00 \\
11 & 37.4 & 7.4 & 27 & 1.00 \\
44 & 72.3 & 28 & 7 & 1.00 \\
\hline\end{tabular}
\vspace{\baselineskip}\\
$^\mathrm{a}$~The listed counts were corrected for \ccor \  (C$_\mathrm{compl}$), as well as false detection fractions (F$_\mathrm{false-det}$), by multiplying the raw counts by the correction factor given in the last column, and equal to (1-F$_\mathrm{false-det}$)/C$_\mathrm{compl}$ (see \t{tab:comp} \ and \t{tab:false_det} ).\\
$^\mathrm{b}$~ The source count errors take into account only the Poissonian errors and completeness and bias correction uncertainties (see text for details).
\\
\end{table}

\subsection{Comparison with 3~GHz counts from the literature }
\label{sec:count3}

In the middle panel of Fig.~\ref{fig:counts} we compare our 3~GHz source counts with other 3~GHz counts available in the literature \citep{condon12, vernstrom14}. 
\citet{condon12} performed a $P(D)$ analysis using 3 GHz confusion-limited data based on 50 hours of on-source C-array configuration observations of one VLA pointing targeting the Lockman hole and reaching an $rms$ of $1~\mu$Jy~beam$^{-1}$. Fitting single power-law  models to the data the analysis allowed these investigators to constrain the counts of discrete sources in the $1-10~\mu$Jy range, also shown in Fig.~\ref{fig:counts}. 
\citet{vernstrom14} performed a more complex $P(D)$ analysis on the same data  fitting various (modified power-law, and node-based) models  to the data allowing them to probe the counts down to $0.1~\mu$Jy. 
In Fig.~\ref{fig:counts} we show the counts based on the fit of a phenomenological parametric model of multiple joined power laws (their node-based model) applied to the inner circular area with a $5\arcmin$ radius (their Zone 1; see \citealt{vernstrom14} for details). 

The counts derived here are in very good agreement with those derived by \citet{condon12}. Fitting the five faintest flux density bins using a power law, $dN/dS\propto S^\gamma$, we find that the slope $\gamma=-1.72$ is perfectly consistent with that inferred by \citet{condon12}, while our normalization is slightly lower.
Our comparison to the \citet{vernstrom14} results shows that
the counts are in agreement down to $\sim30~\mu$Jy with a  discrepancy at fainter flux densitites as our counts are systematically lower than theirs. 

In general, the strength of the $P(D)$ analysis is the ability to probe counts below the nominal noise in the data, while avoiding resolution biases as it is applied on confusion-limited (thus, low resolution) data. However, as the $P(D)$ analyses discussed above were performed on a single VLA pointing, the resulting counts may be subject to cosmic variance due to the small area covered. This could potentially explain the observed discrepancy between the VLA-COSMOS 3~GHz Large Project counts based on a two square degree area and the \citet{vernstrom14} results based on a 0.022 square degree area (their Zone 1). To test this we subdivided the two square degree COSMOS field into 100 square and nonoverlapping subfields, each with an area of 0.020 deg$^2$ roughly corresponding to a circle with a radius of $5\arcmin$. In the middle panel of Fig.~\ref{fig:counts}, we show the range of such obtained counts (corrected for \ccor , and false detection fractions, calculated on the full two square degrees and described in \s{sec:completeness} ).  We find that sample variance that is quantified in this way can introduce a (1$\sigma$) scatter of $^{+0.1}_{-0.2}$~dex in the source counts.   The distribution in counts in the 100 subfields are likely to be an underestimate of the true cosmic variance, which is dominated by cosmic large-scale structure, rather than sample variance, because these fields are likely not fully independent from each other. Thus, cosmic variance may explain the observed discrepancy. 

\subsection{Comparison with 1.4~GHz counts from the literature }
\label{sec:count1.4}

To compare our result with more abundant 1.4~GHz observations and models  (e.g., \citealt{condon84,bondi08,owen08,wilman08,dezotti10,condon12}) 
we scale our flux densities to the 1.4~GHz observed frame using a spectral index of -0.7. This value, which is also in agreement with the spectral index survival analysis described in Sect.~\ref{sec:alpha}, is commonly used and provides the easiest comparison (e.g., \citealt{condon12}). We show the 1.4~GHz source count comparison in the bottom panel of Fig.~\ref{fig:counts}.

The large spread of the 1.4~GHz source counts available in the literature at submillijansky levels (see, e.g., Fig.~1 in \citealt{smo15}) is usually attributed to a combination of i) cosmic variance as often the observed fields are rather small (see \f{fig:surveys} \ and middle panel of \f{fig:counts} ), and ii) resolution bias leading to a loss of sources  in radio continuum surveys conducted at intermediate to high ($\lesssim2\arcsec$) angular resolution (as described in more detail in \s{sec:bias} ). The large, two square degree area of the COSMOS field minimizes the effect of cosmic variance, and in \s{sec:completeness} \ we performed extensive Monte Carlo simulations to account for potential resolution biases. Our source counts agree well with those derived by \cite{condon12} based on the $P(D)$ analysis at the faint end (see previous section). 

The counts derived here are in good agreement with those derived from the VLA-COSMOS 1.4~GHz Large Project (\citealt{schinnerer07,bondi08}; red diamonds in \f{fig:counts} ) at flux densities higher than $\sim200~\mu$Jy,  but are slightly lower in the flux density range of $100-200~\mu$Jy. 
As this is the same field, cosmic variance cannot explain the discrepancy. The uncorrected counts from the two surveys are in very good agreement; the difference is the largest in the flux density range where the 1.4~GHz survey is the least complete (about 60\%), and the corrections, thus, are the largest. In the same flux density range the corrections for the 3~GHz survey are not as severe given the much higher sensitivity of the 3~GHz survey. Further reasons that could explain part of the discrepancy are i) 
the effect of BWS on the radio source count corrections that are present in the 1.4~GHz data, but are not present in the 3~GHz data (see \citealt{bondi08}), and ii) a possibly overly simplistic scaling of the 3~GHz counts to 1.4~GHz using  just one spectral index value.
   Source counts at 1.4~GHz depend on the steepness of the counts at 3~GHz and the spread of the spectral indices. We leave the analysis of the potential bias in source counts due to this effect to an upcoming paper (Novak et al., in prep.).
   
The largest discrepancy between the counts derived here and those in the literature is observed relative to the \citet{owen08} results. \citet{owen08} have observed the Lockman hole at 1.4~GHz in A-, B-, C-, and D-array configurations with the VLA reaching an angular resolution of $\sim1.6\arcsec$ and $rms\approx 2.7~\mu$Jy~beam$^{-1}$. To correct for the resolution bias, they assumed a source size distribution with an extended tail at the high end (see their Fig.~8) and that distribution remains constant as a function of flux density. The source count corrections are significant under these assumptions and result in a flat source count distribution at flux densities fainter than $\sim200~\mu$Jy (green points in the bottom panel of \f{fig:counts} ). As already discussed by \citet{condon12} and \citet{vernstrom14}, these corrections are most likely overestimated. 
In contrast, for the corrections applied to the data presented here we assumed a  model for radio source sizes such that the radio size is a function of flux density with a limiting minimum size (see \s{sec:mock} ). The agreement between our source counts and those derived from  confusion-limited data \citep{condon12, vernstrom14} further strengthens the validity of this assumption.

In \f{fig:counts} \ we also compare our results with the models developed by \citet{condon84}, \citet{wilman08}, and \citet{bethermin12}. The faint end of our counts ($\lesssim80~\mu$Jy), combined with the results from \citet{condon12} that appear as an extrapolation of our data, agree the best with the \citet{condon84} model. The model was constrained by source counts, redshift, and spectral-index distributions for various 400~MHz to 5~GHz flux-limited samples as well as the local 1.4~GHz luminosity function for two dominant, spiral and elliptical galaxy populations. The adopted model is not a unique solution and evolves all sources, i.e., ellipticals and spirals, steep- and flat-spectrum sources, in the same way. At flux densities above $\sim80~\mu$Jy the \citet{condon84} model is slightly higher than our derived source counts and is consistent with the counts determined by \citet{vernstrom14}. 

Our derived source counts deviate from those predicted by the \citet{wilman08} and \citet{bethermin12} models. While they agree with the first down to $\sim100~\mu$Jy, they are systematically higher at fainter flux densities.  On the other hand, the \citet{bethermin12} model underpredicts our counts in the flux density range of $\sim50-300~\mu$Jy, while it overpredicts the counts at flux densities $\lesssim30~\mu$Jy. The discrepancies may possibly be understood when considering how AGN and star-forming galaxies were implemented in the models. \citet{bethermin12} implement only models for X-ray selected AGN ($L_\mathrm{2-10~keV}\sim10^{42}-10^{44}$~erg~s$^{-1}$; see \citealt{mullaney11,mullaney12,aird12}), and thus ignore the population of radio-loud AGN hosted by red, quiescent galaxies,  regularly not identified as X-ray AGN, yet still substantial (e.g., \citealt{best06,smo08,bonzini13}; \smo \ et al, submitted; Delvecchio et al., accepted.). This could explain the lack of sources with flux densities in the range of $\sim50-300~\mu$Jy in the model (see, e.g., \citealt{smo08,padovani15}; \smo \ et al., submitted), compared to the observational results.
On the other hand, \citet{bethermin12} model the star-forming galaxy population using the most recent results from \citealt{bouwens07,rodighiero11,magnelli11,karim11,sargent12} by tracing the star-forming galaxy main sequence and the stellar mass function over cosmic time, while also taking into account main-sequence and starburst galaxy spectral energy distribution libraries. Therefore, the excess of \citet{bethermin12} model compared to that of \citet{wilman08} could suggest that the Wilman model carries potential for improvement in modeling the star-forming galaxy population.


\section{Summary and conclusions}
\label{sec:summary}

We presented the VLA-COSMOS 3~GHz Large Project based on 384 hours of observations with the Karl G. Jansky Very Large Array at 3~GHz (10~cm) toward the two square degree COSMOS field. 
Our final mosaic, imaged per pointing with the multiscale multifrequency algorithm and self-calibration, reaches a median $rms$ of 2.3$~\mu$Jy~beam$^{-1}$ over the two square degrees, at an angular resolution of $0.75\arcsec$. 
We further presented a catalog of \nsource \ radio sources. Combining our data with the 1.4~GHz VLA-COSMOS Joint Project data using survival analysis, we found the expected median spectral index $\alpha$ of -0.7. Comparing the positions of our 3~GHz sources with those from the high-resolution VLBA imaging at 1.4~GHz, we estimated that the astrometry is accurate to $0.01\arcsec$ at the bright end.
Radio source count corrections were calculated for the central two square degrees and used to infer radio source counts. 
The radio angular size parametrization adopted based on the comparison of mock versus real source total over peak flux density ratios suggests that the angular sizes of radio sources at these flux density levels can be modeled as a power law in flux density ($\theta\propto S^n$) with a minimal, flux-dependent size cutoff (eq.~\ref{eq:cut-off}).
Our corrected radio counts with direct detections down to $20~\mu$Jy (at 1.4~GHz) are consistent with those derived based on P(D) analyses \citep{condon12}, and agree best with the \citet{condon84} model, while they are systematically higher than those predicted by the SKADS (Square Kilometer Array Design Studies) simulations \citep{wilman08}.

The VLA-COSMOS 3~GHz Large Project simultaneously provides the largest and deepest radio continuum survey  to date, bridging the gap between radio continuum surveys conducted with past generation and those planned with the next generation facilities. These radio data, in conjunction with the vast panchromatic COSMOS data sets, will allow for the exploration of  various cosmologically relevant topics, such as i) the characteristics of the microJansky radio population, ii) radio-quiet
QSOs by direct detection in the radio band, iii) modes of star formation at early cosmic epochs, and iv) studying stellar mass growth in typical galaxies since early cosmic epochs and star formation quenching via AGN feedback.

\begin{acknowledgements}

We thank the referee, Jim Condon, for insightful comments that helped improve the manuscript. 
Based on observations with the National
Radio Astronomy Observatory which is a facility of the National Science
Foundation operated under cooperative agreement by Associated Universities,
Inc. 
 This research was funded by the European Union's Seventh Frame-work 
programs under grant agreements 333654 (CIG, 'AGN feedback') and 337595 (ERC Starting Grant, 'CoSMass'). MB and PC acknowlege support from the PRIN-INAF 2014. AK and FB acknowledge support by the Collaborative Research Council 956, sub-project A1, funded by the Deutsche Forschungsgemeinschaft (DFG). BM and FB acknowledge support through DFG priority program 1573 founded by the DFG.  
\end{acknowledgements}

\bibliographystyle{apj} 
\bibliography{bibtex.bib}

\newcommand{\noop}[1]{}
\begin{thebibliography}{}
\expandafter\ifx\csname natexlab\endcsname\relax\def\natexlab#1{#1}\fi

\bibitem[{{Afonso} {et~al.}(2005){Afonso}, {Georgakakis}, {Almeida}, {Hopkins},
  {Cram}, {Mobasher}, \& {Sullivan}}]{afonso05}
{Afonso}, J., {Georgakakis}, A., {Almeida}, C., {et~al.} 2005, \apj, 624, 135

\bibitem[{{Aihara} {et~al.}(2011){Aihara}, {Allende Prieto}, {An}, {Anderson},
  {Aubourg}, {Balbinot}, {Beers}, {Berlind}, {Bickerton}, {Bizyaev}, {Blanton},
  {Bochanski}, {Bolton}, {Bovy}, {Brandt}, {Brinkmann}, {Brown}, {Brownstein},
  {Busca}, {Campbell}, {Carr}, {Chen}, {Chiappini}, {Comparat}, {Connolly},
  {Cortes}, {Croft}, {Cuesta}, {da Costa}, {Davenport}, {Dawson}, {Dhital},
  {Ealet}, {Ebelke}, {Edmondson}, {Eisenstein}, {Escoffier}, {Esposito},
  {Evans}, {Fan}, {Femen{\'{\i}}a Castell{\'a}}, {Font-Ribera}, {Frinchaboy},
  {Ge}, {Gillespie}, {Gilmore}, {Gonz{\'a}lez Hern{\'a}ndez}, {Gott}, {Gould},
  {Grebel}, {Gunn}, {Hamilton}, {Harding}, {Harris}, {Hawley}, {Hearty}, {Ho},
  {Hogg}, {Holtzman}, {Honscheid}, {Inada}, {Ivans}, {Jiang}, {Johnson},
  {Jordan}, {Jordan}, {Kazin}, {Kirkby}, {Klaene}, {Knapp}, {Kneib},
  {Kochanek}, {Koesterke}, {Kollmeier}, {Kron}, {Lampeitl}, {Lang}, {Le Goff},
  {Lee}, {Lin}, {Long}, {Loomis}, {Lucatello}, {Lundgren}, {Lupton}, {Ma},
  {MacDonald}, {Mahadevan}, {Maia}, {Makler}, {Malanushenko}, {Malanushenko},
  {Mandelbaum}, {Maraston}, {Margala}, {Masters}, {McBride}, {McGehee},
  {McGreer}, {M{\'e}nard}, {Miralda-Escud{\'e}}, {Morrison}, {Mullally},
  {Muna}, {Munn}, {Murayama}, {Myers}, {Naugle}, {Neto}, {Nguyen}, {Nichol},
  {O'Connell}, {Ogando}, {Olmstead}, {Oravetz}, {Padmanabhan},
  {Palanque-Delabrouille}, {Pan}, {Pandey}, {P{\^a}ris}, {Percival},
  {Petitjean}, {Pfaffenberger}, {Pforr}, {Phleps}, {Pichon}, {Pieri}, {Prada},
  {Price-Whelan}, {Raddick}, {Ramos}, {Reyl{\'e}}, {Rich}, {Richards}, {Rix},
  {Robin}, {Rocha-Pinto}, {Rockosi}, {Roe}, {Rollinde}, {Ross}, {Ross},
  {Rossetto}, {S{\'a}nchez}, {Sayres}, {Schlegel}, {Schlesinger}, {Schmidt},
  {Schneider}, {Sheldon}, {Shu}, {Simmerer}, {Simmons}, {Sivarani}, {Snedden},
  {Sobeck}, {Steinmetz}, {Strauss}, {Szalay}, {Tanaka}, {Thakar}, {Thomas},
  {Tinker}, {Tofflemire}, {Tojeiro}, {Tremonti}, {Vandenberg}, {Vargas
  Maga{\~n}a}, {Verde}, {Vogt}, {Wake}, {Wang}, {Weaver}, {Weinberg}, {White},
  {White}, {Yanny}, {Yasuda}, {Yeche}, \& {Zehavi}}]{aihara11}
{Aihara}, H., {Allende Prieto}, C., {An}, D., {et~al.} 2011, \apjs, 193, 29

\bibitem[{{Aird} {et~al.}(2012){Aird}, {Coil}, {Moustakas}, {Blanton},
  {Burles}, {Cool}, {Eisenstein}, {Smith}, {Wong}, \& {Zhu}}]{aird12}
{Aird}, J., {Coil}, A.~L., {Moustakas}, J., {et~al.} 2012, \apj, 746, 90

\bibitem[{{Aretxaga} {et~al.}(2011){Aretxaga}, {Wilson}, {Aguilar}, {Alberts},
  {Scott}, {Scoville}, {Yun}, {Austermann}, {Downes}, {Ezawa}, {Hatsukade},
  {Hughes}, {Kawabe}, {Kohno}, {Oshima}, {Perera}, {Tamura}, \&
  {Zeballos}}]{aretxaga11}
{Aretxaga}, I., {Wilson}, G.~W., {Aguilar}, E., {et~al.} 2011, \mnras, 415,
  3831

\bibitem[{{Becker} {et~al.}(1995){Becker}, {White}, \& {Helfand}}]{becker95}
{Becker}, R.~H., {White}, R.~L., \& {Helfand}, D.~J. 1995, \apj, 450, 559

\bibitem[{{Bertoldi} {et~al.}(2007){Bertoldi}, {Carilli}, {Aravena},
  {Schinnerer}, {Voss}, {Smolcic}, {Jahnke}, {Scoville}, {Blain}, {Menten},
  {Lutz}, {Brusa}, {Taniguchi}, {Capak}, {Mobasher}, {Lilly}, {Thompson},
  {Aussel}, {Kreysa}, {Hasinger}, {Aguirre}, {Schlaerth}, \&
  {Koekemoer}}]{bertoldi07}
{Bertoldi}, F., {Carilli}, C., {Aravena}, M., {et~al.} 2007, \apjs, 172, 132

\bibitem[{{Best} {et~al.}(2006){Best}, {Kaiser}, {Heckman}, \&
  {Kauffmann}}]{best06}
{Best}, P.~N., {Kaiser}, C.~R., {Heckman}, T.~M., \& {Kauffmann}, G. 2006,
  \mnras, 368, L67

\bibitem[{{B{\'e}thermin} {et~al.}(2012){B{\'e}thermin}, {Daddi}, {Magdis},
  {Sargent}, {Hezaveh}, {Elbaz}, {Le Borgne}, {Mullaney}, {Pannella}, {Buat},
  {Charmandaris}, {Lagache}, \& {Scott}}]{bethermin12}
{B{\'e}thermin}, M., {Daddi}, E., {Magdis}, G., {et~al.} 2012, \apjl, 757, L23

\bibitem[{{Bock} {et~al.}(1999){Bock}, {Large}, \& {Sadler}}]{bock99}
{Bock}, D.~C.-J., {Large}, M.~I., \& {Sadler}, E.~M. 1999, \aj, 117, 1578

\bibitem[{{Bondi} {et~al.}(2008){Bondi}, {Ciliegi}, {Schinnerer}, {Smol{\v
  c}i{\'c}}, {Jahnke}, {Carilli}, \& {Zamorani}}]{bondi08}
{Bondi}, M., {Ciliegi}, P., {Schinnerer}, E., {et~al.} 2008, \apj, 681, 1129

\bibitem[{{Bondi} {et~al.}(2003){Bondi}, {Ciliegi}, {Zamorani}, {Gregorini},
  {Vettolani}, {Parma}, {de Ruiter}, {Le Fevre}, {Arnaboldi}, {Guzzo},
  {Maccagni}, {Scaramella}, {Adami}, {Bardelli}, {Bolzonella}, {Bottini},
  {Cappi}, {Foucaud}, {Franzetti}, {Garilli}, {Gwyn}, {Ilbert}, {Iovino}, {Le
  Brun}, {Marano}, {Marinoni}, {McCracken}, {Meneux}, {Pollo}, {Pozzetti},
  {Radovich}, {Ripepi}, {Rizzo}, {Scodeggio}, {Tresse}, {Zanichelli}, \&
  {Zucca}}]{bondi03}
{Bondi}, M., {Ciliegi}, P., {Zamorani}, G., {et~al.} 2003, \aap, 403, 857

\bibitem[{{Bondi} {et~al.}(2007){Bondi}, {Ciliegi}, {Venturi}, {Dallacasa},
  {Bardelli}, {Zucca}, {Athreya}, {Gregorini}, {Zanichelli}, {Le F{\`e}vre},
  {Contini}, {Garilli}, {Iovino}, {Temporin}, \& {Vergani}}]{bondi07}
{Bondi}, M., {Ciliegi}, P., {Venturi}, T., {et~al.} 2007, \aap, 463, 519

\bibitem[{{Bonzini} {et~al.}(2013){Bonzini}, {Padovani}, {Mainieri},
  {Kellermann}, {Miller}, {Rosati}, {Tozzi}, \& {Vattakunnel}}]{bonzini13}
{Bonzini}, M., {Padovani}, P., {Mainieri}, V., {et~al.} 2013, \mnras, 436, 3759

\bibitem[{{Bonzini} {et~al.}(2012){Bonzini}, {Mainieri}, {Padovani},
  {Kellermann}, {Miller}, {Rosati}, {Tozzi}, {Vattakunnel}, {Balestra},
  {Brandt}, {Luo}, \& {Xue}}]{bonzini12}
{Bonzini}, M., {Mainieri}, V., {Padovani}, P., {et~al.} 2012, \apjs, 203, 15

\bibitem[{{Bourke} {et~al.}(2014){Bourke}, {Mooley}, \& {Hallinan}}]{bourke14}
{Bourke}, S., {Mooley}, K., \& {Hallinan}, G. 2014, in Astronomical Society of
  the Pacific Conference Series, Vol. 485, Astronomical Data Analysis Software
  and Systems XXIII, ed. N.~{Manset} \& P.~{Forshay}, 367

\bibitem[{{Bouwens} {et~al.}(2007){Bouwens}, {Illingworth}, {Franx}, \&
  {Ford}}]{bouwens07}
{Bouwens}, R.~J., {Illingworth}, G.~D., {Franx}, M., \& {Ford}, H. 2007, \apj,
  670, 928

\bibitem[{{Bower} {et~al.}(2006){Bower}, {Benson}, {Malbon}, {Helly}, {Frenk},
  {Baugh}, {Cole}, \& {Lacey}}]{bower06}
{Bower}, R.~G., {Benson}, A.~J., {Malbon}, R., {et~al.} 2006, \mnras, 370, 645

\bibitem[{{Capak} {et~al.}(2007){Capak}, {Aussel}, {Ajiki}, {McCracken},
  {Mobasher}, {Scoville}, {Shopbell}, {Taniguchi}, {Thompson}, {Tribiano},
  {Sasaki}, {Blain}, {Brusa}, {Carilli}, {Comastri}, {Carollo}, {Cassata},
  {Colbert}, {Ellis}, {Elvis}, {Giavalisco}, {Green}, {Guzzo}, {Hasinger},
  {Ilbert}, {Impey}, {Jahnke}, {Kartaltepe}, {Kneib}, {Koda}, {Koekemoer},
  {Komiyama}, {Leauthaud}, {Le Fevre}, {Lilly}, {Liu}, {Massey}, {Miyazaki},
  {Murayama}, {Nagao}, {Peacock}, {Pickles}, {Porciani}, {Renzini}, {Rhodes},
  {Rich}, {Salvato}, {Sanders}, {Scarlata}, {Schiminovich}, {Schinnerer},
  {Scodeggio}, {Sheth}, {Shioya}, {Tasca}, {Taylor}, {Yan}, \&
  {Zamorani}}]{capak07}
{Capak}, P., {Aussel}, H., {Ajiki}, M., {et~al.} 2007, \apjs, 172, 99

\bibitem[{{Ciliegi} {et~al.}(1999){Ciliegi}, {McMahon}, {Miley}, {Gruppioni},
  {Rowan-Robinson}, {Cesarsky}, {Danese}, {Franceschini}, {Genzel}, {Lawrence},
  {Lemke}, {Oliver}, {Puget}, \& {Rocca-Volmerange}}]{ciliegi99}
{Ciliegi}, P., {McMahon}, R.~G., {Miley}, G., {et~al.} 1999, \mnras, 302, 222

\bibitem[{{Civano} {et~al.}(2016){Civano}, {Marchesi}, {Comastri}, {Urry},
  {Elvis}, {Cappelluti}, {Puccetti}, {Brusa}, {Zamorani}, {Hasinger},
  {Aldcroft}, {Alexander}, {Allevato}, {Brunner}, {Capak}, {Finoguenov},
  {Fiore}, {Fruscione}, {Gilli}, {Glotfelty}, {Griffiths}, {Hao}, {Harrison},
  {Jahnke}, {Kartaltepe}, {Karim}, {LaMassa}, {Lanzuisi}, {Miyaji}, {Ranalli},
  {Salvato}, {Sargent}, {Scoville}, {Schawinski}, {Schinnerer}, {Silverman},
  {Smolcic}, {Stern}, {Toft}, {Trakhenbrot}, {Treister}, \&
  {Vignali}}]{civano16}
{Civano}, F., {Marchesi}, S., {Comastri}, A., {et~al.} 2016, \apj, 819, 62

\bibitem[{{Condon}(1984)}]{condon84}
{Condon}, J.~J. 1984, \apj, 287, 461

\bibitem[{{Condon}(1992)}]{condon92}
---. 1992, \araa, 30, 575

\bibitem[{{Condon}(2015)}]{condon15}
---. 2015, ArXiv e-prints, arXiv:1502.05616

\bibitem[{{Condon} {et~al.}(1998){Condon}, {Cotton}, {Greisen}, {Yin},
  {Perley}, {Taylor}, \& {Broderick}}]{condon98}
{Condon}, J.~J., {Cotton}, W.~D., {Greisen}, E.~W., {et~al.} 1998, \aj, 115,
  1693

\bibitem[{{Condon} {et~al.}(2003){Condon}, {Cotton}, {Yin}, {Shupe},
  {Storrie-Lombardi}, {Helou}, {Soifer}, \& {Werner}}]{condon03}
{Condon}, J.~J., {Cotton}, W.~D., {Yin}, Q.~F., {et~al.} 2003, \aj, 125, 2411

\bibitem[{{Condon} {et~al.}(2012){Condon}, {Cotton}, {Fomalont}, {Kellermann},
  {Miller}, {Perley}, {Scott}, {Vernstrom}, \& {Wall}}]{condon12}
{Condon}, J.~J., {Cotton}, W.~D., {Fomalont}, E.~B., {et~al.} 2012, \apj, 758,
  23

\bibitem[{{Croton} {et~al.}(2006){Croton}, {Springel}, {White}, {De Lucia},
  {Frenk}, {Gao}, {Jenkins}, {Kauffmann}, {Navarro}, \& {Yoshida}}]{croton06}
{Croton}, D.~J., {Springel}, V., {White}, S.~D.~M., {et~al.} 2006, \mnras, 365,
  11

\bibitem[{{de Zotti} {et~al.}(2010){de Zotti}, {Massardi}, {Negrello}, \&
  {Wall}}]{dezotti10}
{de Zotti}, G., {Massardi}, M., {Negrello}, M., \& {Wall}, J. 2010, \aapr, 18,
  1

\bibitem[{{Dickinson} {et~al.}(2003){Dickinson}, {Giavalisco}, \& {GOODS
  Team}}]{dickinson03}
{Dickinson}, M., {Giavalisco}, M., \& {GOODS Team}. 2003, in The Mass of
  Galaxies at Low and High Redshift, ed. R.~{Bender} \& A.~{Renzini}, 324

\bibitem[{{Driver} {et~al.}(2009){Driver}, {Norberg}, {Baldry}, {Bamford},
  {Hopkins}, {Liske}, {Loveday}, {Peacock}, {Hill}, {Kelvin}, {Robotham},
  {Cross}, {Parkinson}, {Prescott}, {Conselice}, {Dunne}, {Brough}, {Jones},
  {Sharp}, {van Kampen}, {Oliver}, {Roseboom}, {Bland-Hawthorn}, {Croom},
  {Ellis}, {Cameron}, {Cole}, {Frenk}, {Couch}, {Graham}, {Proctor}, {De
  Propris}, {Doyle}, {Edmondson}, {Nichol}, {Thomas}, {Eales}, {Jarvis},
  {Kuijken}, {Lahav}, {Madore}, {Seibert}, {Meyer}, {Staveley-Smith},
  {Phillipps}, {Popescu}, {Sansom}, {Sutherland}, {Tuffs}, \&
  {Warren}}]{driver09}
{Driver}, S.~P., {Norberg}, P., {Baldry}, I.~K., {et~al.} 2009, Astronomy and
  Geophysics, 50, 12

\bibitem[{{Driver} {et~al.}(2011){Driver}, {Hill}, {Kelvin}, {Robotham},
  {Liske}, {Norberg}, {Baldry}, {Bamford}, {Hopkins}, {Loveday}, {Peacock},
  {Andrae}, {Bland-Hawthorn}, {Brough}, {Brown}, {Cameron}, {Ching}, {Colless},
  {Conselice}, {Croom}, {Cross}, {de Propris}, {Dye}, {Drinkwater}, {Ellis},
  {Graham}, {Grootes}, {Gunawardhana}, {Jones}, {van Kampen}, {Maraston},
  {Nichol}, {Parkinson}, {Phillipps}, {Pimbblet}, {Popescu}, {Prescott},
  {Roseboom}, {Sadler}, {Sansom}, {Sharp}, {Smith}, {Taylor}, {Thomas},
  {Tuffs}, {Wijesinghe}, {Dunne}, {Frenk}, {Jarvis}, {Madore}, {Meyer},
  {Seibert}, {Staveley-Smith}, {Sutherland}, \& {Warren}}]{driver11}
{Driver}, S.~P., {Hill}, D.~T., {Kelvin}, L.~S., {et~al.} 2011, \mnras, 413,
  971

\bibitem[{{Elvis} {et~al.}(2009){Elvis}, {Civano}, {Vignali}, {Puccetti},
  {Fiore}, {Cappelluti}, {Aldcroft}, {Fruscione}, {Zamorani}, {Comastri},
  {Brusa}, {Gilli}, {Miyaji}, {Damiani}, {Koekemoer}, {Finoguenov}, {Brunner},
  {Urry}, {Silverman}, {Mainieri}, {Hasinger}, {Griffiths}, {Carollo}, {Hao},
  {Guzzo}, {Blain}, {Calzetti}, {Carilli}, {Capak}, {Ettori}, {Fabbiano},
  {Impey}, {Lilly}, {Mobasher}, {Rich}, {Salvato}, {Sanders}, {Schinnerer},
  {Scoville}, {Shopbell}, {Taylor}, {Taniguchi}, \& {Volonteri}}]{elvis09}
{Elvis}, M., {Civano}, F., {Vignali}, C., {et~al.} 2009, \apjs, 184, 158

\bibitem[{{Evans} {et~al.}(2006){Evans}, {Worrall}, {Hardcastle}, {Kraft}, \&
  {Birkinshaw}}]{evans06}
{Evans}, D.~A., {Worrall}, D.~M., {Hardcastle}, M.~J., {Kraft}, R.~P., \&
  {Birkinshaw}, M. 2006, \apj, 642, 96

\bibitem[{{Feigelson} \& {Nelson}(1985)}]{feigelson85}
{Feigelson}, E.~D., \& {Nelson}, P.~I. 1985, \apj, 293, 192

\bibitem[{{Fixsen} {et~al.}(2009){Fixsen}, {Kogut}, {Levin}, {Limon}, {Lubin},
  {Mirel}, {Seiffert}, {Singal}, {Wollack}, {Villela}, \&
  {Wuensche}}]{fixsen09}
{Fixsen}, D.~J., {Kogut}, A., {Levin}, S., {et~al.} 2009, ArXiv e-prints,
  arXiv:0901.0555

\bibitem[{{Georgakakis} {et~al.}(1999){Georgakakis}, {Mobasher}, {Cram},
  {Hopkins}, {Lidman}, \& {Rowan-Robinson}}]{georgakakis99}
{Georgakakis}, A., {Mobasher}, B., {Cram}, L., {et~al.} 1999, \mnras, 306, 708

\bibitem[{{Grogin} {et~al.}(2011){Grogin}, {Kocevski}, {Faber}, {Ferguson},
  {Koekemoer}, {Riess}, {Acquaviva}, {Alexander}, {Almaini}, {Ashby}, {Barden},
  {Bell}, {Bournaud}, {Brown}, {Caputi}, {Casertano}, {Cassata}, {Castellano},
  {Challis}, {Chary}, {Cheung}, {Cirasuolo}, {Conselice}, {Roshan Cooray},
  {Croton}, {Daddi}, {Dahlen}, {Dav{\'e}}, {de Mello}, {Dekel}, {Dickinson},
  {Dolch}, {Donley}, {Dunlop}, {Dutton}, {Elbaz}, {Fazio}, {Filippenko},
  {Finkelstein}, {Fontana}, {Gardner}, {Garnavich}, {Gawiser}, {Giavalisco},
  {Grazian}, {Guo}, {Hathi}, {H{\"a}ussler}, {Hopkins}, {Huang}, {Huang},
  {Jha}, {Kartaltepe}, {Kirshner}, {Koo}, {Lai}, {Lee}, {Li}, {Lotz}, {Lucas},
  {Madau}, {McCarthy}, {McGrath}, {McIntosh}, {McLure}, {Mobasher},
  {Moustakas}, {Mozena}, {Nandra}, {Newman}, {Niemi}, {Noeske}, {Papovich},
  {Pentericci}, {Pope}, {Primack}, {Rajan}, {Ravindranath}, {Reddy}, {Renzini},
  {Rix}, {Robaina}, {Rodney}, {Rosario}, {Rosati}, {Salimbeni}, {Scarlata},
  {Siana}, {Simard}, {Smidt}, {Somerville}, {Spinrad}, {Straughn}, {Strolger},
  {Telford}, {Teplitz}, {Trump}, {van der Wel}, {Villforth}, {Wechsler},
  {Weiner}, {Wiklind}, {Wild}, {Wilson}, {Wuyts}, {Yan}, \& {Yun}}]{grogin11}
{Grogin}, N.~A., {Kocevski}, D.~D., {Faber}, S.~M., {et~al.} 2011, \apjs, 197,
  35

\bibitem[{{Haarsma} {et~al.}(2000){Haarsma}, {Partridge}, {Windhorst}, \&
  {Richards}}]{haarsma00}
{Haarsma}, D.~B., {Partridge}, R.~B., {Windhorst}, R.~A., \& {Richards}, E.~A.
  2000, \apj, 544, 641

\bibitem[{{Hales} {et~al.}(2012){Hales}, {Murphy}, {Curran}, {Middelberg},
  {Gaensler}, \& {Norris}}]{hales12}
{Hales}, C.~A., {Murphy}, T., {Curran}, J.~R., {et~al.} 2012, \mnras, 425, 979

\bibitem[{{Hales} {et~al.}(2014){Hales}, {Norris}, {Gaensler}, {Middelberg},
  {Chow}, {Hopkins}, {Huynh}, {Lenc}, \& {Mao}}]{hales14}
{Hales}, C.~A., {Norris}, R.~P., {Gaensler}, B.~M., {et~al.} 2014, \mnras, 441,
  2555

\bibitem[{{Hardcastle} {et~al.}(2007){Hardcastle}, {Evans}, \&
  {Croston}}]{hardcastle07}
{Hardcastle}, M., {Evans}, D., \& {Croston}, J. 2007, \mnras, 376, 1849

\bibitem[{{Hasinger} {et~al.}(2007){Hasinger}, {Cappelluti}, {Brunner},
  {Brusa}, {Comastri}, {Elvis}, {Finoguenov}, {Fiore}, {Franceschini}, {Gilli},
  {Griffiths}, {Lehmann}, {Mainieri}, {Matt}, {Matute}, {Miyaji}, {Molendi},
  {Paltani}, {Sanders}, {Scoville}, {Tresse}, {Urry}, {Vettolani}, \&
  {Zamorani}}]{hasinger07}
{Hasinger}, G., {Cappelluti}, N., {Brunner}, H., {et~al.} 2007, \apjs, 172, 29

\bibitem[{{Hopkins} {et~al.}(2000){Hopkins}, {Georgakakis}, {Cram}, {Afonso},
  \& {Mobasher}}]{hopkins00}
{Hopkins}, A., {Georgakakis}, A., {Cram}, L., {Afonso}, J., \& {Mobasher}, B.
  2000, \apjs, 128, 469

\bibitem[{{Hopkins} {et~al.}(2003){Hopkins}, {Afonso}, {Chan}, {Cram},
  {Georgakakis}, \& {Mobasher}}]{hopkins03}
{Hopkins}, A.~M., {Afonso}, J., {Chan}, B., {et~al.} 2003, \aj, 125, 465

\bibitem[{{Ilbert} {et~al.}(2013){Ilbert}, {McCracken}, {Le F{\`e}vre},
  {Capak}, {Dunlop}, {Karim}, {Renzini}, {Caputi}, {Boissier}, {Arnouts},
  {Aussel}, {Comparat}, {Guo}, {Hudelot}, {Kartaltepe}, {Kneib}, {Krogager},
  {Le Floc'h}, {Lilly}, {Mellier}, {Milvang-Jensen}, {Moutard}, {Onodera},
  {Richard}, {Salvato}, {Sanders}, {Scoville}, {Silverman}, {Taniguchi},
  {Tasca}, {Thomas}, {Toft}, {Tresse}, {Vergani}, {Wolk}, \& {Zirm}}]{ilbert13}
{Ilbert}, O., {McCracken}, H.~J., {Le F{\`e}vre}, O., {et~al.} 2013, \aap, 556,
  A55

\bibitem[{{Jarvis}(2012)}]{jarvis12}
{Jarvis}, M.~J. 2012, African Skies, 16, 44

\bibitem[{{Karim} {et~al.}(2011){Karim}, {Schinnerer},
  {Mart{\'{\i}}nez-Sansigre}, {Sargent}, {van der Wel}, {Rix}, {Ilbert},
  {Smol{\v c}i{\'c}}, {Carilli}, {Pannella}, {Koekemoer}, {Bell}, \&
  {Salvato}}]{karim11}
{Karim}, A., {Schinnerer}, E., {Mart{\'{\i}}nez-Sansigre}, A., {et~al.} 2011,
  \apj, 730, 61

\bibitem[{{Kimball} \& {Ivezi{\'c}}(2008)}]{kimball08}
{Kimball}, A.~E., \& {Ivezi{\'c}}, {\v Z}. 2008, \aj, 136, 684

\bibitem[{{Koekemoer} {et~al.}(2007){Koekemoer}, {Aussel}, {Calzetti}, {Capak},
  {Giavalisco}, {Kneib}, {Leauthaud}, {Le F{\`e}vre}, {McCracken}, {Massey},
  {Mobasher}, {Rhodes}, {Scoville}, \& {Shopbell}}]{koekoemor07}
{Koekemoer}, A.~M., {Aussel}, H., {Calzetti}, D., {et~al.} 2007, \apjs, 172,
  196

\bibitem[{{Koekemoer} {et~al.}(2011){Koekemoer}, {Faber}, {Ferguson}, {Grogin},
  {Kocevski}, {Koo}, {Lai}, {Lotz}, {Lucas}, {McGrath}, {Ogaz}, {Rajan},
  {Riess}, {Rodney}, {Strolger}, {Casertano}, {Castellano}, {Dahlen},
  {Dickinson}, {Dolch}, {Fontana}, {Giavalisco}, {Grazian}, {Guo}, {Hathi},
  {Huang}, {van der Wel}, {Yan}, {Acquaviva}, {Alexander}, {Almaini}, {Ashby},
  {Barden}, {Bell}, {Bournaud}, {Brown}, {Caputi}, {Cassata}, {Challis},
  {Chary}, {Cheung}, {Cirasuolo}, {Conselice}, {Roshan Cooray}, {Croton},
  {Daddi}, {Dav{\'e}}, {de Mello}, {de Ravel}, {Dekel}, {Donley}, {Dunlop},
  {Dutton}, {Elbaz}, {Fazio}, {Filippenko}, {Finkelstein}, {Frazer}, {Gardner},
  {Garnavich}, {Gawiser}, {Gruetzbauch}, {Hartley}, {H{\"a}ussler},
  {Herrington}, {Hopkins}, {Huang}, {Jha}, {Johnson}, {Kartaltepe},
  {Khostovan}, {Kirshner}, {Lani}, {Lee}, {Li}, {Madau}, {McCarthy},
  {McIntosh}, {McLure}, {McPartland}, {Mobasher}, {Moreira}, {Mortlock},
  {Moustakas}, {Mozena}, {Nandra}, {Newman}, {Nielsen}, {Niemi}, {Noeske},
  {Papovich}, {Pentericci}, {Pope}, {Primack}, {Ravindranath}, {Reddy},
  {Renzini}, {Rix}, {Robaina}, {Rosario}, {Rosati}, {Salimbeni}, {Scarlata},
  {Siana}, {Simard}, {Smidt}, {Snyder}, {Somerville}, {Spinrad}, {Straughn},
  {Telford}, {Teplitz}, {Trump}, {Vargas}, {Villforth}, {Wagner}, {Wandro},
  {Wechsler}, {Weiner}, {Wiklind}, {Wild}, {Wilson}, {Wuyts}, \&
  {Yun}}]{koekemoer11}
{Koekemoer}, A.~M., {Faber}, S.~M., {Ferguson}, H.~C., {et~al.} 2011, \apjs,
  197, 36

\bibitem[{{Krishna} {et~al.}(2014){Krishna}, {Sirothia}, {Mhaskey}, {Ranadive},
  {Wiita}, {Goyal}, {Kantharia}, \& {Ishwara-Chandra}}]{krishna14}
{Krishna}, G., {Sirothia}, S.~K., {Mhaskey}, M., {et~al.} 2014, \mnras, 443,
  2824

\bibitem[{{Laigle} {et~al.}(2016){Laigle}, {McCracken}, {Ilbert}, {Hsieh},
  {Davidzon}, {Capak}, {Hasinger}, {Silverman}, {Pichon}, {Coupon}, {Aussel},
  {Le Borgne}, {Caputi}, {Cassata}, {Chang}, {Civano}, {Dunlop}, {Fynbo},
  {Kartaltepe}, {Koekemoer}, {Le F{\`e}vre}, {Le Floc'h}, {Leauthaud}, {Lilly},
  {Lin}, {Marchesi}, {Milvang-Jensen}, {Salvato}, {Sanders}, {Scoville},
  {Smolcic}, {Stockmann}, {Taniguchi}, {Tasca}, {Toft}, {Vaccari}, \&
  {Zabl}}]{laigle16}
{Laigle}, C., {McCracken}, H.~J., {Ilbert}, O., {et~al.} 2016, \apjs, 224, 24

\bibitem[{{Le F{\`e}vre} {et~al.}(2015){Le F{\`e}vre}, {Tasca}, {Cassata},
  {Garilli}, {Le Brun}, {Maccagni}, {Pentericci}, {Thomas}, {Vanzella},
  {Zamorani}, {Zucca}, {Amorin}, {Bardelli}, {Capak}, {Cassar{\`a}},
  {Castellano}, {Cimatti}, {Cuby}, {Cucciati}, {de la Torre}, {Durkalec},
  {Fontana}, {Giavalisco}, {Grazian}, {Hathi}, {Ilbert}, {Lemaux}, {Moreau},
  {Paltani}, {Ribeiro}, {Salvato}, {Schaerer}, {Scodeggio}, {Sommariva},
  {Talia}, {Taniguchi}, {Tresse}, {Vergani}, {Wang}, {Charlot}, {Contini},
  {Fotopoulou}, {L{\'o}pez-Sanjuan}, {Mellier}, \& {Scoville}}]{lefevre15}
{Le F{\`e}vre}, O., {Tasca}, L.~A.~M., {Cassata}, P., {et~al.} 2015, \aap, 576,
  A79

\bibitem[{{Lilly} {et~al.}(2007){Lilly}, {Le F{\`e}vre}, {Renzini}, {Zamorani},
  {Scodeggio}, {Contini}, {Carollo}, {Hasinger}, {Kneib}, {Iovino}, {Le Brun},
  {Maier}, {Mainieri}, {Mignoli}, {Silverman}, {Tasca}, {Bolzonella},
  {Bongiorno}, {Bottini}, {Capak}, {Caputi}, {Cimatti}, {Cucciati}, {Daddi},
  {Feldmann}, {Franzetti}, {Garilli}, {Guzzo}, {Ilbert}, {Kampczyk}, {Kovac},
  {Lamareille}, {Leauthaud}, {Borgne}, {McCracken}, {Marinoni}, {Pello},
  {Ricciardelli}, {Scarlata}, {Vergani}, {Sanders}, {Schinnerer}, {Scoville},
  {Taniguchi}, {Arnouts}, {Aussel}, {Bardelli}, {Brusa}, {Cappi}, {Ciliegi},
  {Finoguenov}, {Foucaud}, {Franceschini}, {Halliday}, {Impey}, {Knobel},
  {Koekemoer}, {Kurk}, {Maccagni}, {Maddox}, {Marano}, {Marconi}, {Meneux},
  {Mobasher}, {Moreau}, {Peacock}, {Porciani}, {Pozzetti}, {Scaramella},
  {Schiminovich}, {Shopbell}, {Smail}, {Thompson}, {Tresse}, {Vettolani},
  {Zanichelli}, \& {Zucca}}]{lilly07}
{Lilly}, S.~J., {Le F{\`e}vre}, O., {Renzini}, A., {et~al.} 2007, \apjs, 172,
  70

\bibitem[{{Lilly} {et~al.}(2009){Lilly}, {Le Brun}, {Maier}, {Mainieri},
  {Mignoli}, {Scodeggio}, {Zamorani}, {Carollo}, {Contini}, {Kneib}, {Le
  F{\`e}vre}, {Renzini}, {Bardelli}, {Bolzonella}, {Bongiorno}, {Caputi},
  {Coppa}, {Cucciati}, {de la Torre}, {de Ravel}, {Franzetti}, {Garilli},
  {Iovino}, {Kampczyk}, {Kovac}, {Knobel}, {Lamareille}, {Le Borgne}, {Pello},
  {Peng}, {P{\'e}rez-Montero}, {Ricciardelli}, {Silverman}, {Tanaka}, {Tasca},
  {Tresse}, {Vergani}, {Zucca}, {Ilbert}, {Salvato}, {Oesch}, {Abbas},
  {Bottini}, {Capak}, {Cappi}, {Cassata}, {Cimatti}, {Elvis}, {Fumana},
  {Guzzo}, {Hasinger}, {Koekemoer}, {Leauthaud}, {Maccagni}, {Marinoni},
  {McCracken}, {Memeo}, {Meneux}, {Porciani}, {Pozzetti}, {Sanders},
  {Scaramella}, {Scarlata}, {Scoville}, {Shopbell}, \& {Taniguchi}}]{lilly09}
{Lilly}, S.~J., {Le Brun}, V., {Maier}, C., {et~al.} 2009, \apjs, 184, 218

\bibitem[{{Lisenfeld} \& {V{\"o}lk}(2000)}]{lisenfeld00}
{Lisenfeld}, U., \& {V{\"o}lk}, H.~J. 2000, \aap, 354, 423

\bibitem[{{Magnelli} {et~al.}(2011){Magnelli}, {Elbaz}, {Chary}, {Dickinson},
  {Le Borgne}, {Frayer}, \& {Willmer}}]{magnelli11}
{Magnelli}, B., {Elbaz}, D., {Chary}, R.~R., {et~al.} 2011, \aap, 528, A35

\bibitem[{{McCracken} {et~al.}(2012){McCracken}, {Milvang-Jensen}, {Dunlop},
  {Franx}, {Fynbo}, {Le F{\`e}vre}, {Holt}, {Caputi}, {Goranova}, {Buitrago},
  {Emerson}, {Freudling}, {Hudelot}, {L{\'o}pez-Sanjuan}, {Magnard}, {Mellier},
  {M{\o}ller}, {Nilsson}, {Sutherland}, {Tasca}, \& {Zabl}}]{mccracken12}
{McCracken}, H.~J., {Milvang-Jensen}, B., {Dunlop}, J., {et~al.} 2012, \aap,
  544, A156

\bibitem[{{McMullin} {et~al.}(2007){McMullin}, {Waters}, {Schiebel}, {Young},
  \& {Golap}}]{mcmullin07}
{McMullin}, J.~P., {Waters}, B., {Schiebel}, D., {Young}, W., \& {Golap}, K.
  2007, in Astronomical Society of the Pacific Conference Series, Vol. 376,
  Astronomical Data Analysis Software and Systems XVI, ed. R.~A. {Shaw},
  F.~{Hill}, \& D.~J. {Bell}, 127

\bibitem[{{Miettinen} {et~al.}(2015){Miettinen}, {Smol{\v c}i{\'c}}, {Novak},
  {Aravena}, {Karim}, {Masters}, {Riechers}, {Bussmann}, {McCracken}, {Ilbert},
  {Bertoldi}, {Capak}, {Feruglio}, {Halliday}, {Kartaltepe}, {Navarrete},
  {Salvato}, {Sanders}, {Schinnerer}, \& {Sheth}}]{miettinen15}
{Miettinen}, O., {Smol{\v c}i{\'c}}, V., {Novak}, M., {et~al.} 2015, \aap, 577,
  A29

\bibitem[{{Miller} {et~al.}(2008){Miller}, {Fomalont}, {Kellermann},
  {Mainieri}, {Norman}, {Padovani}, {Rosati}, \& {Tozzi}}]{miller08}
{Miller}, N.~A., {Fomalont}, E.~B., {Kellermann}, K.~I., {et~al.} 2008, \apjs,
  179, 114

\bibitem[{{Miller} {et~al.}(2013){Miller}, {Bonzini}, {Fomalont}, {Kellermann},
  {Mainieri}, {Padovani}, {Rosati}, {Tozzi}, \& {Vattakunnel}}]{miller13}
{Miller}, N.~A., {Bonzini}, M., {Fomalont}, E.~B., {et~al.} 2013, \apjs, 205,
  13

\bibitem[{{Mooley} {et~al.}(2016){Mooley}, {Hallinan}, {Bourke}, {Horesh},
  {Myers}, {Frail}, {Kulkarni}, {Levitan}, {Kasliwal}, {Cenko}, {Cao}, {Bellm},
  \& {Laher}}]{mooley16}
{Mooley}, K.~P., {Hallinan}, G., {Bourke}, S., {et~al.} 2016, \apj, 818, 105

\bibitem[{{Mullaney} {et~al.}(2011){Mullaney}, {Alexander}, {Goulding}, \&
  {Hickox}}]{mullaney11}
{Mullaney}, J.~R., {Alexander}, D.~M., {Goulding}, A.~D., \& {Hickox}, R.~C.
  2011, \mnras, 414, 1082

\bibitem[{{Mullaney} {et~al.}(2012){Mullaney}, {Daddi}, {B{\'e}thermin},
  {Elbaz}, {Juneau}, {Pannella}, {Sargent}, {Alexander}, \&
  {Hickox}}]{mullaney12}
{Mullaney}, J.~R., {Daddi}, E., {B{\'e}thermin}, M., {et~al.} 2012, \apjl, 753,
  L30

\bibitem[{{Norris} {et~al.}(2015){Norris}, {Basu}, {Brown}, {Carretti},
  {Kapinska}, {Prandoni}, {Rudnick}, \& {Seymour}}]{norris15}
{Norris}, R., {Basu}, K., {Brown}, M., {et~al.} 2015, Advancing Astrophysics
  with the Square Kilometre Array (AASKA14), 86

\bibitem[{{Norris} {et~al.}(2005){Norris}, {Huynh}, {Jackson}, {Boyle},
  {Ekers}, {Mitchell}, {Sault}, {Wieringa}, {Williams}, {Hopkins}, \&
  {Higdon}}]{norris05}
{Norris}, R.~P., {Huynh}, M.~T., {Jackson}, C.~A., {et~al.} 2005, \aj, 130,
  1358

\bibitem[{{Norris} {et~al.}(2011){Norris}, {Hopkins}, {Afonso}, {Brown},
  {Condon}, {Dunne}, {Feain}, {Hollow}, {Jarvis}, {Johnston-Hollitt}, {Lenc},
  {Middelberg}, {Padovani}, {Prandoni}, {Rudnick}, {Seymour}, {Umana},
  {Andernach}, {Alexander}, {Appleton}, {Bacon}, {Banfield}, {Becker}, {Brown},
  {Ciliegi}, {Jackson}, {Eales}, {Edge}, {Gaensler}, {Giovannini}, {Hales},
  {Hancock}, {Huynh}, {Ibar}, {Ivison}, {Kennicutt}, {Kimball}, {Koekemoer},
  {Koribalski}, {L{\'o}pez-S{\'a}nchez}, {Mao}, {Murphy}, {Messias},
  {Pimbblet}, {Raccanelli}, {Randall}, {Reiprich}, {Roseboom},
  {R{\"o}ttgering}, {Saikia}, {Sharp}, {Slee}, {Smail}, {Thompson}, {Urquhart},
  {Wall}, \& {Zhao}}]{norris11}
{Norris}, R.~P., {Hopkins}, A.~M., {Afonso}, J., {et~al.} 2011, PASA, 28, 215

\bibitem[{{Norris} {et~al.}(2013){Norris}, {Afonso}, {Bacon}, {Beck}, {Bell},
  {Beswick}, {Best}, {Bhatnagar}, {Bonafede}, {Brunetti}, {Budav{\'a}ri},
  {Cassano}, {Condon}, {Cress}, {Dabbech}, {Feain}, {Fender}, {Ferrari},
  {Gaensler}, {Giovannini}, {Haverkorn}, {Heald}, {Van der Heyden}, {Hopkins},
  {Jarvis}, {Johnston-Hollitt}, {Kothes}, {Van Langevelde}, {Lazio}, {Mao},
  {Mart{\'{\i}}nez-Sansigre}, {Mary}, {Mcalpine}, {Middelberg}, {Murphy},
  {Padovani}, {Paragi}, {Prandoni}, {Raccanelli}, {Rigby}, {Roseboom},
  {R{\"o}ttgering}, {Sabater}, {Salvato}, {Scaife}, {Schilizzi}, {Seymour},
  {Smith}, {Umana}, {Zhao}, \& {Zinn}}]{norris13}
{Norris}, R.~P., {Afonso}, J., {Bacon}, D., {et~al.} 2013, PASA, 30, 20

\bibitem[{{Novak} {et~al.}(2015){Novak}, {Smol{\v c}i{\'c}}, {Civano}, {Bondi},
  {Ciliegi}, {Wang}, {Loeb}, {Banfield}, {Bourke}, {Elvis}, {Hallinan},
  {Intema}, {Kl{\"o}ckner}, {Mooley}, \& {Navarrete}}]{novak15}
{Novak}, M., {Smol{\v c}i{\'c}}, V., {Civano}, F., {et~al.} 2015, \mnras, 447,
  1282

\bibitem[{{Owen} \& {Morrison}(2008)}]{owen08}
{Owen}, F.~N., \& {Morrison}, G.~E. 2008, \aj, 136, 1889

\bibitem[{{Owen} {et~al.}(2009){Owen}, {Morrison}, {Klimek}, \&
  {Greisen}}]{owen09}
{Owen}, F.~N., {Morrison}, G.~E., {Klimek}, M.~D., \& {Greisen}, E.~W. 2009,
  \aj, 137, 4846

\bibitem[{{Padovani}(2011)}]{padovani11a}
{Padovani}, P. 2011, \mnras, 411, 1547

\bibitem[{{Padovani} {et~al.}(2015){Padovani}, {Bonzini}, {Kellermann},
  {Miller}, {Mainieri}, \& {Tozzi}}]{padovani15}
{Padovani}, P., {Bonzini}, M., {Kellermann}, K.~I., {et~al.} 2015, \mnras, 452,
  1263

\bibitem[{{Padovani} {et~al.}(2009){Padovani}, {Mainieri}, {Tozzi},
  {Kellermann}, {Fomalont}, {Miller}, {Rosati}, \& {Shaver}}]{padovani09}
{Padovani}, P., {Mainieri}, V., {Tozzi}, P., {et~al.} 2009, \apj, 694, 235

\bibitem[{{Prandoni} {et~al.}(2001){Prandoni}, {Gregorini}, {Parma}, {de
  Ruiter}, {Vettolani}, {Wieringa}, \& {Ekers}}]{prandoni01}
{Prandoni}, I., {Gregorini}, L., {Parma}, P., {et~al.} 2001, \aap, 365, 392

\bibitem[{{Prandoni} \& {Seymour}(2015)}]{prandoni15}
{Prandoni}, I., \& {Seymour}, N. 2015, Advancing Astrophysics with the Square
  Kilometre Array (AASKA14), 67

\bibitem[{{Prescott} {et~al.}(2006){Prescott}, {Impey}, {Cool}, \&
  {Scoville}}]{prescott06}
{Prescott}, M.~K.~M., {Impey}, C.~D., {Cool}, R.~J., \& {Scoville}, N.~Z. 2006,
  \apj, 644, 100

\bibitem[{{Rau} \& {Cornwell}(2011)}]{rau11}
{Rau}, U., \& {Cornwell}, T.~J. 2011, \aap, 532, A71

\bibitem[{{Rees}(1967)}]{rees67}
{Rees}, M.~J. 1967, \mnras, 136, 279

\bibitem[{{Rodighiero} {et~al.}(2011){Rodighiero}, {Daddi}, {Baronchelli},
  {Cimatti}, {Renzini}, {Aussel}, {Popesso}, {Lutz}, {Andreani}, {Berta},
  {Cava}, {Elbaz}, {Feltre}, {Fontana}, {F{\"o}rster Schreiber},
  {Franceschini}, {Genzel}, {Grazian}, {Gruppioni}, {Ilbert}, {Le Floch},
  {Magdis}, {Magliocchetti}, {Magnelli}, {Maiolino}, {McCracken}, {Nordon},
  {Poglitsch}, {Santini}, {Pozzi}, {Riguccini}, {Tacconi}, {Wuyts}, \&
  {Zamorani}}]{rodighiero11}
{Rodighiero}, G., {Daddi}, E., {Baronchelli}, I., {et~al.} 2011, \apjl, 739,
  L40

\bibitem[{{Sanders} {et~al.}(2007){Sanders}, {Salvato}, {Aussel}, {Ilbert},
  {Scoville}, {Surace}, {Frayer}, {Sheth}, {Helou}, {Brooke}, {Bhattacharya},
  {Yan}, {Kartaltepe}, {Barnes}, {Blain}, {Calzetti}, {Capak}, {Carilli},
  {Carollo}, {Comastri}, {Daddi}, {Ellis}, {Elvis}, {Fall}, {Franceschini},
  {Giavalisco}, {Hasinger}, {Impey}, {Koekemoer}, {Le F{\`e}vre}, {Lilly},
  {Liu}, {McCracken}, {Mobasher}, {Renzini}, {Rich}, {Schinnerer}, {Shopbell},
  {Taniguchi}, {Thompson}, {Urry}, \& {Williams}}]{sanders07}
{Sanders}, D.~B., {Salvato}, M., {Aussel}, H., {et~al.} 2007, \apjs, 172, 86

\bibitem[{{Sargent} {et~al.}(2012){Sargent}, {B{\'e}thermin}, {Daddi}, \&
  {Elbaz}}]{sargent12}
{Sargent}, M.~T., {B{\'e}thermin}, M., {Daddi}, E., \& {Elbaz}, D. 2012, \apjl,
  747, L31

\bibitem[{{Schinnerer} {et~al.}(2004){Schinnerer}, {Carilli}, {Scoville},
  {Bondi}, {Ciliegi}, {Vettolani}, {Le F{\`e}vre}, {Koekemoer}, {Bertoldi}, \&
  {Impey}}]{schinnerer04}
{Schinnerer}, E., {Carilli}, C.~L., {Scoville}, N.~Z., {et~al.} 2004, \aj, 128,
  1974

\bibitem[{{Schinnerer} {et~al.}(2007){Schinnerer}, {Smol{\v c}i{\'c}},
  {Carilli}, {Bondi}, {Ciliegi}, {Jahnke}, {Scoville}, {Aussel}, {Bertoldi},
  {Blain}, {Impey}, {Koekemoer}, {Le Fevre}, \& {Urry}}]{schinnerer07}
{Schinnerer}, E., {Smol{\v c}i{\'c}}, V., {Carilli}, C.~L., {et~al.} 2007,
  \apjs, 172, 46

\bibitem[{{Schinnerer} {et~al.}(2010){Schinnerer}, {Sargent}, {Bondi}, {Smol{\v
  c}i{\'c}}, {Datta}, {Carilli}, {Bertoldi}, {Blain}, {Ciliegi}, {Koekemoer},
  \& {Scoville}}]{schinnerer10}
{Schinnerer}, E., {Sargent}, M.~T., {Bondi}, M., {et~al.} 2010, \apjs, 188, 384

\bibitem[{{Schmitt}(1985)}]{schmitt85}
{Schmitt}, J.~H.~M.~M. 1985, \apj, 293, 178

\bibitem[{{Scott} {et~al.}(2008){Scott}, {Austermann}, {Perera}, {Wilson},
  {Aretxaga}, {Bock}, {Hughes}, {Kang}, {Kim}, {Mauskopf}, {Sanders},
  {Scoville}, \& {Yun}}]{scott08}
{Scott}, K.~S., {Austermann}, J.~E., {Perera}, T.~A., {et~al.} 2008, \mnras,
  385, 2225

\bibitem[{{Scoville} {et~al.}(2007){Scoville}, {Aussel}, {Brusa}, {Capak},
  {Carollo}, {Elvis}, {Giavalisco}, {Guzzo}, {Hasinger}, {Impey}, {Kneib},
  {LeFevre}, {Lilly}, {Mobasher}, {Renzini}, {Rich}, {Sanders}, {Schinnerer},
  {Schminovich}, {Shopbell}, {Taniguchi}, \& {Tyson}}]{scoville07}
{Scoville}, N., {Aussel}, H., {Brusa}, M., {et~al.} 2007, \apjs, 172, 1

\bibitem[{{Seymour} {et~al.}(2008){Seymour}, {Dwelly}, {Moss}, {McHardy},
  {Zoghbi}, {Rieke}, {Page}, {Hopkins}, \& {Loaring}}]{seymour08}
{Seymour}, N., {Dwelly}, T., {Moss}, D., {et~al.} 2008, \mnras, 386, 1695

\bibitem[{{Smol{\v c}i{\'c}}(2009)}]{smolcic09c}
{Smol{\v c}i{\'c}}, V. 2009, \apjl, 699, L43

\bibitem[{{Smol{\v c}i{\'c}} \& {Riechers}(2011)}]{smolcic11}
{Smol{\v c}i{\'c}}, V., \& {Riechers}, D.~A. 2011, \apj, 730, 64

\bibitem[{{Smol{\v c}i{\'c}} {et~al.}(2008{\natexlab{a}}){Smol{\v c}i{\'c}},
  {Schinnerer}, {Scodeggio}, {Franzetti}, {Aussel}, {Bondi}, {Brusa},
  {Carilli}, {Capak}, {Charlot}, {Ciliegi}, {Ilbert}, {Ivezi{\'c}}, {Jahnke},
  {McCracken}, {Obri{\'c}}, {Salvato}, {Sanders}, {Scoville}, {Trump},
  {Tremonti}, {Tasca}, {Walcher}, \& {Zamorani}}]{smolcic08}
{Smol{\v c}i{\'c}}, V., {Schinnerer}, E., {Scodeggio}, M., {et~al.}
  2008{\natexlab{a}}, \apjs, 177, 14

\bibitem[{{Smol{\v c}i{\'c}} {et~al.}(2008{\natexlab{b}}){Smol{\v c}i{\'c}},
  {Schinnerer}, {Scodeggio}, {Franzetti}, {Aussel}, {Bondi}, {Brusa},
  {Carilli}, {Capak}, {Charlot}, {Ciliegi}, {Ilbert}, {Ivezi{\'c}}, {Jahnke},
  {McCracken}, {Obri{\'c}}, {Salvato}, {Sanders}, {Scoville}, {Trump},
  {Tremonti}, {Tasca}, {Walcher}, \& {Zamorani}}]{smo08}
---. 2008{\natexlab{b}}, \apjs, 177, 14

\bibitem[{{Smol{\v c}i{\'c}} {et~al.}(2009{\natexlab{a}}){Smol{\v c}i{\'c}},
  {Zamorani}, {Schinnerer}, {Bardelli}, {Bondi}, {B{\^i}rzan}, {Carilli},
  {Ciliegi}, {Elvis}, {Impey}, {Koekemoer}, {Merloni}, {Paglione}, {Salvato},
  {Scodeggio}, {Scoville}, \& {Trump}}]{smolcic09b}
{Smol{\v c}i{\'c}}, V., {Zamorani}, G., {Schinnerer}, E., {et~al.}
  2009{\natexlab{a}}, \apj, 696, 24

\bibitem[{{Smol{\v c}i{\'c}} {et~al.}(2009{\natexlab{b}}){Smol{\v c}i{\'c}},
  {Schinnerer}, {Zamorani}, {Bell}, {Bondi}, {Carilli}, {Ciliegi}, {Mobasher},
  {Paglione}, {Scodeggio}, \& {Scoville}}]{smolcic09a}
{Smol{\v c}i{\'c}}, V., {Schinnerer}, E., {Zamorani}, G., {et~al.}
  2009{\natexlab{b}}, \apj, 690, 610

\bibitem[{{Smol{\v c}i{\'c}} {et~al.}(2012){Smol{\v c}i{\'c}}, {Aravena},
  {Navarrete}, {Schinnerer}, {Riechers}, {Bertoldi}, {Feruglio}, {Finoguenov},
  {Salvato}, {Sargent}, {McCracken}, {Albrecht}, {Karim}, {Capak}, {Carilli},
  {Cappelluti}, {Elvis}, {Ilbert}, {Kartaltepe}, {Lilly}, {Sanders}, {Sheth},
  {Scoville}, \& {Taniguchi}}]{smo12}
{Smol{\v c}i{\'c}}, V., {Aravena}, M., {Navarrete}, F., {et~al.} 2012, \aap,
  548, A4

\bibitem[{{Smol{\v c}i{\'c}} {et~al.}(2014){Smol{\v c}i{\'c}}, {Ciliegi},
  {Jeli{\'c}}, {Bondi}, {Schinnerer}, {Carilli}, {Riechers}, {Salvato},
  {Brkovi{\'c}}, {Capak}, {Ilbert}, {Karim}, {McCracken}, \&
  {Scoville}}]{smolcic14}
{Smol{\v c}i{\'c}}, V., {Ciliegi}, P., {Jeli{\'c}}, V., {et~al.} 2014, \mnras,
  443, 2590

\bibitem[{{Smol{\v c}i{\'c}} {et~al.}(2015{\natexlab{a}}){Smol{\v c}i{\'c}},
  {Padovani}, {Delhaize}, {Prandoni}, {Seymour}, {Jarvis}, {Afonso},
  {Magliocchetti}, {Huynh}, {Vaccari}, \& {Karim}}]{smolcic15}
{Smol{\v c}i{\'c}}, V., {Padovani}, P., {Delhaize}, J., {et~al.}
  2015{\natexlab{a}}, Advancing Astrophysics with the Square Kilometre Array
  (AASKA14), 69

\bibitem[{{Smol{\v c}i{\'c}} {et~al.}(2015{\natexlab{b}}){Smol{\v c}i{\'c}},
  {Padovani}, {Delhaize}, {Prandoni}, {Seymour}, {Jarvis}, {Afonso},
  {Magliocchetti}, {Huynh}, {Vaccari}, \& {Karim}}]{smo15}
---. 2015{\natexlab{b}}, Advancing Astrophysics with the Square Kilometre Array
  (AASKA14), 69

\bibitem[{{Tasse} {et~al.}(2007){Tasse}, {R{\"o}ttgering}, {Best}, {Cohen},
  {Pierre}, \& {Wilman}}]{tasse07}
{Tasse}, C., {R{\"o}ttgering}, H.~J.~A., {Best}, P.~N., {et~al.} 2007, \aap,
  471, 1105

\bibitem[{{Trump} {et~al.}(2007){Trump}, {Impey}, {McCarthy}, {Elvis},
  {Huchra}, {Brusa}, {Hasinger}, {Schinnerer}, {Capak}, {Lilly}, \&
  {Scoville}}]{trump07}
{Trump}, J.~R., {Impey}, C.~D., {McCarthy}, P.~J., {et~al.} 2007, \apjs, 172,
  383

\bibitem[{{Vernstrom} {et~al.}(2011){Vernstrom}, {Scott}, \&
  {Wall}}]{vernstrom11}
{Vernstrom}, T., {Scott}, D., \& {Wall}, J.~V. 2011, \mnras, 415, 3641

\bibitem[{{Vernstrom} {et~al.}(2014){Vernstrom}, {Scott}, {Wall}, {Condon},
  {Cotton}, {Fomalont}, {Kellermann}, {Miller}, \& {Perley}}]{vernstrom14}
{Vernstrom}, T., {Scott}, D., {Wall}, J.~V., {et~al.} 2014, \mnras, 440, 2791

\bibitem[{{Wilman} {et~al.}(2008){Wilman}, {Miller}, {Jarvis}, {Mauch},
  {Levrier}, {Abdalla}, {Rawlings}, {Kl{\"o}ckner}, {Obreschkow}, {Olteanu}, \&
  {Young}}]{wilman08}
{Wilman}, R.~J., {Miller}, L., {Jarvis}, M.~J., {et~al.} 2008, \mnras, 388,
  1335

\end{thebibliography}

\begin{table}
\caption{ Catalog sample page}
\centering
\label{tab:cat}
\begin{sideways}
\resizebox{24.5cm}{!}{

\begin{tabular}{c c c c c c c c c c c c c }
\hline
 ID & NAME & RA & RA\_ERR & DEC & DEC\_ERR & FLUX & FLUX\_ERR & RMS & SNR & NPIX & RES & MULTI \\
 & & (J2000, deg) & (arcsec) & (J2000, deg) & (arcsec) & ($\mu$Jy) & ($\mu$Jy) & ($\mu$Jy~beam$^{-1}$) & &  &   &   \\
\hline
  78 & COSMOSVLA3 J095709.33+020940.7 & 149.288886 & 0.01 & 2.161331 & 0.01 & 13400.0 & 670.0 & 28.7 & 385.0 & 104 & 1 & 0\\
  1110 & COSMOSVLA3 J095709.83+015457.4 & 149.290996 & 0.021 & 1.915946 & 0.021 & 1190.0 & 64.0 & 22.1 & 29.0 & 102 & 1 & 0\\
  5144 & COSMOSVLA3 J095710.49+013644.7 & 149.29372 & 0.065 & 1.612418 & 0.065 & 145.0 & 19.0 & 17.4 & 8.32 & 20 & 0 & 0\\
  3749 & COSMOSVLA3 J095710.53+025132.2 & 149.29389 & 0.053 & 2.858966 & 0.053 & 345.0 & 32.0 & 26.8 & 10.4 & 31 & 1 & 0\\
  10099 & COSMOSVLA3 J095710.57+022657.0 & 149.294072 & 0.1 & 2.449174 & 0.1 & 92.0 & 18.0 & 17.4 & 5.3 & 11 & 0 & 0\\
  4979 & COSMOSVLA3 J095710.88+020929.8 & 149.295339 & 0.076 & 2.158297 & 0.076 & 2820.0 & 140.0 & 21.4 & 7.13 & 332 & 1 & 0\\
  3366 & COSMOSVLA3 J095711.09+023031.3 & 149.296231 & 0.048 & 2.508722 & 0.048 & 318.0 & 27.0 & 21.5 & 11.4 & 35 & 1 & 0\\
  8753 & COSMOSVLA3 J095711.15+021104.2 & 149.296459 & 0.095 & 2.184514 & 0.095 & 121.0 & 22.0 & 21.2 & 5.68 & 18 & 0 & 0\\
  4046 & COSMOSVLA3 J095711.64+021236.8 & 149.298529 & 0.055 & 2.210248 & 0.055 & 223.0 & 26.0 & 22.6 & 9.83 & 22 & 0 & 0\\
  6546 & COSMOSVLA3 J095711.67+021401.3 & 149.298655 & 0.077 & 2.233706 & 0.077 & 135.0 & 21.0 & 19.3 & 7.02 & 18 & 0 & 0\\
  ... &  &  &  &  &  &  &  &  &  & & & \\
  10942 & COSMOSVLA3 J100034.76+014635.7 & 150.144846 & -99.0 & 1.776607 & -99.0 & 374.0 & -99.0 & 2.4 & -99.0 & 828 & 1 & 1\\
  9229 & COSMOSVLA3 J100034.78+025027.4 & 150.144947 & 0.1 & 2.840947 & 0.1 & 25.6 & 2.7 & 2.4 & 5.18 & 25 & 1 & 0\\
  8777 & COSMOSVLA3 J100034.80+021421.1 & 150.145009 & 0.095 & 2.239195 & 0.095 & 13.0 & 2.4 & 2.27 & 5.7 & 14 & 0 & 0\\
  489 & COSMOSVLA3 J100034.81+025515.6 & 150.145042 & 0.013 & 2.921023 & 0.013 & 229.0 & 13.0 & 3.66 & 62.5 & 49 & 0 & 0\\
  439 & COSMOSVLA3 J100034.83+014247.1 & 150.145157 & 0.013 & 1.713091 & 0.013 & 162.0 & 9.0 & 2.27 & 71.3 & 51 & 0 & 0\\
  2850 & COSMOSVLA3 J100034.94+020234.9 & 150.145599 & 0.042 & 2.043054 & 0.042 & 29.9 & 2.8 & 2.29 & 13.1 & 29 & 0 & 0\\
  2439 & COSMOSVLA3 J100034.99+024524.5 & 150.145815 & 0.038 & 2.756821 & 0.038 & 32.7 & 2.8 & 2.22 & 14.7 & 29 & 0 & 0\\
  2837 & COSMOSVLA3 J100035.00+024614.3 & 150.145836 & 0.041 & 2.770643 & 0.041 & 39.6 & 2.9 & 2.18 & 13.3 & 41 & 1 & 0\\
  4092 & COSMOSVLA3 J100035.05+024154.6 & 150.146065 & 0.055 & 2.69852 & 0.055 & 28.3 & 2.6 & 2.23 & 9.93 & 31 & 1 & 0\\
  4148 & COSMOSVLA3 J100035.07+020350.5 & 150.14614 & 0.057 & 2.064044 & 0.057 & 23.4 & 2.7 & 2.43 & 9.64 & 23 & 0 & 0\\
  ... &  &  &  &  &  &  &  &  &  & & & \\
  5351 & COSMOSVLA3 J100345.81+015420.5 & 150.940886 & 0.067 & 1.905719 & 0.067 & 117.0 & 16.0 & 14.6 & 8.05 & 18 & 0 & 0\\
  5225 & COSMOSVLA3 J100346.48+023458.3 & 150.943703 & 0.067 & 2.582882 & 0.067 & 127.0 & 17.0 & 15.5 & 8.14 & 21 & 0 & 0\\
  8447 & COSMOSVLA3 J100346.52+022031.2 & 150.943867 & 0.11 & 2.342001 & 0.11 & 290.0 & 20.0 & 14.4 & 5.04 & 48 & 1 & 0\\
  10677 & COSMOSVLA3 J100346.56+015500.1 & 150.944027 & 0.11 & 1.916707 & 0.11 & 75.7 & 15.0 & 14.9 & 5.1 & 11 & 0 & 0\\
  6533 & COSMOSVLA3 J100346.63+022415.8 & 150.944306 & 0.076 & 2.404402 & 0.076 & 143.0 & 16.0 & 14.2 & 7.09 & 28 & 1 & 0\\
  2195 & COSMOSVLA3 J100347.12+022510.6 & 150.946345 & 0.036 & 2.419631 & 0.036 & 548.0 & 31.0 & 15.3 & 15.6 & 83 & 1 & 0\\
  10084 & COSMOSVLA3 J100347.27+020117.6 & 150.94699 & 0.1 & 2.021575 & 0.1 & 73.1 & 15.0 & 14.1 & 5.18 & 16 & 0 & 0\\
  9695 & COSMOSVLA3 J100347.51+023539.7 & 150.947979 & 0.1 & 2.594384 & 0.1 & 93.2 & 18.0 & 17.6 & 5.3 & 10 & 0 & 0\\
  10652 & COSMOSVLA3 J100347.80+024951.9 & 150.949175 & 0.11 & 2.831096 & 0.11 & 80.7 & 17.0 & 16.1 & 5.01 & 14 & 0 & 0\\
  10484 & COSMOSVLA3 J100348.53+021102.3 & 150.952242 & 0.11 & 2.183987 & 0.11 & 82.3 & 17.0 & 16.3 & 5.05 & 13 & 0 & 0\\
 \hline
 \end{tabular}

}

\end{sideways}
\end{table}

\end{document}